\documentclass[]{amsart}
\usepackage[utf8]{inputenc}
\usepackage[a4paper, margin=1in]{geometry}
\usepackage{bbm,amsmath,amsfonts,amssymb,amsthm,amscd,amsbsy,mathabx,epsfig,array,mathrsfs,tikz,tikz-cd,url, graphicx, subcaption,csquotes,algorithm,autobreak}
\usepackage[colorlinks=false,citecolor=green,linktoc=page]{hyperref}

\usepackage{tikz}
\usetikzlibrary{trees,calc,decorations.markings,decorations.pathmorphing}

\theoremstyle{thmstyleone}%
\newtheorem{theorem}{Theorem}
\newtheorem{proposition}[theorem]{Proposition}%
\newtheorem{lemma}[theorem]{Lemma}
\newtheorem{corollary}[theorem]{Corollary}

\theoremstyle{thmstyletwo}%
\newtheorem{example}{Example}%

\theoremstyle{thmstylethree}%
\newtheorem{definition}{Definition}%

\numberwithin{equation}{section}

\title[Cutkosky's Theorem for Massive One-Loop Feynman Integrals - Part 1]{Cutkosky's Theorem for Massive One-Loop Feynman Integrals - Part 1}

\author{Maximilian M{\"u}hlbauer}

\begin{document}
\tikzset{
    photon/.style={decorate, decoration={snake}, draw=black},
    electron/.style={draw=black, postaction={decorate},
        decoration={markings,mark=at position .55 with {\arrow[draw=black]{>}}}},
    gluon/.style={decorate, draw=black,
        decoration={coil,amplitude=4pt, segment length=5pt}}
}

\begin{abstract}
    We formulate and prove Cutkosky's Theorem regarding the discontinuity of Feynman integrals in the massive one-loop case up to the involved intersection index. This is done by applying the techniques to treat singular integrals developed in \cite{app-iso}. We write one-loop integrals as an integral of a holomorphic family of holomorphic forms over a compact cycle. Then, we determine at which points simple pinches occur and explicitly compute a representative of the corresponding vanishing sphere. This also yields an algorithm to compute the Landau surface of a one-loop graph without explicitly solving the Landau equations. We also discuss the bubble, triangle and box graph in detail.
\end{abstract}

\maketitle
\tableofcontents

\section{Introduction}\label{sec1}

One of the primary tasks in perturbative quantum field theory is the computation of Feynman integrals. These are integrals associated with graphs, essentially given by quadratic functions
\begin{equation}
    Q_1(p,m),\ldots,Q_n(p,m):\mathbb{C}^{LD}\to\mathbb{C},
\end{equation}
with $n$ the number of edges and $L$ the number of independent cycles of the underlying graph, which themselves depend (quadratically) on physical parameters (the external momenta $p$ and masses $m$). The integrals under consideration are then of the form
\begin{equation}\label{eq:feynman_integral_intro}
    \left(\prod_{i=1}^L\int_{i\mathbb{R}\times\mathbb{R}^{D-1}}\textrm{d}^Dk_i\right)\frac{1}{\prod_{i=1}^n(Q_i(p,m)(k_1,\ldots,k_L))^{\lambda_i}}
\end{equation}
and, if they are well-defined at all,\footnote{We refer here to the convergence of the integral in question. It should be remarked that, as it stands, the expression \eqref{eq:feynman_integral_intro} is generally ill-defined, even when it converges: The integral need not be absolutely convergent so that Fubini's Theorem does not apply. Thus, one needs to agree on the order of integration to make sense of \eqref{eq:feynman_integral_intro}.} define functions in the parameters $p$ and $m$. Even in simple cases, it is unfortunately very difficult (though sometimes possible, see for example the database Loopedia \cite{loopedia} for a collection of available results or the articles \cite{scalaroneloop} and \cite{twoloopsunrise} for some calculations in action) to express such an integral in terms of well-understood mathematical functions. Therefore, there is a growing need to understand the properties of functions defined by Feynman integrals in their own right. To this end, the benefits of considering a Feynman integral as a function of general complex momenta instead of just physical Minkowski momenta (living in $i\mathbb{R}\times\mathbb{R}^{D-1}$ in our setup) have been discovered long ago. Among many other advantages, this allows us to apply the plethora of elegant techniques from complex analysis to the problem.\\
Leaving aside the problem of renormalization (which we conveniently sidestep in this work by using analytic regulators $\lambda_1,\ldots,\lambda_n\in\mathbb{C}$) posed by the fact that the integral \eqref{eq:feynman_integral_intro} might be ill-defined for every value of $(p,m)$ even when integrating over $(\mathbb{R}^D)^L$ instead of $(i\mathbb{R}\times\mathbb{R}^{D-1})^L$, the expression \eqref{eq:feynman_integral_intro} does certainly not make sense for all possible complex values of $(p,m)$. Leaving the masses $m$ fixed and positive (as we do throughout this text), we can, however, find an open neighborhood $U$ in the complex space of external momenta where the expression is well-defined and yields in fact a holomorphic function on $U$. This immediately leads to the question along which paths this function can be analytically continued and what the result of such a continuation is. In other words, we seek to understand the analytic structure of functions defined by Feynman integrals.\\
To answer questions of this nature, the authors of \cite{app-iso} developed a framework, later expanded by Pham in \cite{pham_old} and \cite{pham}, to deal with integrals of the form
\begin{equation}\label{eq:standard_form_intro}
    \int_{\Gamma}\omega(t).
\end{equation}
\newline
This is to be understood as follows: Suppose $X$ and $T$ are two complex analytic manifolds. Denote by $\pi:X\times T\twoheadrightarrow T$ the canonical projection. We assume there is an analytic subset $S\subset X\times T$ such that every fiber $S_t:=\pi^{-1}(t)\cap S$ (naturally viewed as a subset of $X$) is a codimension 1 analytic subset of $X$. We take $\omega(t)$ to be a holomorphic $n$-form on $X\backslash S_t$ holomorphically dependent on $t\in T$ (a notion which we define precisely in the preliminaries in Section \ref{sec:preliminaries}) and $\Gamma$ to be a compact $n$-cycle in $X\backslash S_{t_0}$ for some fixed $t_0\in T$. The authors of \cite{app-iso} were able to show that in good cases the integral \eqref{eq:standard_form_intro} defines a holomorphic function in an open neighborhood of $t_0$ which can be analytically continued along any path in $T$ which does not meet a certain analytic subset $L\subset T$ of codimension 1. This $L$ is called the \textit{Landau surface} of the integral \eqref{eq:standard_form_intro}. The discontinuity of such functions along simple loops around points $t\in L$ of codimension 1 at certain $t'\in T\backslash L$ close to $t$ can then be computed to be
\begin{equation}
    N\int_{\tilde{e}}\omega(t')=(2\pi i)^mN\int_e\text{Res}^m\,\omega(t'),
\end{equation}
where $\tilde{e}$ (resp. $e$) is an $n$-cycle in $X\backslash S_{t'}$ (resp. ($n-m$)-cycle in $S_{t'}$) called the vanishing sphere (resp. vanishing cycle) which can be computed by means of the local geometry of $S_{t'}$ alone and $\text{Res}^m$ is the iterated Leray residue. Here, $N$ is an integer determined by the intersection index of the integration cycle at $t'$ with a homology class in $X$ relative to $S_{t'}$ associated with $\tilde{e}$ and $e$.\\
We want to apply this program to a class of simple integrals of the form \eqref{eq:feynman_integral_intro}, namely integrals associated with one-loop Feynman graphs. The goal is to prove Cutkosky's Theorem, first stated in \cite{cutkosky}, in this case. It describes the discontinuity of functions defined by Feynman integrals in terms of simpler integrals. Suppose we consider the case $\lambda_1=\cdots=\lambda_n=1$. Then, conjecturally the discontinuity along certain loops around a problematic point $p$ in Minkowski space evaluated at points $p'$  close to $p$ is given by the formula
\begin{equation}
    (2\pi i)^{\vert C \vert}\left(\prod_{i=1}^L\int_{i\mathbb{R}\times\mathbb{R}^{D-1}}\textrm{d}^Dk_i\right)\frac{\prod_{e\in C}\delta_+(Q_e(p',m)(k))}{\prod_{e\in E(G)-C}Q_e(p',m)(k)}
\end{equation}
with $C$ a certain subset of edges. Here, the integration over $\delta_+(Q_e(p',m)(k))$ means integrating the residue of $Q_e(p',m)$ along its positive energy part. Unfortunately, this is not yet a mathematically precise statement and correspondingly there does not seem to be a rigorous proof of this statement anywhere in the literature. The only attempt the author is aware of is the paper \cite{outer-space} by Bloch and Kreimer which is still work in progress and has not been published yet. The paper does not discuss the compactifiction of the integration cycle and the ambient space which is necessary to apply the techniques from \cite{app-iso}. But even in the one-loop case, where this task does not immediately lead to substantial problems, achieving this compactification in a satisfying way is not at all a trivial task as is extensively discussed here. It is far from obvious how this can be done in the multi-loop case, if it can be done at all. Furthermore, the discussion in \cite{outer-space} chooses a kinematic configuration of Minkowski momenta as the starting point for an analytic continuation. While from the perspective of physics, this might be more satisfying than starting at Euclidean momenta (as is done here), potentially severe mathematical issues arise from the Minkowski setup: To move the poles of the propagators off the integration domain, the $i\epsilon$-prescription must be employed. But \enquote{at infinity,} the $i\epsilon$-term vanishes, so any naive compactification fails immediately and something more has to be done.\\
It should also be mentioned that while \cite{cutkosky} discusses the general case of discontinuities across arbitrary branch cuts (equation (6) in the paper in question), which we aim to prove in this work, the paper also contains a separate discussion of a special case (equation (17) in \cite{cutkosky}).\footnote{The author would like to thank the reviewer who pointed this out to him.} This later case is predominantly used in physics and concerns the discontinuity along so called \enquote{normal thresholds}, arising from sets of edges who's removal results in a graph with two connected components. It is used to compute the imaginary part of transition amplitudes and can also be derived from more physical considerations relating to the $S$-matrix.\\
The plan of this paper is as follows: First, we modify Feynman integrals to fit the form \eqref{eq:standard_form_intro}. The main obstacle is the non-compactness of the integration domain and the complex ambient space $\mathbb{C}^{LD}$ it lives in. From this new representation of one-loop Feynman integrals, we derive the well-known fact that the Landau surface $L$ is given by the Landau equations.\footnote{It should be remarked that although this is common knowledge for physicists, a mathematical proof does not seem to be available. For an as of now unpublished attempt in the multiloop case, see \cite{max}.} We then proceed to establish that, outside of a small set of pathological external momenta, the integral is behaved well enough for the techniques above to apply (more precisely: all relevant pinches are simple pinches). Therefore, the vanishing sphere and cell are defined and we compute them explicitly. Finally, putting all these ingredients together, we prove Cutkosky's Theorem for one-loop graphs up to the yet undetermined intersection index. The computation of the latter is postponed to the second part of this work as it involves an array of quite different techniques than the ones employed here.\\
The paper has the following structure: In Section \ref{sec:preliminaries} we review all the necessary concepts and techniques needed to state and proof our version of Cutkosky's Theorem. This includes remarks on real and complex projective space, some basics from the theory of sheaves (Subsection \ref{subsec:sheaves}) and the theory of singular integrals as developed by Pham et. al. (Subsection \ref{subsec:singular_integrals}). The theory of Feynman integrals is quickly reviewed in Section \ref{sec:feynman_integrals}. The subsequent Section \ref{sec:one_loop} contains the main part of this paper, leading to a statement and proof of Cutkosky's Theorem for one-loop graphs. The proof also yields an algorithm to compute the Landau surface without solving the Landau equations explicitly. To the authors knowledge, this algorithm is new. After the results for general one-loop graphs are established, we look at two advanced examples, the triangle and the box graph, in detail in Section \ref{sec:examples}. In the concluding Section \ref{sec:conclusion}, we comment on the relevance of these results and ideas for future continuation of this work.

\section{Preliminaries}\label{sec:preliminaries}
Before diving into the details regarding Feynman integrals and Cutkosky's Theorem, we introduce some amount of known theory for the convenience of the reader. In Subsection \ref{subsec:projective_space}, we start with some elementary properties of real (resp. complex) projective space viewed as a real analytic (resp. complex analytic) manifold. After that, we recap some basics from the theory of sheaves, in particular the theory of local systems, in Subsection \ref{subsec:sheaves}. These provide a convenient language to talk about the theory of singular integrals as initiated in \cite{app-iso} and developed further by Pham in \cite{pham_old} and \cite{pham}, which we review in some detail in Subsection \ref{subsec:singular_integrals}.

\subsection{Some Aspects of Projective Space}\label{subsec:projective_space}
A considerable amount of this work's content relies on various properties of the real and complex projective spaces $\mathbb{R}\mathbb{P}^n$ and $\mathbb{C}\mathbb{P}^n$. Therefore, in this subsection we recall some of their properties that we make use of frequently throughout this work.\\
Let $\mathbb{K}\in\{\mathbb{R},\mathbb{C}\}$. First, recall that the $n$-dimensional projective space $\mathbb{K}\mathbb{P}^n$ viewed as a set can be defined as the quotient of $\mathbb{K}^{n+1}\backslash\{0\}$ by the equivalence relation
\begin{equation}
    x\sim y\quad:\Leftrightarrow\quad \exists\lambda\in\mathbb{K}^\times\;:\;x=\lambda\cdot y.
\end{equation}
The equivalence class of any $x=(x_0,\ldots,x_n)\in\mathbb{K}^{n+1}\backslash\{0\}$ with respect to this equivalence relation is denoted by $[x]=[x_0:\cdots:x_n]$. The quotient space comes with a natural projection
\begin{equation}\label{eq:projective_space_projection}
    \pi_\mathbb{K}:\mathbb{K}^{n+1}\backslash\{0\}\twoheadrightarrow\mathbb{K}\mathbb{P}^n,\quad (x_0,\ldots,x_n)\mapsto[x_0:\cdots:x_n].
\end{equation}
As a topological space $\mathbb{K}\mathbb{P}^n$ is equipped with the induced quotient topology. Additionally, $\mathbb{R}\mathbb{P}^n$ is a real analytic manifold of dimension $n$ and $\mathbb{C}\mathbb{P}^n$ is a complex analytic manifold of (complex) dimension $n$. For convenience of language, we simply use the term \textit{$\mathbb{K}$-analytic} to mean real analytic if $\mathbb{K}=\mathbb{R}$ and complex analytic if $\mathbb{K}=\mathbb{C}$. Both $\mathbb{R}\mathbb{P}^n$ and $\mathbb{C}\mathbb{P}^n$ can be covered by $n+1$ charts, namely
\begin{equation}
    \varphi_{i,\mathbb{K}}:U_{i,\mathbb{K}}:=\{[x]\in\mathbb{K}\mathbb{P}^n \;\vert\; x_i\neq0\}\overset{\sim}{\to}\mathbb{K}^n,\quad [x]\mapsto\left(\frac{x_0}{x_i},\ldots,\widehat{\frac{x_i}{x_i}},\ldots,\frac{x_n}{x_i}\right)
\end{equation}
for $i\in\{0,\ldots,n\}$. The transition maps are $\mathbb{K}$-analytic, and the projection $\pi_\mathbb{K}$ is $\mathbb{K}$-analytic with respect to the induced manifold structure. While $\mathbb{C}\mathbb{P}^n$ is always orientable (as it is a complex manifold), the manifold $\mathbb{R}\mathbb{P}^n$ is orientable if and only if $n$ is odd. We denote
\begin{equation}
    H_{\infty,\mathbb{K}}:=\mathbb{K}\mathbb{P}^n\backslash U_{0,\mathbb{K}}
\end{equation}
for $\mathbb{K}\in\{\mathbb{R},\mathbb{C}\}$ which we call the \textit{hyperplane at infinity}.\footnote{Of course, our choice to single out the 0th coordinate here is arbitrary and simply a matter of taste. In principal, any hyperplane could be chosen to be the one \enquote{at infinity.}} The reason for this terminology is the fact that $\mathbb{K}^n$ embeds into $\mathbb{K}\mathbb{P}^n$ via the inclusion
\begin{equation}
    i:\mathbb{K}^n\hookrightarrow\mathbb{K}\mathbb{P}^n,\quad z\mapsto[1:z].
\end{equation}
This defines a diffeomorphism (for $\mathbb{K}=\mathbb{R}$) or a biholomorphic map (for $\mathbb{K}=\mathbb{C}$) $\mathbb{K}^n\overset{\sim}\to\mathbb{K}\mathbb{P}^n\backslash H_\infty=U_{0,\mathbb{K}}$ with inverse $\varphi_{0,\mathbb{K}}$. The remaining points in $H_\infty$ not in the image of $i$ can be thought of as additional points added \enquote{at infinity}. If it is clear from context if we mean $\mathbb{K}=\mathbb{R}$ or $\mathbb{K}=\mathbb{C}$, we simply write $\varphi_i:=\varphi_{i,\mathbb{K}}$, $H_\infty:=H_{\infty,\mathbb{K}}$ and so on.\\
It is often useful to work in $\mathbb{K}^{n+1}\backslash\{0\}$ instead of $\mathbb{K}\mathbb{P}^n$ and then, infer desired results by passing to the quotient using the projection $\pi_\mathbb{K}$. The following useful proposition is an example of a result which allows such an inference.
\begin{proposition}\label{prop:projective_submanifold}
    Let $f:\mathbb{K}^{n+1}\backslash\{0\}\to\mathbb{K}$ be a $\mathbb{K}$-analytic, homogeneous function and $p\in\mathbb{K}$ a regular value of $f$. Then, $\pi_\mathbb{K}(f^{-1}(p))$ is a $\mathbb{K}$-analytic submanifold of $\mathbb{K}\mathbb{P}^n$.
\end{proposition}
\begin{proof}
    The Lie group $\mathbb{K}^\times$ acts on $\mathbb{K}^{n+1}\backslash\{0\}$ by sending $x$ to $\lambda x$ for every $x\in\mathbb{K}^{n+1}\backslash\{0\}$ and $\lambda\in\mathbb{K}^\times$. It is not difficult to see that this action is $\mathbb{K}$-analytic, free and proper. Thus, by the Quotient Manifold Theorem, we see that the quotient of $f^{-1}(p)$ (which is a $\mathbb{K}$-analytic manifold by the Regular Value Theorem) by this action is a $\mathbb{K}$-analytic manifold itself.
\end{proof}
For our purposes, we need the pull-back of the canonical $n$-form $d^nz$ on $\mathbb{C}^n$ by $\varphi_{0,\mathbb{C}}$. This is computed in the following
\begin{lemma}\label{lem:diff_form_lemma_1}
    We have
    \begin{equation}\label{eq:diff_form_lemma_1}
        \varphi_{0,\mathbb{C}}^\ast\mathrm{d}^nz=\frac{1}{z_0^{n+1}}\sum_{i=0}^n(-1)^iz_i\mathrm{d}z_0\wedge\cdots\wedge\widehat{\mathrm{d}z_i}\wedge\cdots\wedge \mathrm{d}z_n.
    \end{equation}
\end{lemma}
\begin{proof}
    Obtained by a straightforward calculation.
\end{proof}
Note that \eqref{eq:diff_form_lemma_1} extends only to a meromorphic $n$-form on $\mathbb{C}\mathbb{P}^n$. The pullback of $\mathrm{d}^nz$ by $\varphi_{0,\mathbb{C}}$ cannot be extended to a holomorphic form: The manifold $\mathbb{R}\mathbb{P}^n$ is compact $n$-cycle in $\mathbb{C}\mathbb{P}^n$, so integrating a holomorphic $n$-form over it yields a finite result. But since removing the set $H_\infty$ of measure zero from $\mathbb{R}\mathbb{P}^n$ and choosing inhomogeneous coordinates yields $\int_{\mathbb{R}\mathbb{P}^n-H_\infty}\varphi_{0,\mathbb{C}}^\ast \mathrm{d}^nz=\int_{\mathbb{R}^n}\mathrm{d}^nz=\infty$, this cannot be correct.

\subsection{Some Basics on Sheaves and Monodromy}\label{subsec:sheaves}
The central aim of this work is to understand the multivaluedness of holomorphic functions defined by Feynman integrals. The theory of sheaves, particularly the theory of local systems, provides a convenient framework to formulate and investigate such questions. We start by recalling some of the basic notions of sheaf theory together with the theorems that are relevant to us. Everything contained in this subsection is rather elementary and well-understood. All statements presented here can be found in standard textbooks on the subject, and we recommend \cite{tennison} or \cite{sheavesonmanifolds}. For details on local systems in particular, the reader is referred to \cite{localsystems}. But since the objects of investigation in this text are Feynman integrals which are mainly studied by physicists, the author decided to include this material for the convenience of the reader.\\
First recall that a \textit{pre-ordered set} or \textit{poset} $(X,\prec)$ is a set $X$ together with a relation $\prec\subset X\times X$ on $X$ such that $x\prec x$ as well as $x\prec y\land y\prec x\;\Rightarrow\; x=y$ and $x\prec y\land y\prec z\;\Rightarrow\;x\prec z$ for all $x,y,z\in X$ (i.e. $\prec$ is reflexive, anti-symmetric and transitive). Now, recall that to any pre-ordered set $(X,\prec)$ we can associate a category by taking the objects to be the elements of $X$ and for any $x,y\in X$ taking the set of morphisms $\text{Hom}(x,y)$ to contain one element if $x\prec y$ and be empty otherwise. For any topological space $X$, the open subsets of $X$ form a pre-ordered set with respect to the subset-relation. The associated category of this pre-ordered set is denoted by $\text{Ouv}_X$ and can be thought of as a subcategory of $\textbf{Top}$, the category of topological spaces and continuous maps, with objects the open subsets $U$ of $X$ and morphisms the natural inclusion maps. A \textit{pre-sheaf} on $X$ with values in a category $\mathcal{C}$ is a functor $\mathcal{F}:\text{Ouv}_X^\text{op}\to\mathcal{C}$, where $\mathcal{D}^\text{op}$ denotes the opposite category of $\mathcal{D}$ for any category $\mathcal{D}$. For any two open sets $V\subset U$ with $i:V\hookrightarrow U$ the natural inclusion and any $s\in\mathcal{F}(U)$, it is customary to denote $s\vert_V:=\mathcal{F}(i)(s)$ called the \textit{restriction} of $s$ to $V$. A morphism of pre-sheaves is just a natural transformation between the functors. The \textit{stalk} $\mathcal{F}_x$ of $\mathcal{F}$ at $x\in X$ is
\begin{equation}
    \mathcal{F}_x:=\lim_{U\ni x}\mathcal{F}(U)
\end{equation}
where the limit is over all open neighborhoods $U$ of $x$. The elements of $\mathcal{F}_x$ are called \textit{germs} of $\mathcal{F}$ at $x$. Given an element $s\in\mathcal{F}(U)$, we can thus speak of the germ of $s$ at $x$. A pre-sheaf $\mathcal{F}$ on $X$ is called a \textit{sheaf} on $X$ if for every $U\in X$ and every open cover $\{U_i\}_{i\in I}$ of $U$ the diagram
\begin{equation}
    \mathcal{F}(U)\overset{\prod_{i\in I}\mathcal{F}(\tau_i)}\to\prod_{i\in I}\mathcal{F}(U_i)\overset{\prod_{i,j\in I}\mathcal{F}(\tau_j^i)}{\underset{\prod_{i,j\in I}\mathcal{F}(\tau_i^j)}\rightrightarrows}\prod_{i,j\in I}\mathcal{F}(U_i\cap U_j)
\end{equation}
is an equaliser diagram. Here, $\tau_i:U_i\hookrightarrow U$ and $\tau^i_j:U_i\cap U_j\hookrightarrow U_i$ are the natural inclusions. Recall that this means that $\prod_{i\in I}\mathcal{F}(\tau_i)$ is injective and that
\begin{equation}
    \text{im}\left(\prod_{i\in I}\mathcal{F}(\tau_i)\right)=\left\{x\in\prod_{i\in I}\mathcal{F}(U_i) \;\vert\; \prod_{i,j\in I}\mathcal{F}(\tau^i_j)(x)=\prod_{i,j\in I}\mathcal{F}(\tau^j_i)(x)\right\}.
\end{equation}
More concretely, this means two things: First, for any open covering $\{U_i\}_{i\in I}$ of an open set $U$, if $s,t\in\mathcal{F}(U)$ satisfy $s\vert_{U_i}=t\vert_{U_i}$ for all $i\in I$ we have $s=t$ (\textit{Locality}). Second, if $\{s_i\}_{i\in I}$ are elements $s_i\in\mathcal{F}(U_i)$ such that $s_i\vert_{U_i\cap U_j}=s_j\vert_{U_i\cap U_j}$ for all $i,j\in I$, there exists a section $s\in\mathcal{F}(U)$ such that $s\vert_{U_i}=s_i$ for all $i\in I$ (\textit{Gluing}). For any open set $U\subset X$, the elements of $\mathcal{F}(U)$ are called the \textit{(local) sections} over $U$. In particular, the elements of $\mathcal{F}(X)$ are called \textit{global sections}. A morphism of sheaves is a morphism of the underlying pre-sheaves.\\
For every pre-sheaf $\mathcal{F}$ on $X$, we can construct a sheaf $\mathcal{F}^+$ on $X$ called the \textit{sheafification} of $\mathcal{F}$ together with a morphism $\theta:\mathcal{F}\to\mathcal{F}^+$, which satisfies the following universal property: Given any sheaf $\mathcal{G}$ on $X$ and any morphism of sheaves $\phi:\mathcal{F}\to\mathcal{G}$, there exists a unique morphism $\psi:\mathcal{F}^+\to\mathcal{G}$ such that $\phi=\psi\circ\theta$. One possibility to construct $\mathcal{F}^+$ is to set
\begin{equation}
    \mathcal{F}^+(U):=\left\{f:U\to\bigcup_{x\in U}\mathcal{F}_x \;\vert\; \substack{\forall x\in X\;:\;f(x)\in\mathcal{F}_x \\ \forall x\in U\;:\;\exists V\text{ ngh. of }x\;:\;\exists g\in\mathcal{F}(V)\;:\;\forall y\in V\;:\;g_y=f(y)}\right\}
\end{equation}
for every open $U\subset X$. Let $Y$ be another topological space and $f:X\to Y$ a continuous map. Given a sheaf $\mathcal{F}$ on $Y$, the \textit{inverse image sheaf} $f^{-1}\mathcal{F}$ of $\mathcal{F}$ by $f$ is the sheafification of the pre-sheaf given by
\begin{equation}
    U\mapsto\lim_{V\supset f(U)}\mathcal{F}(V),
\end{equation}
where the limit is taken over all open $V\subset X$ containing $f(U)$. Let $X$ be a topological space and $\mathcal{C}$ a category. Let $\mathcal{F}$ be a sheaf on $X$ with values in $\mathcal{C}$. Then, for any subspace $Y\subset X$, there is a sheaf $\mathcal{F}\vert_Y$ called the \textit{restriction of $\mathcal{F}$ to $Y$} defined by $\mathcal{F}\vert_Y:=i^{-1}\mathcal{F}$, where $i:Y\hookrightarrow X$ is the natural inclusion map.
For any object $A$ in $\mathcal{C}$, there is a \textit{constant pre-sheaf associated with $A$} defined by sending each open set to $A$ and each morphism to $\text{id}_A$. The sheafification of this pre-sheaf is called the \textit{constant sheaf associated with $A$}. A sheaf $\mathcal{F}$ on $X$ is called a \textit{local system} or a \textit{locally constant sheaf} if for any $x\in X$ there is an open neighborhood $U\subset X$ of $x$ such that $\mathcal{F}\vert_U$ is a constant sheaf. The inverse image sheaf of any (locally) constant sheaf is (locally) constant \cite{localsystems}.
\begin{proposition}[\cite{localsystems}]\label{prop:locally_constant_sheaf}
    Any locally constant sheaf on a contractible space is constant.
\end{proposition}
Local systems on a path-connected space $X$ with fiber $F$ (i.e., a local system which is locally isomorphic to the constant sheaf associated with $F$) are in a bijective correspondence to homomorphisms from the fundamental group of $X$ to the automorphism group of $F$. This correspondence can be established as follows: Let $\mathcal{F}$ be a local system on $X$, let $x\in X$ and let $\gamma:[0,1]\to X$ be a loop based at $x$. Then, the inverse image sheaf $\gamma^{-1}\mathcal{F}$ is a local system on $[0,1]$ and since $[0,1]$ is contractible, this is a constant sheaf by Proposition \ref{prop:locally_constant_sheaf}. Thus,
\begin{equation}
    F\simeq(\gamma^{-1}\mathcal{F})_0\simeq\gamma^{-1}\mathcal{F}([0,1])\simeq(\gamma^{-1}\mathcal{F})_1\simeq F
\end{equation}
and we obtain an automorphism of $F$, called the \textit{monodromy along $\gamma$}, by composing the above isomorphisms. It can be shown that this automorphism depends only on the homotopy class $[\gamma]\in\pi_1(X,x)$ of $\gamma$ (see \cite{localsystems}). The other way around, suppose we are given a homomorphism $\rho:\pi_1(X,x)\to\text{Aut}(F)$. Let $\tilde{\mathcal{F}}$ be the constant sheaf associated with $F$ on the universal covering space $\tilde{X}$ of $X$. Then, the sections of $\tilde{\mathcal{F}}$ invariant under deck-transformations form a local system on $X$. It is not difficult to see that these two operations are inverse to each other.\\
We also need the notion of a multivalued section:
\begin{definition}
    Let $\mathcal{F}$ be a sheaf on a topological space $X$ and $\pi:\tilde{X}\to X$ the universal covering. A \textit{multivalued section} of $\mathcal{F}$ is a section in the (constant) inverse image sheaf $\pi^{-1}\mathcal{F}$.
\end{definition}
On well-behaved topological spaces, local sections of local systems can always be extended to multivalued global sections:
\begin{proposition}[\cite{pham}]\label{prop:extension_of_local_sections}
    Let $X$ be a locally connected topological space and $\mathcal{F}$ a sheaf on $X$. If $\mathcal{F}$ is locally constant, then every local section of $\mathcal{F}$ can be extended to a multi-valued global section of $\mathcal{F}$.
\end{proposition}
A class of sheaves which bares particular importance to us is given in the following 
\begin{definition}\label{defn:homology_sheaf_over_T}
    Let $Y$ and $T$ be smooth manifolds and $\pi:Y\to T$ a smooth map. The \textit{homology sheaf in degree $p$ of $Y$ over $T$}, denoted by $\mathcal{F}^p_{Y/T}$, is the sheafification of the pre-sheaf
    \begin{equation}
        U\mapsto H_{p+\dim T}(Y,\pi^{-1}(T-U)).
    \end{equation}
    For any section $h$ of $\mathcal{F}_{Y/T}^p$, we denote its germ at $t$ by $h(t)$.
\end{definition} 
\begin{proposition}[\cite{pham}]\label{prop:homology_sheaf}
     Let $Y,T$ and $\pi:Y\to T$ as in Definition \ref{defn:homology_sheaf_over_T}. Then, for every $t\in T$ and every $p\in\mathbb{N}$, there are isomorphisms
    \begin{equation}
        \begin{split}
            H_p(\pi^{-1}(t))&\simeq H_{p+\dim T}(Y,\pi^{-1}(T-\{t\}))\\
            &\simeq\lim_{U\ni t}H_{p+\dim T}(Y,\pi^{-1}(T-U))=(\mathcal{F}_{Y/T}^p)_t,
        \end{split}
    \end{equation}
    where the limit is taken over all open neighborhoods of $t$.
\end{proposition}
When we consider germs of sections of $\mathcal{F}^p_{Y/T}$, we usually apply the isomorphism from Proposition \ref{prop:homology_sheaf} above implicitly without mention as long as no confusion can arise.\\
The sheaves from Definition \ref{defn:homology_sheaf_over_T} play an important role for the analytical continuation of functions defined by integrals in our setup. In fact, the extension of local sections to larger domains in these sheaves corresponds directly to analytic continuations as we will see in the next Subsection.

\subsection{Singularities of Integrals}\label{subsec:singular_integrals}
We want to understand Feynman integrals, revisited in some detail in the following Section \ref{sec:feynman_integrals}, as holomorphic functions in the external momenta (and possibly the masses of the virtual particles). To do so, we employ the framework from \cite{app-iso} which deals with the analytic properties of functions defined by integrals in a sufficiently general manner. We begin by revisiting the basic ideas of this framework. Let $Y$ and $T$ be two complex analytic manifolds of dimension $n+m$ and $m$, respectively, and let $\pi:Y\to T$ be a smooth submersion. We denote the fibers by $Y_t:=\pi^{-1}(t)$ for all $t\in T$. By the Implicit Function Theorem, there exists for every $y\in Y$ a coordinate neighborhood $U\subset Y$ of $y$ with coordinates
\begin{equation}\label{eq:local_coordinates}
    \varphi:=(x_1,\ldots,x_n,t_1,\ldots,t_m)\;:\;U\to\mathbb{C}^{n+m}
\end{equation}
such that there is a chart $(V,\psi)$ of $T$ around $\pi(y)$ with $\psi\circ\pi=(t_1,\ldots,t_m)$. In particular, the fibers $Y_t$ are smooth manifolds for all $t\in T$. This can be used to define families of differential forms on the fibers $Y_t$ which depend holomorphically on $t$. We use the common multi-index notation for wedge products: For any
\begin{equation}
    I=\{i_1,\ldots,i_p\}\subset\{1,\ldots,n\}
\end{equation}
with $i_1<\cdots<i_p$, we write
\begin{equation}
    dx^I:=dx_{i_1}\wedge\cdots\wedge dx_{i_p}.
\end{equation}
The following definition and notation is adapted from \cite{pham}:
\begin{definition}
    We say that a differential $p$-form $\omega$ on $Y$ is a \textit{holomorphic $p$-form relative to $T$} if it can be expressed in the local coordinates \eqref{eq:local_coordinates} as 
    \begin{equation}
        \omega=\sum_{\substack{I\subset\{1,\ldots,n\} \\ \vert I\vert=p}}f_I(x,t)dx^I,
    \end{equation}
     with holomorphic functions $f_I$. We denote the space of all holomorphic $p$-forms relative to $T$ by $\Omega^p(Y/T)$. The germ of $\omega$ at $t\in T$ is denoted by $\omega(t)$.
\end{definition}
It makes sense to define a codifferential which acts only on the $x$-part of a differential form relative to $T$. Thus, we define linear maps
\begin{equation}
    d_{Y/T,p}:\Omega^p(Y/T)\to\Omega^{p+1}(Y/T)
\end{equation}
for all $p\in\mathbb{N}$ whose action on $\omega\in\Omega^p(Y/T)$ in local coordinates reads
\begin{equation}
    d_{Y/T,p}\;\omega=\sum_{\substack{I\subset\{1,\ldots,n\} \\ \vert I\vert=p}}\sum_{i=1}^n\frac{\partial f_I}{\partial x_i}(x,t)dx_i\wedge dx^I.
\end{equation}
As usual, we simply write $d_{Y/T}$ for the corresponding endomorphism of $\bigoplus_{p\in\mathbb{N}}\Omega^p(Y/T)$. It is easy to check that $d_{Y/T}\circ d_{Y/T}=0$ and thus,
\begin{equation}
    0\to\Omega^0(Y/T)\overset{d_{Y/T,0}}\to\Omega^1(Y/T)\overset{d_{Y/T,1}}\to\cdots\overset{d_{Y/T,p-1}}\to\Omega^p(Y/T)\overset{d_{Y/T,p}}\to\cdots
\end{equation}
is a cochain complex. Correspondingly, just as in the regular de Rham complex of differential forms, we say that $\omega$ is a \textit{closed} differential $p$-form relative to $T$ if $d_{Y/T}\omega=0$.\\
Note that if $h$ is a section of $\mathcal{F}_{Y/T}^p$, then according to Proposition \ref{prop:homology_sheaf} we have
\begin{equation}
    h(t)\in\lim_{U\ni t}H_{p+\dim T}(Y,\pi^{-1}(T-U))\simeq H_p(Y_t).
\end{equation}
Thus, if $\omega\in\Omega^p(Y/T)$, it makes sense to integrate $\omega(t)$ over $h(t)$. The next proposition shows why this is a good idea.
\begin{proposition}[\cite{pham}]\label{prop:holomorphic_integral}
    Let $\omega\in\Omega^p(Y/T)$ and $h\in\mathcal{F}^p_{Y/T}(T)$. If $\omega$ is closed then
    \begin{equation}\label{eq:holomorphic_integral}
        T\to\mathbb{C},\qquad t\mapsto \int_{h(t)}\omega(t)
    \end{equation}
    defines a holomorphic function on $T$.
\end{proposition}
This proposition is the basis for our investigation of holomorphic functions defined by integrals. In practice, however, the situation is scarcely as nice as in Proposition \ref{prop:holomorphic_integral}. Usually, we are not provided with a global section $h$ but only with a local section. More specifically, often times (in particular for one-loop Feynman integrals) the situation is as follows: The manifold $Y$ has a product structure minus some analytic subset of problematic points, i.e., there is a compact complex analytic manifold $X$ of dimension $n$ and an analytic subset $S\subset X\times T$ such that $Y=(X\times T)\backslash S$. In this case, the map $\pi:Y\twoheadrightarrow T$ is taken to be the canonical projection. We denote the fiber of $S$ at $t$ by $S_t:=Y_t\cap S$ for all $t\in T$. Suppose we are only provided with a fixed $p$-cycle $\Gamma$ in the fiber $Y_{t_0}$ for some given $t_0\in T$. This is already sufficient to define at least a local section:
\begin{lemma}\label{lem:cycle_is_section}
    There is a $\dim T$-simplex $\sigma$ in $T$ such that $t_0\in\overset{\circ}\sigma$ and $\Gamma\times\sigma$ defines an element in $\mathcal{F}_{Y/T}^p(\overset{\circ}\sigma)$ (where $\overset{\circ}\sigma$ denotes the interior of $\sigma$).
\end{lemma}
\begin{proof}
    Since $Y$ is a $T_4$-space (every metric space is a $T_4$-space \cite{querenburg}) and $\Gamma$ as well as $S_{t_0}$ are closed sets, there exists open neighborhoods $V_1\subset X$ of $\Gamma$ and $V_2\subset X$ of $S_{t_0}$ such that $V_1\cap V_2=\emptyset$. Due to continuity and the compactness of the fibers, there exists an open neighborhood $U\subset T$ of $t_0$ such that $S_t\subset V_2$ for all $t\in U$. Then, $\Gamma\cap S_t\subset V_1\cap V_2=\emptyset$ for all $t\in U$. Thus, $\Gamma\times U$ is a subset of $Y$. We may assume that there exists a $\dim T$-simplex $\sigma\subset U$ in $T$. Hence, viewing $\Gamma\times\sigma$ as a $(p+\dim T)$-chain in $X\times T$, we compute
    \begin{equation}
        \partial(\Gamma\times\sigma)=\underbrace{\partial\Gamma}_{=0}\times U+(-1)^p\cdot\Gamma\times\partial\sigma.
    \end{equation}
    Now, $\Gamma\times\partial\sigma=0$ in $H_{p+\dim T}(Y,\pi^{-1}(T-\overset{\circ}\sigma))$ (this is already true on the level of chains) since it is contained within $\pi^{-1}(T-\overset{\circ}\sigma)$. Thus, we indeed have $\partial(\Gamma\times\sigma)=0$  in $H_{p+\dim T}(Y,\pi^{-1}(T-\overset{\circ}\sigma))$ as claimed.
    We conclude that $\Gamma\times\sigma$ defines a local section in the homology sheaf in degree $p$ of $Y$ relative to $T$.
\end{proof}
The situation described above applies for example to one-loop Feynman integrals (defined further below, see for example equation \eqref{eq:one_loop_2}), where $t_0$ corresponds to a Euclidean (i.e., real) configuration of momenta. Now, the question is if this local section can be extended to a (possibly multivalued) global section, generally relative to a slightly smaller base space $T^\ast\subset T$. A central result by Pham in this direction is the following
\begin{proposition}[\cite{pham}]\label{prop:trivial_fibration}
    Let $T^\ast\subset T$ be an open subset. If $\pi\vert_{\pi^{-1}(T^\ast)}:\pi^{-1}(T^\ast)\to T^\ast$ defines a locally trivial $C^\infty$-fibration, then $\int_{h(t_0)}\omega(t_0)$ defines a multi-valued holomorphic function on $T^\ast$.
\end{proposition}
The idea for the proof of Proposition \ref{prop:trivial_fibration} is rather simple: By Lemma \ref{lem:cycle_is_section}, the initial integration domain $\Gamma$ defines a local section of $\mathcal{F}^p_{Y/T^\ast}$. If $\pi\vert_{\pi^{-1}(T^\ast)}:\pi^{-1}(T^\ast)\to T^\ast$ defines a locally trivial $C^\infty$-fibration, then the sheaf $\mathcal{F}^p_{Y/T^\ast}$ is locally constant. Hence, every local section can be extended to a multivalued global section according to Proposition \ref{prop:extension_of_local_sections}.\\
As it stands, this is a little bit too abstract for our purposes. Much more useful to us is a more detailed discussion of the situation where $Y$ has a product structure minus an analytic set and $\pi$ is the canonical projection as described above. For this purpose, we recall the following notions:
\begin{definition}[\cite{app-iso}]
    A \textit{fiber bundle of pairs} is a tuple $(E,S,B,\pi,F)$ consisting of topological spaces $E$, $B$, $F$, a subset $S\subset E$ and a continuous surjection $\pi:E\twoheadrightarrow B$ such that $(E,B,\pi,F)$ is a fiber bundle which has local trivializations at every point $b\in B$ whose restriction to $S$ are local trivializations of $(S,B,\pi\vert_S,\pi^{-1}(\{\text{pt.}\})\cap S)$ (where $\text{pt.}$ denotes any point in $B$).\\
    Furthermore, we shall say that $(E,S,B,\pi,F)$ is \textit{smooth} as a fiber bundle of pairs if the following conditions are satisfied:
    \begin{itemize}
        \item $E$, $B$ and $F$ are smooth manifolds.
        \item $\pi$ is a smooth map.
        \item For any local trivialization $g$, the inverse $g^{-1}$ is differentiable with respect to $t$.
        \item For any local trivialization $g$, the inverse $g^{-1}$ lifts smooth vector fields on $B$ to locally Lipschitzian vector fields on $E$.
    \end{itemize}
    To simplify the notation, we denote a smooth fiber bundle of pairs $(E,S,B,\pi,F)$ by $\pi:(E,S)\to B$.
\end{definition}
\begin{definition}[\cite{hirsch}]
    Let $V$ and $X$ be smooth manifolds. An \textit{isotopy} from $V$ to $X$ is a continuous map $\sigma:V\times[0,1]\to X$ such that $\sigma(\cdot,t)$ is an embedding for all $t\in[0,1]$. If $V\subset X$ and $\sigma(\cdot,0)$ is the natural inclusion, we say that $\sigma$ is an \textit{isotopy of $V$ in $X$}. If $V=X$, $\sigma(\cdot,0)=\text{id}_X$ and $\sigma(\cdot,t)$ is a diffeomorphism for all $t\in[0,1]$, we say that $\sigma$ is an \textit{ambient isotopy}.
\end{definition}
For an isotopy $\sigma:V\times[0,1]\to X$, we denote its \textit{track} by
\begin{equation}
    \hat{\sigma}:V\times[0,1]\to X\times[0,1],\quad (x,t)\mapsto(\sigma(x,t),t)
\end{equation}
and its \textit{support} by
\begin{equation}
    \text{Supp}\,\sigma:=\text{cl}\{x\in X \;\vert\; \exists t\in[0,1]\,:\,\sigma(x,t)\neq \sigma(x,0)\},
\end{equation}
where $\text{cl}$ denotes the topological closure. Under reasonable circumstances, isotopies in $X$ can be extended to ambient isotopies of $X$.
\begin{theorem}[\cite{hirsch}]\label{thm:isotopy_extension}
    Let $X$ be a manifold, $A\subset X$ a compact subset and $U\subset X$ an open neighborhood of $A$. If $\sigma:U\times[0,1]\to X$ is an isotopy of $U$ in $X$ such that $\hat{\sigma}(U\times[0,1])$ is open in $X\times I$, then there exists an ambient isotopy $\tilde{\sigma}$ of $X$ with compact support such that $\tilde{\sigma}$ agrees with $\sigma$ in a neighborhood of $A\times[0,1]$.
\end{theorem}
Suppose again that we are in the situation above where $Y=(X\times T)\backslash S$ with an analytic subset $S\subset X\times T$ and suppose that $\pi:(X\times T,S)\to T$ is a (locally trivial) smooth fiber bundle of pairs, i.e., $\pi:X\to T$ is a fiber bundle which has local trivializations which also locally trivialize $\pi\vert_S:S\to T$. Then, in particular $\pi:Y\to T$ defines a locally trivial $C^\infty$-fibration and Proposition \ref{prop:trivial_fibration} applies. This can be thought of quite intuitively: If $\gamma:[0,1]\to T$ is any path from $t_0\in T$ to $t_1\in T$, then the pull-back bundle of pairs of $\pi:(X\times T,S)\to T$ by $\gamma$ has a contractible base space and is thus globally trivial. A trivialization of this pull-back bundle of pairs is an ambient isotopy
\begin{equation}
    \sigma:X\times[0,1]\to X
\end{equation}
such that $\sigma(\Gamma,s)\cap S_{\gamma(s)}=\emptyset$ for all $s\in[0,1]$. In the context of our problem, this can be viewed as a continuous deformation of the integration cycle $\Gamma$ away from the points were the differential form is singular. Phrased differently, the ambient isotopy $\sigma$ allows us to extend the local section defined by $\Gamma$ along $\gamma$:
\begin{proposition}\label{prop:extension_along_path}
    Let $\gamma:[0,1]\to T$ be a path, $\Gamma\subset X\backslash S_{\gamma(0)}$ a $p$-cycle and $\sigma:X\times[0,1]\to X$
    an ambient isotopy such that
    \begin{equation}
        \sigma(\Gamma,s)\cap S_{\gamma(s)}=\emptyset
    \end{equation}
    for all $s\in[0,1]$. Then, there is an open neighborhood $U$ of $\gamma([0,1])$ such that $\Gamma$ defines a section of $\mathcal{F}_{Y/T}(U)$.
\end{proposition}
\begin{proof}
    We know from Lemma \ref{lem:cycle_is_section} that for every $s\in[0,1]$ there exists a $\dim T$-simplex $\alpha_s$ in $T$ such that $\gamma(s)\in\overset{\circ}\alpha_s$ and such that $\sigma(\Gamma,s)\times\alpha_s$ defines an element in $\mathcal{F}_{Y/T}^p(\overset{\circ}\alpha_s)$. We want to show that these can be glued together to yield a local section of $\mathcal{F}_{Y/T}^p$ on $U:=\bigcup_{s\in[0,1]}\overset{\circ}\alpha_s$. By the sheaf axioms, it suffices to show that for any $s_1,s_2\in[0,1]$ such that $\overset{\circ}\alpha_{s_1}\cap\overset{\circ}\alpha_{s_2}\neq\emptyset$, the two sets $\sigma(\Gamma,s_1)\times(\alpha_{s_1}\cap\alpha_{s_2})$ and $\sigma(\Gamma,s_2)\times(\alpha_{s_1}\cap\alpha_{s_2})$ define the same element in $\mathcal{F}_{Y/T}^p(\overset{\circ}\alpha_{s_1}\cap\overset{\circ}\alpha_{s_2})$. We can assume $s_1<s_2$ without loss of generality. Now, let
    \begin{equation}
        \tau:=\sigma^\ast(\Gamma\times[s_1,s_2])\times(\alpha_{s_1}\cap\alpha_{s_2})
    \end{equation}
    where $\sigma^\ast$ denotes the map on the level of cycles induced by $\sigma$. Then, the boundary of $\tau$ is
    \begin{equation}
        \partial\tau=(\partial\sigma^\ast(\Gamma\times[s_1,s_2]))\times(\alpha_{s_1}\cap\alpha_{s_2})+(-1)^{p+1}\sigma^\ast(\Gamma\times[s_1,s_2])\times\partial(\alpha_{s_1}\cap\alpha_{s_2}).
    \end{equation}
    The second term $\sigma^\ast(\Gamma\times[s_1,s_2])\times\partial(\alpha_{s_1}\cap\alpha_{s_2})$ is 0 in $\mathcal{F}_{Y/T}(\overset{\circ}\alpha_{s_1}\cap\overset{\circ}\alpha_{s_2})$, and we compute the first factor of the first term to be
    \begin{equation}
        \begin{split}
            \partial\sigma^\ast(\Gamma\times[s_1,s_2])&=\sigma^\ast(\partial(\Gamma\times[s_1,s_2]))\\
            &=\sigma^\ast(\underbrace{\partial\Gamma}_{=0}\times[s_1,s_2]+(-1)^p(\Gamma\times s_2-\Gamma\times s_1))\\
            &=(-1)^p(\sigma^\ast(\Gamma,s_2)-\sigma^\ast(\Gamma,s_1)).
        \end{split}
    \end{equation}
    Thus, we conclude that
    \begin{equation}
        \sigma(\Gamma,s_2)\times(\alpha_{s_1}\cap\alpha_{s_2})-\sigma(\Gamma,s_1)\times(\alpha_{s_1}\cap\alpha_{s_2})=(\sigma(\Gamma,s_2)-\sigma(\Gamma,s_1))\times(\alpha_{s_1}\cap\alpha_{s_2})
    \end{equation}
    is the boundary of $(-1)^p\cdot\tau$ which completes the proof.
\end{proof}
\begin{corollary}\label{cor:extension_along_path}
    In the same situation as in Proposition \ref{prop:extension_along_path}, the integral $\int_\Gamma\omega(\gamma(0))$ defines a holomorphic function on some open neighborhood of $\gamma(0)$ which can be analytically continued along $\gamma$ by
    \begin{equation}
        \int_{\sigma(\Gamma,s)}\omega(\gamma(s)).
    \end{equation}
\end{corollary}
\begin{proof}
    Follows directly from Propositions \ref{prop:holomorphic_integral} and \ref{prop:extension_along_path}.
\end{proof}
Due to this result, we make the following definition:
\begin{definition}
    Let $\gamma:[0,1]\to T$ be a path, $\Gamma\subset X$ a submanifold and $\sigma:X\times[0,1]\to X$ an ambient isotopy of $\Gamma$ in $X$. We say that $\sigma$ is \textit{adapted to $\gamma$} if
    \begin{equation}
        \sigma(\Gamma,s)\cap S_{\gamma(s)}=\emptyset
    \end{equation}
    for all $s\in[0,1]$.
\end{definition}
With this definition in place, Corollary \ref{cor:extension_along_path} can be rephrased as
\begin{proposition}\label{prop:adapted}
    Let $\gamma:[0,1]\to T$ be a path from $t_0\in T$ to some $t\in T$ and $\sigma:X\times [0,1]\to X$ an ambient isotopy adapted to $\gamma$. Then, the integral $\int_\Gamma\omega(t_0)$ can be analytically continued along $\gamma$ via
    \begin{equation}
        \int_{\sigma(\Gamma,s)}\omega(\gamma(s)).
    \end{equation}
\end{proposition}
We need some criteria to determine if an appropriate ambient isotopy for such an analytic continuation exists. Above, we discussed how it is sufficient for $\pi:(X\times T,S)\to T$ to be a fiber bundle of pairs. A special case that is of particular interest to us is the case where the fibers $S_t$ are finite unions of complex analytic manifolds in general position. Recall that for smooth submanifolds $S_1,\ldots,S_k\subset X$ of a smooth manifold $X$, we say that $S_1,\ldots,S_k$ are in \textit{general position} if the normal vectors at every intersection point are linearly independent. More concretely, for every $x\in\bigcup_{i=1}^kS_i$ set
\begin{equation}
    I_x:=\{i\in\{1,\ldots,k\} \;\vert\; x\in S_i\}
\end{equation}
and let $s_{1,x},\ldots,s_{k,x}:X\to\mathbb{R}$ be local equations for the manifolds $S_1,\ldots,S_k$ around $x\in\bigcup_{i=1}^kS_i$. Then, $S_1,\ldots,S_k$ are in general position if and only if for every $x\in\bigcup_{i=1}^kS_i$ there are coordinates $\phi$ around $x$ such that the gradients of $\{s_{i,x}\circ\varphi^{-1}\}_{i\in I_x}$ at $\phi(x)$ are linearly independent. We have the following result on the analytic continuation in this case:
\begin{proposition}[\cite{app-iso}]\label{prop:trivial_bundle}
    Suppose that for all $t\in T$ the fiber $S_t=\bigcup_{i=1}^k(S_i)_t$ is the union of finitely many complex analytic manifolds $(S_i)_t$, depending smoothly on $t$. If the $(S_1)_t,\ldots,(S_k)_t$ are in general position for all $t\in T$, then $\pi:(X\times T,S)\to T$ is a smooth fiber bundle of pairs. In particular, $\int_\Gamma\omega(t_0)$ can be analytically continued along any path in $T$. Furthermore, the analytic continuation to $t\in T$ can be written as
    \begin{equation}
        \int_{\Gamma'}\omega(t)
    \end{equation}
    for an appropriate cycle $\Gamma'$.
\end{proposition}
For our purposes, this is unfortunately not enough. It turns out that after compactification, the relevant $(S_1)_t,\ldots,(S_k)_t$ in the case of Feynman integrals are generally not in general position at any $t\in T$. A more refined criterion can be obtained by equipping $S$ with a Whitney stratification and we introduce the necessary theory here.
\begin{definition}[\cite{stratified_morse}]
    Let $X$ be a topological space and $(\mathcal{S},\prec)$ a partially ordered set. An \textit{$\mathcal{S}$-decomposition} of $X$ is a locally finite collection of disjoint locally closed sets $A_i\subset X$ such that the following hold:
    \begin{enumerate}
        \item $X=\bigcup_{i\in\mathcal{S}}A_i$.
        \item For all $i,j\in\mathcal{S}$, we have
        \begin{equation}
            A_i\cap \bar{A}_j\neq\emptyset\;\Leftrightarrow\; A_i\subset\bar{A}_j\;\Leftrightarrow\; i\prec j.
        \end{equation}
    \end{enumerate}
    The $A_i$ are called the \textit{pieces}, their connected components the \textit{strata} of the $\mathcal{S}$-decomposition.
\end{definition}
To apply techniques from differential topology, it is useful to require that the pieces $A_i$ fit together nicely.
\begin{definition}[\cite{stratified_morse}]\label{defn:whitneys_condition_B}
    Let $M$ be a smooth manifold and $X,Y\subset M$ two smooth submanifolds. We say that the pair $(X,Y)$ satisfies \textit{Whitney's condition B} if the following holds: Let $y\in Y$ and suppose $x_n$ is a sequence in $X$ converging to $y$ and $y_n$ is a sequence in $Y$ converging to $y$. Then, if the sequence of secants $\overline{x_ny_n}$ converges to some line $l$ and the sequence of tangent planes $T_{x_n}X$ converges to some plane $T$,\footnote{This is to be understood as convergence in the relevant Graßmannians.} we have $l\subset T$.
\end{definition}
A few remarks are in order: First, note that if $X$ and $Y$ are submanifolds of $M$ such that $\bar{X}\cap Y=\emptyset$, the above condition is vacuous. If $Y\subset\bar{X}$ and $\bar{X}$ is a smooth manifold, it is also evident that Whitney's condition B is satisfied for the pair $(X,Y)$ (simply check the condition in coordinates on $\bar{X}$).
\begin{definition}[\cite{stratified_morse}]
    Let $Y$ be a smooth manifold, $S\subset Y$ a closed subset and $\mathcal{S}$ a partially ordered set. A \textit{Whitney stratification} of $S$ is an $\mathcal{S}$-decomposition such that all pairs of strata satisfy Whitney's condition B. We call the set $S$ together with a Whitney stratification of $S$ a \textit{stratified set}.
\end{definition}
Suppose that $S\subset Y$ is a closed subset which decomposes into a finite union of smooth manifolds $S_1,\ldots,S_n\subset Y$ in general position. Then, there is a natural Whitney stratification of $S$ with strata given by the connected components of
\begin{equation}\label{eq:canonical_stratification}
    S_i-\bigcup_{i<j}S_i\cap S_j,\quad S_i\cap S_j-\bigcup_{i<j<k} S_i\cap S_j\cap S_k,\quad \cdots
\end{equation}
The pieces are the union of all strata with equal codimension (i.e., equally many manifolds involved in the intersection), and the underlying partial order is given by the dimension of the pieces.
\begin{definition}[\cite{pham}]
    Let $Y$ be a smooth manifold equipped with a Whitney stratification. We say that $Y$ is a \textit{stratified bundle} if there exists a Whitney stratified set $X$ such that $Y$ is locally homeomorphic to $X\times T$ by homeomorphisms which map every stratum in $Y$ to the product of a stratum in $X$ with $T$.
\end{definition}
Now, we can formulate the following criterion:
\begin{theorem}[Thom's Isotopy Theorem, \cite{pham}]\label{thm:isotopy_theorem}
    Let $Y$ and $T$ be two differentiable manifolds with $T$ connected and let $\pi:Y\to T$ be a proper differential map. Suppose that $Y$ is a stratified set and the restriction of $\pi$ to each stratum is a submersion. Then, $\pi:Y\to T$ is a stratified bundle.
\end{theorem}
Suppose $S\subset Y$ is a closed subset equipped with a Whitney stratification. Then, the stratification of $S$ together with the connected components of $Y\backslash S$ yield a Whitney stratification of $Y$ \cite{pham}. If this makes $\pi:Y\to T$ a stratified bundle, then in particular $\pi:(Y,S)\to T$ is a fiber bundle of pairs.\\
Supposing again that we are in the situation where $Y=(X\times T)\backslash S$ as above, we can equip $S$ with a Whitney stratification (every analytic set admits such a stratification, see \cite{whitney}). Denote by $\{A_i\}_{i\in I}$ the collection of strata and for every $i\in I$ denote by $cA_i\subset A_i$ the set of all points at which the restriction of $\pi$ to $A_i$ fails to be a submersion. Then, we can conclude that the integral of interest defines a (multivalued) holomorphic function outside of the following set:
\begin{definition}\label{defn:landau_surface}
    In the same situation as above, we call the set
    \begin{equation}
        L:=\pi(\bigcup_{i\in I}cA_i)
    \end{equation}
    the \textit{Landau surface} of the integral $\int_\Gamma\omega(t_0)$.
\end{definition}
It should be remarked that it can be shown that the Landau surface is an analytic set \cite{pham} (a consequence of Remmert's Proper Mapping Theorem). Thus, if we assume $T$ to be connected, $L$ is either all of $T$ or $T\backslash L$ is a non-empty open subset. We conclude that $\int_\Gamma\omega(t_0)$ defines a holomorphic function on $T^\ast:=T\backslash L$. Furthermore, it is well-known among physicists that in the case of Feynman integrals, the Landau surface is given by the solutions to a set of algebraic equations (see equations \eqref{eq:landau_equations_1} and \eqref{eq:landau_equations_2} for the one-loop case). These equations are famously known as the \textit{Landau equations} and informally detect points at which the singular loci of the integrand are in non-general position at finite distance.\\
The fact that a given local section may only be extended to a global section in the multivalued sense immediately gives rise to the question of the discontinuity of the function defined by a given integral. This can be easily expressed in terms of the monodromy of the section $h$ in question. Let $t_0\in T^\ast$ and let $\gamma:[0,1]\to T^\ast$ be a loop based at $t_0$. Then, continuing the germ $h(t_0)$ along $\gamma$ yields a class $\gamma^\ast h(t_0)\in H_p(Y_{t_0})$, which can be shown to only depend on the homotopy class of $\gamma$ \cite{pham}. Then, analytically continuing the integral along $\gamma$ yields
\begin{equation}
    \int_{\gamma^\ast h(t_0)}\omega(t_0).
\end{equation}
We define the \textit{variation} of $h(t_0)$ along $\gamma$ by
\begin{equation}
    \text{Var}_{[\gamma]}h(t_0)=\gamma^\ast h(t_0)-h(t_0)=(\gamma^\ast-1)h(t_0)
\end{equation}
and the \textit{discontinuity} of $\int_{h(t_0)}\omega(t_0)$ along $\gamma$ by
\begin{equation}
    \text{Disc}_{[\gamma]}\int_{h(t_0)}\omega(t_0)=\int_{\text{Var}_{[\gamma]h(t_0)}}\omega(t_0).
\end{equation}
The variation can be understood as a homomorphism
\begin{equation}
        \text{Var}:\pi_1(T^\ast,t_0)\to\text{Aut}(H_\bullet(Y_{t_0})).
\end{equation}
The task is now to compute $\text{Var}_{[\gamma]}h(t_0)$. This is generally a difficult problem. A rather simple case which can be applied in particular to one-loop Feynman integrals, however, is extensively studied in \cite{pham}. These are the so-called \textit{simple pinches} we discuss further below.\\
The idea to understand these simple pinches is to localize the problem such that the variation yields a homology class which has a representative with support contained in an arbitrarily small neighborhood of a pinch point. We cannot expect that the problem can be localized like this for any arbitrary loop. But there is a class of loops whose homotopy classes span the fundamental group of $T\backslash L$ if $T$ is simply connected which allow for such a localization:\\
Let $u\in L$ be a point of codimension 1. Then, there are coordinates $(t_1,\ldots,t_m)$ of $T$ defined in a neighborhood $U\subset T$ of $u$ such that $L$ locally looks like the set $\{t_1=0\}$. Let $\theta:[0,1]\to T\backslash L$ be a path from some basepoint $u_0\in T\backslash L$ to some $u_1\in U-L$. Let $\omega:[0,1]\to U-L$ be a loop based at $u_1$ that traces a circle in the coordinate $t_1$ with $t_2,\ldots,t_m$ fixed. Then, $\gamma:=\theta\omega\theta^{-1}$ is a loop in $T\backslash L$ based at $u_0$. A loop constructed like this is called a \textit{simple loop}. In an abuse of language, we shall call elements of the fundamental group which can be represented by a simple loop also simple loops. In the cases we study here, it is always possible to restrict our attention to simple loops without sacrificing any amount of generality due to the following
\begin{proposition}[\cite{pham}]\label{prop:simple_loops}
    The fundamental group $\pi_1(T\backslash L,u_0)$ of $T\backslash L$ is spanned by simple loops if and only if $T$ is simply connected.
\end{proposition}

\subsubsection{\textbf{Leray's Calculus of Residues}}
In the preceding subsubsection, we have seen how to compute the discontinuity of a function defined by an integral of the form \eqref{eq:standard_form_intro} by extending a local section of $\mathcal{F}_{Y/T}^n$ to a multivalued global section. To obtain Cutkosky's Theorem in Section \ref{sec:one_loop}, we also need to employ the multi-variant version of the Residue Theorem, going back to Leray \cite{leray}, as already eluded to in the introduction.\\
First recall the regular Residue Theorem from complex analysis: Let $U\subset\mathbb{C}$ be a simply connected open set and let $a\in U$ be a point. Then, the theorem states that for a holomorphic function $f:U\backslash\{a\}\to\mathbb{C}$ and a positively oriented simple closed curve $\gamma:[0,1]\to U\backslash\{a\}$ we have
\begin{equation}
    \oint_\gamma f(z)\mathrm{d}z=2\pi i\cdot\text{Res}(f,a),
\end{equation}
where $\text{Res}(f,a)$ is the \textit{residue of $f$ at $a$}.\footnote{Of course there are more general formulations of the Residue Theorem (e.g., for a finite list of points $(a_1,\ldots,a_n)$). But this version suffices here for the illustration of the concept.} To generalize this to multiple dimensions, we essentially have to answer two questions: What is the multi-dimensional analog for the curve $\gamma$ along which we integrate and what is the multi-dimensional analog of the residue?\\
First, we attend to the first question. We follow \cite{pham}. Recall that for a smooth manifold $M$ and a smooth submanifold $S\subset M$ a tubular neighborhood of $S$ in $M$ is a vector bundle $\pi:E\to S$ (with $S$ the base space, $E$ the total space and $\pi$ the bundle projection of the bundle) together with a smooth map $J:E\to M$ such that:
\begin{itemize}
    \item If $i:S\hookrightarrow M$ is the natural embedding and $0_E$ is the zero section, we have $J\circ 0_E=i$.
    \item There exist $U\subset E$ and $V\subset M$ with $0_E[S]\subset U$ and $S\subset V$ such that $J\vert_U:U\overset{\sim}\to V$ is a diffeomorphism.
\end{itemize}
We are mainly interested in $V$ and in an abuse of language call this a tubular neighborhood of $S$ as well. Note that there is an associated retraction $\mu:V\to S$ by applying $(J\vert_U)^{-1}$, retracting to the zero section $0_E$ and going back to $M$ via $J\vert_U$. Now, suppose $X$ is a complex analytic manifold and $S\subset X$ a closed analytic submanifold of codimension 1. Fix a closed tubular neighborhood $V\subset X$ of $S$. The associated retraction $\mu: V\to S$ induces a disk bundle structure over $S$. Then, if $\sigma$ is a simplex in $S$, its preimage $\mu^{-1}(\sigma)$ is homeomorphic to $\sigma\times D$, where $D$ is the unit disk. On the level of chains, we obtain
\begin{equation}
    \mu^\ast\sigma=D\otimes\sigma.
\end{equation}
The boundary of $\mu^\ast\sigma$ is thus
\begin{equation}
    \partial\mu^\ast\sigma=\partial D\otimes\sigma+D\otimes\partial\sigma.
\end{equation}
Using this result, we define a homomorphism $\delta_\mu:C_p(S)\to C_{p+1}(X-S)$ by setting $\delta_\mu(\sigma):=\partial D\otimes\sigma$ for each simplex $\sigma$ and extending by linearity. Note that $\delta_\mu$ anti-commutes with the boundary $\partial$:
\begin{equation}
    \partial\delta_\mu\sigma=\partial(\partial D\otimes \sigma)=-\partial D\otimes\partial\sigma=-\delta_\mu\partial\sigma
\end{equation}
Hence, $\delta_\mu$ descends to a homomorphism
\begin{equation}
    \delta_\ast:H_p(S)\to H_{p+1}(X\backslash S)
\end{equation}
which we call the \textit{Leray coboundary}. It can be shown that this map does not depend on the choice of $\mu$ which is why we dropped the subscript. Note that this construction gives a simple closed curve $\gamma$ if we consider the one-dimensional case where $S$ is just a point.\\
Now, we turn to the second question. To answer it, let us consider a closed differential $p$-form $\omega$ on $X\backslash S$. For all $x\in S$, let $s_x$ be a local equation for $S$ near $x$. If for all $x\in S$ the form $s_x\cdot\omega$ can be extended to a differential $p$-form on a neighborhood of $x$, we say that $\omega$ has a \textit{polar singularity of order $1$} along $S$.
\begin{proposition}[\cite{pham}]\label{prop:leray_residue_1}
    If $\omega$ has a polar singularity of order 1 along $S$ then for every $x\in X$ there exist differential forms $\psi_x$ and $\theta_x$ defined on a neighborhood of $x$ such that
    \begin{equation}
        \omega=\frac{ds_x}{s_x}\wedge\psi_x+\theta_x.
    \end{equation}
    Furthermore, $\psi_x\vert_S$ is closed and depends only on $\omega$. We call $\psi_x\vert_S$ the \textit{residue} of $\omega$.
\end{proposition}
In the situation of Proposition \ref{prop:leray_residue_1}, we denote
\begin{equation}
    \text{res}[\omega]:=\left.\frac{s_x\omega}{ds_x}\right\vert_S:=\psi_x\vert_S \quad\in\quad \Omega^{p-1}(S).
\end{equation}
The $(p-1)$-form $\text{res}[\omega]$ is called the \textit{residue form} of $\omega$. Now, if $\omega$ is any closed form on $X\backslash S$, then $\omega$ is cohomologous to a closed form $\tilde{\omega}$ on $X\backslash S$ with a simple pole along $S$ \cite{pham}. Thus, it makes sense to define a homology class
\begin{equation}
    \text{Res}[\omega]:=[\text{res}[\tilde{\omega}]] \quad\in\quad H^{p-1}(S)
\end{equation}
and it can be shown that it only depends on the cohomology class of $\tilde{\omega}$ in $X\backslash S$ \cite{pham}.
\begin{theorem}[Residue Theorem, \cite{pham}]\label{thm:residue_theorem}
    Let $\gamma$ be a $(p-1)$-cycle in $S$ and $\omega$ a closed differential $p$-form on $X\backslash S$. Then, the following identity holds:
    \begin{equation}
        \int_{\delta\gamma}\omega=2\pi i\int_\gamma\mathrm{Res}[\omega]
    \end{equation}
\end{theorem}
We also have to deal with situations in which $S$ is not a submanifold but a union of $N$ closed submanifolds $S_1,\ldots,S_N\subset X$ in general position. To achieve this, we can iterate the construction above. For each $k\in\{1,\ldots,N\}$ the manifold $S_1\cap\cdots\cap S_k$ is a closed submanifold in which $S_{k+1},\ldots,S_N$ intersect in general position. Thus, we obtain sequences of maps
\begin{equation}
    \begin{split}
        H^p(X\backslash(S_1\cup\cdots\cup S_N))&\overset{\text{Res}_1}\to H^{p-1}(S_1-S_2\cup\cdots\cup S_N)\\
        &\overset{\text{Res}_2}\to H^{p-2}(S_1\cap S_2-S_3\cup\cdots S_N)\overset{\text{Res}_3}\to\cdots
    \end{split}
\end{equation}
and
\begin{equation}
    \begin{split}
        H_p(X\backslash(S_1\cup\cdots\cup S_N))&\overset{\delta_1}\leftarrow H_{p-1}(S_1-S_2\cup\cdots\cup S_N)\\
        &\overset{\delta_2}\leftarrow H_{p-2}(S_1\cap S_2-S_3\cup\cdots S_N)\overset{\delta_3}\leftarrow\cdots
    \end{split}
\end{equation}
This allows us to define the composite maps
\begin{equation}
    \text{Res}^m:=\text{Res}_m\circ\cdots\circ\text{Res}_1 \quad\text{and}\quad \delta^m:=\delta_1\circ\cdots\circ\delta_m.
\end{equation}
Similarly to the case of one manifold $S$, we obtain the following
\begin{theorem}[Iterated Residue Theorem, \cite{pham}]\label{thm:multi_residue_theorem}
    Let $\gamma$ be a $(p-m)$-cycle in $S$ and $\omega$ a closed differential $p$-form on $X\backslash S$. Then, the following identity holds:
    \begin{equation}
        \int_{\delta^m\gamma}\omega=(2\pi i)^m\int_\gamma\mathrm{Res}^m[\omega]
    \end{equation}
\end{theorem}
The (iterated) Leray residue has the following nice property that we employ later on:
\begin{proposition}\label{prop:residue_commutes_with_pullbacks}
    The Leray residue commutes with pullbacks, i.e., for any differential map $f:Y\to X$ between two smooth manifolds $X$ and $Y$, any submanifold $S\subset X$ of codimension 1 and any differential form $\omega$ on $X\backslash S$ we have
    \begin{equation}
        f^\ast\mathrm{Res}^m[\omega]=\mathrm{Res}^m[f^\ast\omega].
    \end{equation}
\end{proposition}
\begin{proof}
    It suffices to show the case $m=1$ since the case for general $m$ follows by induction. We can assume without loss of generality that $\omega$ has a polar singularity of order 1 along $S$ (otherwise, we replace $\omega$ by a cohomologous form with this property). First, note that $f^\ast\omega$ is a closed differential form on $Y\backslash f^{-1}(S)$. According to Proposition \ref{prop:leray_residue_1}, we can write $\omega=\frac{\mathrm{d}s_x}{s_x}\wedge\psi_x+\theta_x$. We compute
    \begin{equation}
        f^\ast\omega=f^\ast\left(\frac{\mathrm{d}s_x}{s_x}\wedge\psi_x+\theta_x\right)=f^\ast\frac{\mathrm{d}s_x}{s_x}\wedge f^\ast\psi_x+f^\ast\theta_x=\frac{\mathrm{d}(s_x\circ f)}{s_x\circ f}\wedge f^\ast\psi_x+f^\ast\theta_x.
    \end{equation}
    Now, $s_x\circ f$ is a local equation for $f^{-1}(S)$ around any point in $f^{-1}(x)$. Thus, we conclude
    \begin{equation}
        \mathrm{Res}[f^\ast\omega]=[(f^\ast\psi_x)\vert_S]=f^\ast[\psi_x\vert_S]=f^\ast\mathrm{Res}[\omega]
    \end{equation}
    as claimed.
\end{proof}
As mentioned in the introduction, Cutkosky's Theorem involves the use of $\delta$-functions (which are of course not functions in the conventional sense). These are usually defined in the language of functional analysis. But in the context of singular integrals, we can define the $\delta$-function and its derivatives evaluated at a test function by integrating the residue of a differential form induced by the given test function.
\begin{definition}[\cite{pham}]\label{defn:delta}
    For any test function $f$ set $\omega:=f\mathrm{d}x_1\wedge\cdots\mathrm{d}x_n$. Then, we define 
    \begin{equation}
        \delta^{(a)}(S)[f]:=\int_S\delta^{(a)}(f):=\int_S\left.\frac{\mathrm{d}^\alpha\omega}{\mathrm{d}s^{\alpha+1}}\right\vert_S.
    \end{equation}
\end{definition}

\subsubsection{\textbf{Simple Pinches}}
Now, we want to compute the variation of homology classes for a rather simple situation in which the singular points of the differential form is a finite union of manifolds which are in general position except at isolated points. Again, we follow \cite{pham}.
\begin{definition}\label{defn:simple_pinch}
    Let $Y$ and $T$ be two complex analytic manifolds, $\pi:Y\to T$ a smooth map and $S_1,\ldots,S_N\subset Y$ complex analytic submanifolds of codimension 1. Denote by $(S_i)_t:=\pi^{-1}(t)\cap S_i$ the fiber of $S_i$ over $t$ for all $i\in\{1,\ldots,N\}$ and $t\in T$. We say that the system $S_1,\ldots,S_N$ has a \textit{simple pinch} at $y\in Y$ if there is a coordinate chart $(\varphi=(x_1,\ldots,x_n,t_1,\ldots,t_m),U)$ in a neighborhood $U\subset Y$ of $y$ such that there are local equations
    \begin{equation}
        s_1=x_1,\quad\ldots\quad s_{N-1}=x_{N-1}
    \end{equation}
    and
    \begin{equation}
        s_N=t_1-x_1-\cdots-x_{N-1}-x_N^2-\cdots-x_n^2.
    \end{equation}
    for $S_1,\ldots,S_N$ around $y$.
\end{definition}
In should be remarked that the case $N=n+1$ is a little different from the remaining cases $N\leq n$. The first one is called a \textit{linear pinch} while the remaining cases are called \textit{quadratic pinches}.\\
Suppose $S_1,\ldots,S_N$ has a simple pinch at $y\in Y$ and let $(\varphi,U)$ be the coordinate chart from Definition \ref{defn:simple_pinch}. Then, $t:=\pi(y)$ must necessarily be a point of codimension 1 in the Landau surface. Suppose we fix $t\in T$ such that $t_1$ is real and positive and denote $U_t:=Y_t\cap U$. Then,
\begin{equation}\label{eq:complex_sphere}
    \begin{split}
        &U_t\cap(S_1)_t\cap\cdots\cap(S_N)_t\\
        &\quad\simeq\{(0,\ldots,0,x_N,\ldots,x_n)\in\varphi(U)\;\vert\; x_N^2+\cdots+x_n^2=t_1\}\\
        &\quad\simeq S^{n-N}_\mathbb{C},
    \end{split}
\end{equation}
where $S^k_\mathbb{C}$ denotes the complex unit sphere of dimension $k\in\mathbb{N}$.\footnote{Equation \eqref{eq:complex_sphere} also holds in the linear case $N=n+1$ if we agree on the convention that the -1 dimensional sphere is empty.} It is well-known that the complex $k$-sphere deformation retracts to the real $k$-sphere $S^k$. Thus, in case $n>N$, we obtain
\begin{equation}
    H_k(U_t\cap(S_1)_t\cap\cdots\cap(S_N)_t)\simeq H_k(S^{n-N})\simeq\begin{cases} \mathbb{Z} & \text{if }k=0,n-N \\ 0 & \text{otherwise}\end{cases}
\end{equation}
for the homology groups of $U_t\cap(S_1)_t\cap\cdots\cap(S_N)_t$ and the generator of $H_{n-N}(U_t\cap(S_1)_t\cap\cdots\cap(S_N)_t)$ is represented by a real $(n-N)$-sphere contained within $U_t\cap(S_1)_t\cap\cdots\cap(S_N)_t$. On the level of cycles, this is called the \textit{vanishing sphere} denoted by
\begin{equation}
    e:=\{(x_1,\ldots,x_n)\in\mathbb{C}^n \;\vert\; x_1,\ldots,x_n\in\mathbb{R},\; s_1(x,t)=\cdots=s_N(x,t)=0\}.
\end{equation}
It is the iterated boundary of the \textit{vanishing cell}
\begin{equation}
    \mathbf{e}:=\{(x_1,\ldots,x_n)\in\mathbb{C}^n \;\vert\; x_1,\ldots,x_n\in\mathbb{R},\; s_1(x,t),\ldots,s_N(x,t)\geq0\}.
\end{equation}
This means that $e=(\partial_1\circ\cdots\circ\partial_N)(\mathbf{e})$, where $\partial_i$ is the operator taking the boundary within $(S_i)_t$ for all $i\in\{1,\ldots,N\}$. Furthermore, we define the \textit{vanishing cycle}
\begin{equation}
    \tilde{e}:=(\delta_1\circ\cdots\circ\delta_N)(e)
\end{equation}
by taking the iterated Leray coboundary of the vanishing sphere. These define homology classes
\begin{equation}
    \begin{array}{ccl}
        [e]            & \quad\in & \quad H_{n-N}(U_t\cap (S_1)_t\cap\cdots\cap(S_N)_t),\\
        {[}\mathbf{e}] & \quad\in & \quad H_n(U_t,(S_1)_t\cup\cdots\cup(S_N)_t),\\
        {[}\tilde{e}]  & \quad\in & \quad H_n(U_t-(S_1)_t\cup\cdots\cup(S_N)_t).
    \end{array}
\end{equation}
These three classes are the protagonists in the calculation of the variation in the situation of a simple pinch. Due to the following proposition, it suffices in principle to know one of the three.
\begin{proposition}[\cite{app-iso}]\label{prop:homology_iso}
    In the situation above, the maps $\partial_i$ and $\delta_i$ are isomorphisms on the level of homology. In particular, $[\mathbf{e}]$ generates the relative homology group $H_n(U_t,(S_1)_t\cap\cdots\cap(S_N)_t)$ and $[\tilde{e}]$ generates the homology group $H_n(U_t-(S_1)_t\cap\cdots\cap(S_N)_t)$ of the complement.
\end{proposition}
In fact, we can use this generating property as the definition of the three classes in question.\footnote{Of course, this is a slight abuse of language since above, we defined the three objects in question as cycles. However, this does not cause any problems since all computations in this work depend only on the homology classes.}
\begin{definition}
    The \textit{vanishing cell} $[\mathbf{e}]$, \textit{vanishing sphere} $[e]$ and \textit{vanishing cycle} $[\tilde{e}]$ is a generator of $H_{n-N}(U_t\cap (S_1)_t\cap\cdots\cap(S_N)_t)$, $H_n(U_t,(S_1)_t\cup\cdots\cup(S_N)_t)$ and $H_n(U_t-(S_1)_t\cup\cdots\cup(S_N)_t)$, respectively.
\end{definition}
Note that this definition does only determine $[\mathbf{e}]$, $[e]$ and $[\tilde{e}]$ up to orientation ($\mathbb{Z}$ has two generators: 1 and -1). In what follows, it will become evident that this choice of orientation does not matter as all calculation in which these three classes appear are independent of the chosen orientation.\\
There are two marginal cases that we also need to cover: In case $n=N$, we are dealing with the 0-sphere which consists of two points. Thus,
\begin{equation}
    H_k(U_t\cap(S_1)_t\cap\cdots\cap(S_N)_t)\simeq H_k(S^{n-N})\simeq\begin{cases} \mathbb{Z}^2 & \text{if }k=0 \\ 0 & \text{otherwise}\end{cases}
\end{equation}
and the 0th homology group is generated by the two points $p_1,p_2$ of the 0-sphere. In this case, we set $[e]:=[p_2]-[p_1]$. In case linear case $n=N+1$, the intersection $U_t\cap(S_1)_t\cap\cdots\cap(S_N)_t$ is empty and the vanishing sphere does not exist. In accordance with \cite{pham}, we agree on the convention that all associated vanishing classes are 0 in this case.

\subsubsection{\textbf{The Picard-Lefschetz Formula for Simple Pinches}}\label{subsubsec:picard_lefschetz}
We have seen that an element $[\gamma]\in\pi_1(T\backslash L,t_0)$ gives rise to an automorphism $[\gamma]_\ast$ of $H_\bullet(Y_{t_0}\backslash S_{t_0})$. In the situation of a simple pinch, the problem can be localized in the sense that the variation $([\gamma]_\ast-\text{id})[h]$ of a class $[h]\in H_n(Y_{t_0}\backslash S_{t_0})$ can be represented by a cycle with support entirely contained within $U$, the domain of the coordinate chart $(\varphi,U)$ appearing in the definition of a simple pinch. Since the homology group $H_n(U_{t_0}-(S_1)_{t_0}\cap\cdots\cap (S_N)_{t_0})$ is spanned by a single element (the vanishing cycle), it is clear that the variation must thus be some integer multiple $N\cdot[\tilde{e}]$ of the vanishing cycle.\footnote{This also holds in the marginal case $N=n$ (see \cite{pham}), but the argument is different.} The discontinuity of the integral \eqref{eq:holomorphic_integral} corresponding to $[\gamma]$ is thus given by
\begin{equation}\label{eq:discontinuity_1}
    \int_{N\tilde{e}}\omega=N\cdot\int_{\tilde{e}}\omega,
\end{equation}
and the remaining task is to determine the integer $N$. This can be done by a Picard-Lefschetz type formula introduced in \cite{app-iso}. To understand its content, we first need the notion of an \textit{intersection index}. There are quite general definitions of this concept.\footnote{For example in \cite{pham}, the intersection index of two \textit{currents} is defined.} But for our purposes it suffices to consider the intersection index of two closed oriented manifolds with- or without boundary intersecting transversally and with dimensions adding up to the dimension of the whole space.
\begin{proposition}[\cite{pham}]
    Let $M$ be a smooth manifold of dimension $n$ and $S_1,S_2\subset M$ two orientable submanifolds (with or without boundary) of dimensions $p$ and $n-p$ such that
    \begin{equation}
        S_1\cap\partial S_2=S_2\cap\partial S_1=\emptyset.
    \end{equation}
    If $S_1$ and $S_2$ intersect transversally, then $S_1\cap S_2$ consists of a finite number of isolated points and the intersection index $\langle S_2,S_1\rangle$ of $S_1$ and $S_2$ is given by
    \begin{equation}
        \langle S_2\vert S_1\rangle=N_+-N_- \quad\in\quad \mathbb{Z},
    \end{equation}
    where $N_+$ (resp. $N_-$) is the number of points in $S_1\cap S_2$ at which the orientations of $S_1$ and $S_2$ match (resp. do not match).
\end{proposition}
We can express the integer $N$ from \eqref{eq:discontinuity_1} as the intersection index of the vanishing cell with the integration cycle as follows:
\begin{theorem}[Picard-Lefschetz Formula, \cite{pham}]\label{thm:picard_lefschetz}
    The integer $N$ in equation \eqref{eq:discontinuity_1} is given by
    \begin{equation}
        N=(-1)^\frac{(n+1)(n+2)}{2}\cdot\langle\mathbf{e}\vert h\rangle
    \end{equation}
    where $h$ is the integration cycle at $t$.
\end{theorem}
In this paper, we do not explicitly compute the intersection index relevant for one-loop Feynman integrals since it involves a very different array of techniques, mostly from homological algebra. Therefore, this step is postponed to the second part of this work.

\subsubsection{\textbf{Variation and Discontinuity Along Products of Loops}}
So far, we only discussed how to compute the variation and discontinuity for simple pinches along simple loops around points of codimension 1. To fully understand the discontinuities of one-loop Feynman integrals, it is useful to be able to relate the variation along products of loops to the variation along the individual factors. For the case of simple pinches, we have the following:
\begin{lemma}\label{lem:loop_product}
    Let $[\gamma_1],[\gamma_2]\in\pi_1(T\backslash L,t_0)$ be two simple loops. Then,
    \begin{equation}
        \mathrm{Var}_{[\gamma_1]\cdot[\gamma_2]}=\mathrm{Var}_{[\gamma_1]}+\mathrm{Var}_{[\gamma_2]}+\mathrm{Var}_{[\gamma_2]}\mathrm{Var}_{[\gamma_1]}
    \end{equation}
    and
    \begin{equation}
        \mathrm{Disc}_{[\gamma_1]\cdot[\gamma_2]}=\mathrm{Disc}_{[\gamma_1]}+\mathrm{Disc}_{[\gamma_2]}+\mathrm{Disc}_{[\gamma_2]}\mathrm{Disc}_{[\gamma_1]}.
    \end{equation}
\end{lemma}
\begin{proof}
    Let $h\in H_n(Y_{t_0})$. Denote $\text{Var}_{[\gamma_i]}h=:N_i\tilde{e}_i$ for $i=1,2$. We compute
    \begin{equation}
        \begin{split}
            \text{Var}_{[\gamma_1]\cdot[\gamma_2]}h&=\gamma_2^\ast\gamma_1^\ast h-h\\
            &=\gamma_2^\ast(h+N_1[\tilde{e}_1])-h\\
            &=N_2[\tilde{e}_2]+N_1\gamma_2^\ast[\tilde{e}_1]\\
            &=N_1[\tilde{e}_1]+N_2[\tilde{e}_2]+(\gamma_2^\ast-1)(N_1\tilde{e}_1)\\
            &=\text{Var}_{[\gamma_1]}h+\text{Var}_{[\gamma_2]}h+\text{Var}_{[\gamma_2]}\text{Var}_{[\gamma_1]}h.
        \end{split}
    \end{equation}
    The corresponding formula for the discontinuity follows immediately.
\end{proof}
We can easily generalize this to arbitrary finite products of loops:
\begin{proposition}\label{prop:loop_product}
    Let $[\gamma_1],\ldots,[\gamma_k]\in\pi_1(T^\ast,t_0)$ be simple loops. Then,
    \begin{equation}
        \mathrm{Var}_{[\gamma_1]\cdots[\gamma_k]}=\sum_{i=1}^k\sum_{1\leq j_1<\cdots<j_i\leq k}\mathrm{Var}_{[\gamma_{j_i}]}\cdots\mathrm{Var}_{[\gamma_{j_1}]}
    \end{equation}
    and
    \begin{equation}
        \mathrm{Disc}_{[\gamma_1]\cdots[\gamma_k]}=\sum_{i=1}^k\sum_{1\leq j_1<\cdots<j_i\leq k}\mathrm{Disc}_{[\gamma_{j_i}]}\cdots\mathrm{Disc}_{[\gamma_{j_1}]}.
    \end{equation}
\end{proposition}
\begin{proof}
    We proceed by induction on $k$. For $k=1$, the formulas are trivial. For the induction step, we compute
    \begin{equation}
        \begin{split}
            \text{Var}_{[\gamma_1]\cdots[\gamma_{k+1}]}&=\text{Var}_{[\gamma_1]\cdots[\gamma_k]}+\text{Var}_{[\gamma_{k+1}]}+\text{Var}_{[\gamma_{k+1}]}\text{Var}_{[\gamma_1]\cdots[\gamma_k]}\\
            &=\sum_{i=1}^k\sum_{1\leq j_1<\cdots<j_i\leq k}\text{Var}_{[\gamma_{j_i}]}\cdots\text{Var}_{[\gamma_{j_1}]}+\text{Var}_{[\gamma_{k+1}]}\\
            &+\text{Var}_{[\gamma_{k+1}]}\sum_{i=1}^k\sum_{1\leq j_1<\cdots<j_i\leq k}\text{Var}_{[\gamma_{j_i}]}\cdots\text{Var}_{[\gamma_{j_1}]}\\
            &=\sum_{1\leq j_1\leq k+1}\text{Var}_{[\gamma_{j_1}]}+\sum_{i=2}^{k+1}\sum_{1\leq j_1<\ldots<j_i\leq k+1}\text{Var}_{[\gamma_{j_i}]}\cdots\text{Var}_{[\gamma_{j_1}]}\\
            &=\sum_{i=1}^{k+1}\sum_{1\leq j_1<\cdots<j_i\leq k+1}\text{Var}_{[\gamma_{j_i}]}\cdots\text{Var}_{[\gamma_{j_1}]},
        \end{split}
    \end{equation}
    where we used Lemma \ref{lem:loop_product} in the first step and the induction hypothesis in the second step. Again the formula for the discontinuity follows immediately.
\end{proof}
With Proposition \ref{prop:loop_product}, we can compute the discontinuity along any loop around points where codimension 1 parts of the Landau surface intersect in general position.

\section{Feynman Graphs and Integrals}\label{sec:feynman_integrals}
\begin{figure}
\centering
\begin{subfigure}[c]{0.4\textwidth}
\begin{tikzpicture}[thick, scale=0.6]
    \draw (0,2) edge [electron] (3,1);
    \draw (3,1) edge [electron] (6,2);
    \draw (3,-1) edge [electron] (0,-2);
    \draw (6,-2) edge [electron] (3,-1);
    \draw (3,1) edge [photon] (3,-1);
\end{tikzpicture}
\end{subfigure}
\begin{subfigure}[c]{0.4\textwidth}
\begin{tikzpicture}[thick, scale=0.6]
    \draw (-1,4) edge [electron] (0,4);
    \draw (0,4) edge [electron] (4,4);
    \draw (4,4) edge [electron] (5,4);
    \draw (4,0) edge [photon] (4,4);
    \draw (0,0) edge [photon] (0,4);
    \draw (0,0) edge [electron] (-1,0);
    \draw (4,0) edge [electron] (0,0);
    \draw (5,0) edge [electron] (4,0);
\end{tikzpicture}
\end{subfigure}
\caption{Two Feynman graphs in quantum electrodynamics, both contributing to the probability amplitude of electron-positron scattering}
\label{fig:qed_diagrams}
\end{figure}
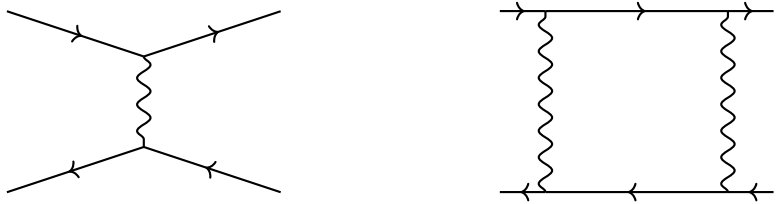

Now, we turn to the study of Feynman integrals as holomorphic functions in the external momenta. To ensure that there are no ambiguities regarding the notation, conventions and terminology employed in this work and to revisit the basic notions of the field for the non-expert reader, we quickly recall the definition of Feynman graphs and integrals. For a more detailed exposition, the reader is referred to \cite{weinberg} or \cite{weinzierl}. A graph always means a finite multi-graph in this text. For a graph $G$, we denote its underlying set of vertices by $V(G)$ and its underlying (multi-)set of edges by $E(G)$. The first Betti number of $G$, which is the maximal number of independent cycles, is denoted by $h_1(G)$. It is customary to say that $G$ has $h_1(G)$ \textit{loops}. For us, a Feynman integral means a graph together with some additional information specifying at which vertices how many particles are in- or outgoing. More specifically, we make the following
\begin{definition}
    A \textit{Feynman graph} $(G,\phi)$ is a graph $G$ together with a map $\phi:V(G)\to\mathbb{N}$ called the \textit{external structure}.
\end{definition}
This is not the most general definition of Feynman graphs. In many physical theories, for example quantum electro- or chromodynamics, one wants to distinguish different types of edges (see Figure \ref{fig:qed_diagrams} for two examples). The essential features of the analytic structure of the parameter-dependent integrals associated with such a graph are, however, already captured in our setting. So we do not go into detail regarding these more general Feynman graphs and integrals.\\
A Feynman graph (or \textit{Feynman diagram} as they are sometimes called) is a representation of a collection of possible ways elementary particles can interact: The external structure represents in- and outgoing particles. It assigns each vertex a number of external momenta. It is common to draw a pictorial representation of a Feynman graph $(G,\phi)$ by drawing the underlying graph $G$ as usual and then attaching $\phi(v)$ lines not connected to a second vertex to each vertex $v\in V(G)$.
Each Feynman graph $G$ is assigned a \textit{Feynman integral} by applying the \textit{Feynman rules} to it, which contributes to the probability amplitude of a given process. There are several equivalent descriptions of this procedure. Here, we focus on the so-called \textit{momentum space representation}, which is obtained from a Feynman graph $(G,\phi)$ in the following manner: First, we equip the graph $G$ with an arbitrary orientation and denote by $\mathcal{E}\in M(\vert V(G)\vert\times\vert E(G)\vert;\mathbb{Z})$ the corresponding incidence matrix. For each edge $e\in E(G)$, we write down a factor of $\frac{1}{(k_e^2+m_e^2)^{\lambda_e}}$ (called the \textit{propagator}), where $k_e\in\mathbb{C}^D$ is called the \textit{internal momentum} and $m_e\in\mathbb{R}_{\geq0}$ the \textit{mass} associated with the edge $e$. The $\lambda_e$ are generally complex numbers with positive real part called \textit{analytic regulators}. For each vertex $v\in V(G)$, we write down a constant factor (which we ignore in this text since it does not influence the analytic structure) and assign $\phi(v)$ external momenta $p_{v,1},\ldots,p_{v,\phi(v)}\in\mathbb{C}^D$ to it. We denote the total external momentum at $v$ by $p_v:=\sum_{i=1}^{\phi(v)}p_{v,i}$. Then, we enforce momentum conservation at every vertex $v\in V(G)$, i.e., we insist on the internal and external momenta satisfying $\sum_{e\in E(G)}\mathcal{E}_{v,e}k_e+p_v=0$ for every $v\in V(G)$.\footnote{Note that this implies that all external momenta are counted as incoming in our convention.} Usually, one wants to factor out the overall momentum conservation, i.e., the condition $\sum_{v\in V(G)}p_v=0$, as it does not depend on the internal momenta $k_e$. To do this, one may fix some vertex $v_0\in V(G)$ and drop the momentum conservation at $v_0$. The Feynman integral $I(G)$ corresponding to a Feynman graph $G$ thus reads
\begin{equation}\label{eq:physicists_feynman_integral_1}
    \begin{split}
        I(G)(p):=\int_{\mathbb{R}^{\vert E(G)\vert D}}&\mathrm{d}^{\vert E(G)\vert D}k\prod_{e\in E(G)}\frac{1}{(k_e^2+m_e^2)^{\lambda_e}}\\
        &\times\prod_{v\in V(G)\backslash\{v_0\}}\delta\left(\sum_{e\in E(G)}\mathcal{E}_{v,e}k_e+p_v\right).
    \end{split}
\end{equation}
Integrating out the $\delta$s leads to a linear system of equations for the $k_e$, and it can be shown that all, but $h_1(G)$ of the internal momenta can be eliminated by solving this system. The remaining internal momenta are then integrated over, and the result reads
\begin{equation}\label{eq:physicists_feynman_integral_2}
    I(G)=\int_{\mathbb{R}^{h_1(G)D}}\frac{\mathrm{d}^{h_1(G)D}k}{\prod_{e\in E(G)}((K_e(k)+P_e(p))^2+m_e^2)^{\lambda_e}},
\end{equation}
where $K_e:\mathbb{C}^{h_1(G)D}\to\mathbb{C}^D$ and $P_e:\mathbb{C}^{(\sum_{v\in V(G)}\phi(v))D}\to\mathbb{C}^D$ are linear maps for all $e\in E(G)$. The resulting propagators and thus, the maps $K_e$ and $P_e$ are not uniquely determined, but the result of the integration (if it is well-defined at all) is independent of the remaining freedom of choice. In physics, one is almost always concerned with Minkowski momenta to adhere to the principals of special relativity. This means in the physics literature the internal momenta are considered to be real, but the momentum-squares are typically defined as $k^2:=-k_0^2+k_1^2+\cdots+k_{D-1}^2$.\footnote{It is customary to start indexing the components of momenta with 0 instead of 1. It should also be mentioned that we employ a different sign convention than most particle physicists (it appears that the author's convention is in fact met with open hatred, see for example the footnote on page 2 in \cite{fun}), who like to set $k^2=k_0^2-k_1^2-\cdots-k_{D-1}^2$. In this case, the propagator needs to be $(k_e^2-m_e^2)^{-\lambda_e}$ instead.} Unfortunately, this immediately leads to problems since the integration domain now includes the poles at $k_e^2+m_e^2=0$. The usual ploy to avoid this issue is to introduce a small complex shift in the propagator by replacing $(k_e^2+m_e^2)^{-\lambda_e}$ with $(k_e^2+m_e^2-i\epsilon)^{\lambda_e}$ for some $0<\epsilon\ll1$. Then, the integration is carried out and the limit $\epsilon\to0^+$ is taken at the very end of the calculation. This is called the \textit{$i\epsilon$-prescription}. We shall see that this is not satisfactory for our purposes and we take a different route: In our setting, it is necessary to consider complex internal momenta. The Minkowski momenta can then be identified with those momenta that have purely imaginary 0th component and all remaining components real. We, however, start with an entirely real integration domain $\mathbb{R}^D$ (in the physics literature this is known as a \textit{Euclidean Feynman integral}) instead of choosing to integrate over all Minkowski momenta $i\mathbb{R}\times\mathbb{R}^{D-1}$, which never meets the zero locus of $k_e^2+m_e^2$ as long as $m_e^2>0$. This serves as the starting point for an analytic continuation. In Section \ref{sec:preliminaries} we saw how an analytic continuation requires us to continuously deform the integration domain as we move along a path in the space of external momenta. In particular, we show in Section \ref{sec:one_loop} how this is done explicitly for one-loop Feynman integrals in the case where we want to continue from Euclidean external momenta to Minkowski external momenta and explain how this is in agreement with the $i\epsilon$-prescription. For now, we simply define our Feynman integrals as Euclidean integrals.
\begin{definition}\label{defn:feynman_integral}
    Let $G$ be a Feynman graph. The corresponding \textit{Feynman integral in momentum space representation} in $D\in\mathbb{N}$ dimensions is
    \begin{equation}\label{eq:Feynman_integral}
        I(G)(p):=\int_{\mathbb{R}^{h_1(G)D}}\mathrm{d}^{h_1(G)D}k\prod_{e\in E(G)}\frac{1}{((K_e(k)+P_e(p))^2+m_e^2)^{\lambda_e}},
    \end{equation}
    where $K_e$ and $P_e$ are the linear maps obtained as described above.
\end{definition}
As mentioned above, we have omitted some constant factors in this definition as they do not play a role in the analytical structure. These factors are only needed to compare numerical values obtained from Feynman integrals with the experiment. Note that it suffices to understand the bridge-less graphs $G$, or one-particle-reducible (1PI) graphs as they are called in the physics literature, to understand all Feynman integrals, as integrals corresponding to graphs with bridges factorize into 1PI contributions.\\
Up to this point, we have not stated what complex manifold we would like to (or even can) consider a Feynman integral to be a function on. In this work, we focus on the dependence on the external momenta. The masses are regarded as fixed and positive (the massless case works differently and we postpone the discussion of this case to future research). Thus, an obvious choice would be $(\mathbb{C}^D)^{\sum_{v\in V(G)}\phi(v)}$. But this can be simplified. First of all, the external momenta are restricted to a hyperplane by overall momentum conservation. In particular, we can express one of the momenta as minus the sum of all the others so that $(\mathbb{C}^D)^{\sum_{v\in V(G)}\phi(v)-1}$ would be a sufficient space to work with. Furthermore, it is well-known in physics that a Feynman integral is Lorentz-invariant, i.e., applying the same Lorentz transformation to all external momenta does not change the value of the integral. In our setup, this statement takes a slightly different form: Since we need to work with arbitrary complex momenta and not just Minkowski momenta, our transformation group is different. But the general idea stays the same: The integral should be invariant under all linear transformations of the external momenta leaving the products $p_ip_j$ unchanged. This group is the Lie group $O(D,\mathbb{C})$ of complex orthogonal $D\times D$-matrices. We define an action of $O(D,\mathbb{C})$ on $(\mathbb{C}^D)^n$ by
\begin{equation}\label{eq:action_orthogonal_group}
    g\cdot (x_1,\ldots,x_n)=(g\cdot x_1,\ldots,g\cdot x_n)
\end{equation}
for all $g\in O(D,\mathbb{C})$ and all $n\in\mathbb{N}^\ast$, where the multiplication on the right is just regular matrix-vector multiplication. The invariance of Feynman integrals under $O(D,\mathbb{C})$ for the one-loop case is proven in the next section.

\section{One-Loop Feynman Integrals}\label{sec:one_loop}
We now begin our investigation of one-loop Feynman graphs, i.e., Feynman graphs $G$ with $h_1(G)=1$. As mentioned in Section \ref{sec:feynman_integrals}, it suffices to consider the 1-particle-irreducible graphs to understand their analytic structure. The 1PI graphs with one loop are the cycle graphs $C_n$ (viewed as Feynman graphs) shown in the following figure:
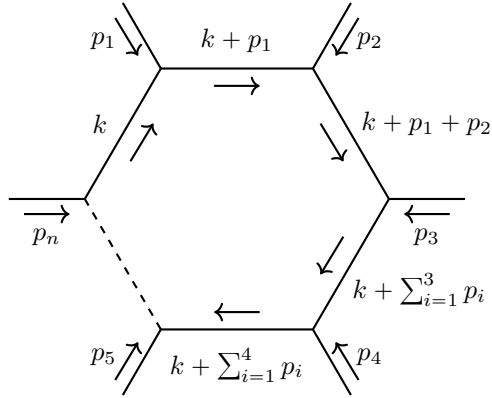
\begin{center}
\begin{tikzpicture}[thick]
\draw (-2,0) to (-1,1.73);
\draw (-3,0) to (-2,0);
\draw [<-] (-2.2,-0.2) -- (-2.8,-0.2);
\draw node at (-2.5,-0.5) {$p_n$};

\draw [<-] (-1.1,1) to (-1.4,0.5);
\draw node at (-1.8,1.0) {$k$};

\draw (-1,1.73) to (1,1.73);
\draw (-1.5,2.6) to (-1,1.73);
\draw [->] (-1.6,2.4) -- (-1.3,1.9);
\draw node at (-1.75,2.1) {$p_1$};

\draw [->] (-0.3,1.5) to (0.3,1.5);
\draw node at (0,2.1) {$k+p_1$};

\draw (1,1.73) to (2,0);
\draw (1.5,2.6) to (1,1.73);
\draw [->] (1.6,2.4) -- (1.3,1.9);
\draw node at (1.75,2.1) {$p_2$};

\draw [->] (1.1,1) to (1.4,0.5);
\draw node at (2.5,1) {$k+p_1+p_2$};

\draw (2,0) to (1,-1.73);
\draw (2,0) to (3,0);
\draw [<-] (2.2,-0.2) -- (2.8,-0.2);
\draw node at (2.5,-0.5) {$p_3$};

\draw [<-] (1.1,-1) to (1.4,-0.5);
\draw node at (2.4,-1.2) {$k+\sum_{i=1}^3p_i$};

\draw (1,-1.73) to (-1,-1.73);
\draw (1.5,-2.6) to (1,-1.73);
\draw [->] (1.6,-2.4) -- (1.3,-1.9);
\draw node at (1.75,-2.1) {$p_4$};

\draw [->] (0.3,-1.5) to (-0.3,-1.5);
\draw node at (0,-2.2) {$k+\sum_{i=1}^4p_i$};

\draw [dashed] (-1,-1.73) to (-2,0);
\draw (-1.5,-2.6) to (-1,-1.73);
\draw [->] (-1.6,-2.4) -- (-1.3,-1.9);
\draw node at (-1.75,-2.1) {$p_5$};

\end{tikzpicture}
\end{center}
On the level of (multi-)sets, this means
\begin{equation}
    V(C_n):=\{1,\ldots,n\},\quad E(C_n):=\{\{1,2\},\{2,3\},\ldots,\{n-1,n\},\{1,n\}\}.
\end{equation}
Note that we assigned exactly one external momentum to each vertex, i.e., the external structure $\phi$ of the graphs under consideration is simply given by $\phi(v)=1$ for all $v\in V(C_n)$.\footnote{For physicists, this means we consider Feynman graphs in $\phi^3$-theory, where $\phi$ stands for a scalar field and the exponent gives the power with which this field occurs in the Lagrangian. The later is also the allowed valency for the vertices of graphs appearing in a perturbative expansion.} We could consider more general external structures, but this bears no relevance to our discussion: If there is more than one line attached to a vertex, the integral depends only on the sum of all external momenta going into that vertex. In the coordinates we chose, the general one-loop Feynman integral in momentum space representation in $D$ dimensions from Definition \ref{defn:feynman_integral} reads
\begin{equation}\label{eq:one_loop_1}
    \begin{split}
        I(C_n)(p)&=\int_{\mathbb{R}^D}\frac{\mathrm{d}^Dk}{\prod_{i=1}^n((k+\sum_{j=1}^{i-1}p_j)^2+m_i^2)^{\lambda_i}}\\
        &=\int_{\mathbb{R}^D}\frac{\mathrm{d}^Dk}{\prod_{i=1}^n((k+P^{(i)}(p))^2+m_i^2)^{\lambda_i}}.
    \end{split}
\end{equation}
To ease notation, we set $P^{(i)}(p):=\sum_{j=1}^{i-1}p_j$ for all $i\in\{1,\ldots,n\}$. Recall that we agreed to fix all masses $m_i$ to be real and positive. Of course, the integral \eqref{eq:one_loop_1} is not well-defined in general. But for $p\in(\mathbb{R}^D)^{n-1}$ the integral \eqref{eq:one_loop_1} converges absolutely if and only if $2\text{Re}(\lambda)>D$. This is an application of Weinberg's famous Power Counting Theorem \cite{power-counting} in its simplest form (where no subdivergencies need to be considered since all proper subgraphs of $C_n$ are forests and hence, trivially \enquote{converge} as there is no integration to be performed).\\
The Feynman graph $C_2$ will accompany us as a running example throughout this section to illustrate all the ideas as they occur:
\begin{example}
    Consider the Feynman graph $C_2$:
    \begin{center}
        $C_2\quad=\quad$
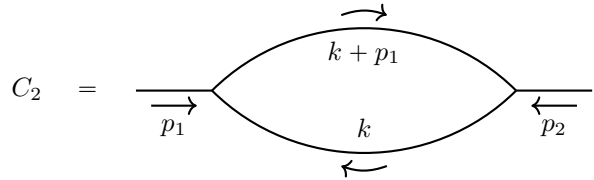
        \begin{tikzpicture}[thick, baseline={([yshift=-.5ex]current bounding box.center)},vertex/.style={anchor=base,
    circle,fill=black!25,minimum size=18pt,inner sep=2pt}]
            \draw (-2,0) to [out=45, in=135] (2,0);
            \draw [->] (-2.8,-0.2) -- (-2.2,-0.2);
            \draw node at (-2.5,-0.5) {$p_1$};
            \draw (-2,0) to [out=-45, in=-135] (2,0);
            \draw [->] (-0.3,1.0) to [out=20, in=160] (0.3,1.0);
            \draw node at (0,0.5) {$k+p_1$};
            \draw [<-] (-0.3,-1.0) to [out=-20, in=-160] (0.3,-1.0);
            \draw node at (0,-0.5) {$k$};
            \draw (-3,0) -- (-2,0);
            \draw (2,0) -- (3,0);
            \draw [<-] (2.2,-0.2) -- (2.8,-0.2);
            \draw node at (2.5,-0.5) {$p_2$};
        \end{tikzpicture}
    \end{center}
    In terms of (multi-)sets, this means $C_2=(G,\varphi)$ with
    \begin{equation}
        G=(\{1,2\},\{\{1,2\},\{1,2\}\}).
    \end{equation}
    and $\varphi(v)=1$ for $v=1,2$. The corresponding Feynman integral in $D$ dimensions reads
    \begin{equation}
        \int_{\mathbb{R}^D}\frac{\mathrm{d}^Dk}{(k^2+m_1^2)^{\lambda_1}((k+p_1)^2+m_2^2)^{\lambda_2}}.
    \end{equation}
    It does not depend on $p_2$ and in fact momentum conservation demands $p_2=-p_1$.
\end{example}
For reasons that become apparent further below, we exclude a certain subset of momenta from our parameter space:
\begin{definition}
    For all $n\in\mathbb{N}^\ast$, we define
    \begin{equation}
        T_n:=\{p\in(\mathbb{C}^D)^{n-1} \;\vert\; \det
        \begin{pmatrix}
            p_1p_1 & p_1p_2 & \cdots & p_1p_{n-1} \\
            p_2p_1 & p_2p_2 & \cdots & p_2p_{n-1} \\
            \vdots & \vdots & \ddots & \vdots \\
            p_{n-1}p_1 & p_{n-1}p_2 & \cdots & p_{n-1}p_{n-1}
        \end{pmatrix}\neq0 \}.
    \end{equation}
\end{definition}
The configurations of external momenta in $(\mathbb{C}^D)^{n-1}\backslash T_n$ are colinear and behave rather differently. Examples have shown that for these momentum configurations, Feynman integrals exhibit poles instead of essential singularities and hence, there is no associated monodromy. A discussion of these points is beyond the scope of this paper and postponed to future research. Note that $p\in T_n$ implies that the momenta $p_1,\ldots,p_{n-1}$ are linearly independent (over $\mathbb{C}$) and in particular $T_n\neq\emptyset$ if and only if $D\geq n-1$. During the course of this section, we will find that for momenta in $T_n$ only simple pinches occur which can be analyzed by the techniques from Subsection \ref{subsec:singular_integrals}.

\subsection{Compactification and Stratification}
In the form \eqref{eq:one_loop_1}, we can not yet apply the techniques from Subsection \ref{subsec:singular_integrals} to $I(C_n)$. We first have to compactify the integration cycle as well as its ambient space. In the one-loop case, this can be done without substantial problems. For details on the problems that occur when working with multiple loops, see \cite{max}. There are various ways to achieve a compactification, but we stick to the arguably simplest one for the purpose of this work. Further, below we show that in the cases we are interested in, genuine pinches which trap the integration cycle appear only at points $k\in\mathbb{C}^D$ at \enquote{finite distance,} i.e., outside of the set of additional points the compactification introduces. So the chosen compactification is not particularly important for our purposes.\\
There is one rather obvious way to achieve the desired compactification in the case of odd $D$. We view the integration domain $\mathbb{R}^D$ as being embedded in the complex analytic manifold $\mathbb{C}^D$. The ambient space $\mathbb{C}^D$ can in turn be viewed as being embedded in the compact complex analytic manifold $\mathbb{C}\mathbb{P}^D$. Applying the pull-back of the inverse of the natural inclusion
\begin{equation}
    i:\mathbb{C}^D\hookrightarrow\mathbb{C}\mathbb{P}^D,\quad z\mapsto[1:z]
\end{equation}
restricted to its image (i.e., viewed as a biholomorphic map $\mathbb{C}^D\overset{\sim}\to i(\mathbb{C}^D)=\mathbb{C}\mathbb{P}^D\backslash H_\infty$) to the integral \eqref{eq:one_loop_1}, we obtain
\begin{equation}\label{eq:one_loop_2}
    I(C_n)(p)=\int_{\mathbb{R}\mathbb{P}^D}\frac{u^{2\lambda-D-1}\cdot\Omega_D}{\prod_{i=1}^n((k+uP^{(i)}(p))^2+u^2m_i^2)^{\lambda_i}}=:\int_{\mathbb{R}\mathbb{P}^D}\omega_{n,D}(p),
\end{equation}
where
\begin{equation}
    \Omega_D:=u\cdot d^Dk-\sum_{i=0}^{D-1}(-1)^{i}k_i\cdot du\wedge dk_0\wedge\cdots\wedge\widehat{dk_i}\wedge\cdots\wedge dk_{D-1}
\end{equation}
is the differential form from Lemma \ref{lem:diff_form_lemma_1}. Here, we denote the additional (homogeneous) coordinate introduced by the inclusion into complex projective space by $u$ instead of $k_0$ since, as mentioned above, it is customary in the physics literature to index the components of $k$ from 0 to $D-1$ instead of from 1 to $D$. This notation also helps to render the conceptual difference between the coordinate $u$ and the coordinates given by the $D$ components of $k$ more visible. Note also that we replaced the integration domain $i(\mathbb{R}^D)$ by its closure $\overline{i(\mathbb{R}^D)}=\mathbb{R}\mathbb{P}^D$. This does not affect the value of the integral as $\mathbb{R}\mathbb{P}^D-i(\mathbb{R}^D)=H_\infty\cap\mathbb{R}\mathbb{P}^D$ has Lebesgue measure 0.\\
The same idea works for even $D$ with a slight modification: In this case, $\mathbb{R}\mathbb{P}^D$ is not orientable, so the integral \eqref{eq:one_loop_2} does not make any sense. However, we can still apply the program from \cite{pham} by lifting to the oriented double cover $S^D$ of $\mathbb{R}\mathbb{P}^D$. The geometric part of the analysis, which does not require the integration cycle to be oriented, can be performed on the level of $\mathbb{R}\mathbb{P}^D$ embedded in the compact space $\mathbb{C}\mathbb{P}^D$ while the actual integration can be performed on the double cover. The details of this can be found in the second part of this work.\\
Now, we need to establish some notation. We write
\begin{equation}
    Q_i(p):\mathbb{C}^{D+1}\to\mathbb{C},\quad (u,k)\mapsto (k+uP^{(i)}(p))^2+u^2m_i^2\qquad\forall i\in\{1,\ldots,n\}
\end{equation}
for all $p\in(\mathbb{C}^D)^{n-1}$. For all $i\in\{1,\ldots,n\}$, we set
\begin{equation}
    S_i:=\{([u:k],p)\in\mathbb{C}\mathbb{P}^D\times T_n\;\vert\; (k+uP^{(i)}(p))^2+u^2m_i^2=0\},
\end{equation}
and the corresponding fiber (with respect to the obvious projection) at $p\in T_n$ by
\begin{equation}
    S_i(p):=\{[u:k]\in\mathbb{C}\mathbb{P}^D \;\vert\; (k+uP^{(i)}(p))^2+u^2m_i^2=0\}.
\end{equation}
Furthermore, we set $S:=\bigcup_{i=1}^nS_i$ and $S(p):=\bigcup_{i=1}^nS_i(p)$ for all $p\in T_n$. Note that $S_i(p)$ (resp. $S_i$) is a complex analytic closed submanifold of $\mathbb{C}\mathbb{P}^D$ (resp. $\mathbb{C}\mathbb{P}^D\times T_n$) of (complex) codimension 1 for all $p\in T_n$ and all $i\in\{1,\ldots,n\}$ by Proposition \ref{prop:projective_submanifold}. Indeed, the gradient of the defining equation $Q_i(p)(u,k)=0$ in $\mathbb{C}^{D+1}\backslash\{0\}$ vanishes nowhere:
\begin{equation}
    \frac{\partial Q_i(p)}{\partial(u,k)}(u,k)=2\begin{pmatrix} k+uP^{(i)}(p) \\ kP^{(i)}+u((P^{(i)}(p))^2+m_i^2)\end{pmatrix}\overset{!}=0
\end{equation}
implies $k=-uP^{(i)}(p)$ by the first $D$ equations and thus, $u^2m_i^2=0$ by the last equation. Since $m_i^2\neq0$, this means $u=0$ which in turn implies $k=0$, a contradiction. Thus, the zero locus $S(p)$ is the union of a finite number of closed complex analytic submanifolds for all $p\in T_n$. The integration domain $\mathbb{R}\mathbb{P}^D$ is now a compact $D$-cycle in $\mathbb{C}\mathbb{P}^D\backslash S(p)$ for any $p\in(\mathbb{R}^D)^{n-1}$. Indeed, for $(u,k)\in\mathbb{R}^{D+1}\backslash\{0\}$, $p\in(\mathbb{R}^D)^{n-1}$ and $i\in\{1,\ldots,n\}$, the equation
\begin{equation}
    Q_i(p)(u,k)=(k+uP^{(i)}(p))^2+u^2m_i^2\overset{!}=0
\end{equation}
implies $u=0$ and thus, $k=0$, again a contradiction. The integrand of \eqref{eq:one_loop_2} is a holomorphic $D$-form on $\mathbb{C}\mathbb{P}^D\backslash S(p)$ for all $p\in T_n$ which depends holomorphically on $p$, i.e.,
\begin{equation}
    \omega_{n,D}\in\Omega^D(((\mathbb{C}\mathbb{P}^D\times T_n)\backslash S)/T_n).
\end{equation}
Therefore, we can conclude that $I(C_n)$ defines a holomorphic function outside of its Landau surface. Now, we would like to compute the Landau surface of $I(C_n)$ as in Definition \ref{defn:landau_surface}. Further below, we introduce a stratification on $S$ which allows us to see that the Landau surface is given precisely by all $p\in T_n$ such that the finite parts of the $S_i(p)$ are not in general position and effectively ignore what happens at infinity. This is the case if and only if there is an index set $I\subset\{1,\ldots,n\}$ and a point $k\in\bigcap_{i\in I}(S_i(p)-H_\infty)$ (where we implicitly use inhomogeneous coordinates) such that the normal vectors of the $S_i(p)-H_\infty$ with $i\in I$ at $k$ are linearly dependent. Thus, the Landau surface consists of all points $p\in T_n$ for which a solution
\begin{equation}
    (\alpha,k,p)\in(\mathbb{C}^n\backslash\{0\})\times\mathbb{C}^D\times T_n
\end{equation}
with
\begin{equation}\label{eq:landau_equations_1}
    \alpha_i=0\quad\text{or}\quad Q_i(p)(1,k)=0\qquad\forall i\in\{1,\ldots,n\}
\end{equation}
and
\begin{equation}\label{eq:landau_equations_2}
    \sum_{i=1}^n\alpha_i\frac{\partial Q_i(p)}{\partial k}(1,k)=0 \qquad\Leftrightarrow\qquad \sum_{i=1}^n\alpha_i(k+P^{(i)}(p))=0
\end{equation}
exists. These equations are the famous \textit{Landau equations} \cite{landau}. As a remark, it should be mentioned that $\sum_{i=1}^n\alpha_i=0$ implies $\sum_{i=1}^n\alpha_iP^{(i)}(p)=0$. This means that the momenta $p_1,\ldots,p_{n-1}$ are linearly dependent so that $p\notin T_n$. Hence, when solving the Landau equations, we can always assume $\sum_{i=1}^n\alpha_i=1$ without loss of generality by dividing $\alpha$ by $\sum_{i=1}^n\alpha_i\neq0$. We denote the set of all $(\alpha,k,p)$ satisfying these equations by
\begin{equation}
    B_n:=\{(\alpha,k,p)\in(\mathbb{C}^n\backslash\{0\})\times\mathbb{C}^D\times T_n \;\vert\; (\alpha,k,p)\text{ satisfies }\eqref{eq:landau_equations_1}\text{ and }\eqref{eq:landau_equations_2}\}.
\end{equation}
Let
\begin{equation}
    \pi:(\mathbb{C}^n\backslash\{0\})\times\mathbb{C}^D\times T_n\twoheadrightarrow T_n, \quad (\alpha,k,p)\mapsto p
\end{equation}
be the canonical projection and for now define the Landau surface of $C_n$ to be $L_n:=\pi(B_n)$. Note that $L_n$ is a closed set. In a moment, we shall see that $L_n$ agrees with the Landau surface of $I(C_n)$ as in Definition \ref{defn:landau_surface}, justifying this terminology.\\
Note that if we were to consider the case $u=0$ as well, the equations to test for linear dependence of the normal vectors become
\begin{equation}\label{eq:landau_second_type}
    k^2=0,\qquad \sum_{i=1}^na_ik=0,\qquad \sum_{i=1}^n\alpha_ikP^{(i)}(p)=0.
\end{equation}
We remark that \eqref{eq:landau_second_type} can only have a solution if $\sum_{i=1}^n\alpha_i=0$. Nevertheless, these equations generally have a solution for every $p\in T_n$ which poses a problem. To circumvent this issue, we introduce a Whitney stratification of $S$ such that some strata lie entirely in the finite part of $\mathbb{C}\mathbb{P}^D$ and the remaining ones entirely in $H_\infty$. Then, we show that the restriction of the canonical projection to the strata at infinity is a submersion.
\begin{example}\label{ex:landau_surface}
    Again considering the simplest example $C_2$, the corresponding Landau equations read
    \begin{equation}\label{eq:landau_example_1}
        \alpha_1 = 0 \quad\text{or}\quad k^2+m_1^2=0,
    \end{equation}
    \begin{equation}\label{eq:landau_example_2}
        \alpha_2 = 0 \quad\text{or}\quad (k+p_1)^2+m_2^2=0,
    \end{equation}
    \begin{equation}\label{eq:landau_example_3}
        \alpha_1k+\alpha_2(k+p_1) = 0.
    \end{equation}
    First assume that $\alpha_1=0$. Then, $\alpha_2\neq0$ and by equation \eqref{eq:landau_example_3} we have $k+p_1=0$. But by \eqref{eq:landau_example_2} this implies $m_2=0$ which we explicitly excluded. A similar argument applies to the case $\alpha_2=0$ so that a solution $(\alpha,k,p)$ to the Landau equations must satisfy $\alpha_1,\alpha_2\neq0$. Now, assume $\alpha_1+\alpha_2=0$. Then, \eqref{eq:landau_example_3} implies $p_1=0$ (so $p\notin T_2$ which we already knew from the discussion above) and from \eqref{eq:landau_example_1} and \eqref{eq:landau_example_2} we get $m_1^2=m_2^2$. Since we assumed the masses to be real and positive, this means $m_1=m_2$. If $\alpha_1+\alpha_2\neq0$ on the other hand, we can assume $\alpha_1+\alpha_2=1$ without loss of generality by dividing by $\alpha_1+\alpha_2$ if necessary. Then, from \eqref{eq:landau_example_3} we obtain $k=-\alpha_2p_1$. Plugging this into \eqref{eq:landau_example_1}, we obtain $\alpha_2^2=-\frac{m_1^2}{p_1^2}$ (note that $p_1^2=0$ is not possible due to $m_1\neq0$), and thus, we get
    \begin{equation}\label{eq:landau_example_4}
        \left(1\pm\sqrt{-\frac{m_1^2}{p_1^2}}\right)^2p_1^2+m_2^2=0 \qquad\Leftrightarrow\qquad -p_1^2=(m_1\pm m_2)^2.
    \end{equation}
    from \eqref{eq:landau_example_2}. Note that equation \eqref{eq:landau_example_4} also covers the case $\alpha_1+\alpha_2=0$. In conclusion, the Landau surface of $C_2$ is given by
    \begin{equation}
        L_2=\{p_1\in T_2 \;\vert\; -p_1^2=(m_1\pm m_2)^2\}
    \end{equation}
    which can be conveniently written as the zero locus of
    \begin{equation}
        T_2\to\mathbb{C},\quad p_1\mapsto\lambda(-p_1^2,m_1^2,m_2^2)=(-p_1^2-(m_1+m_2)^2)(-p_1^2-(m_1-m_2)^2),
    \end{equation}
    with $\lambda:\mathbb{C}^3\to\mathbb{C}$ being the \textit{Källén-function} defined by
    \begin{equation}\label{eq:kallen_fc}
        \lambda(a,b,c):=a^2+b^2+c^2-2(ab+bc+ca)
    \end{equation}
    for all $a,b,c\in\mathbb{C}$.
\end{example}
Let us take a moment to reflect on the combinatorial structure of the Landau surfaces associated with one-loop graphs. Let $n\geq2$ and suppose $I\subset E(C_n)$. Now, we consider solutions $(\alpha,k,p)\in B_n$ to the Landau equations \eqref{eq:landau_equations_1} and \eqref{eq:landau_equations_2} with $\alpha_i\neq0$ for all $i\in I$ and $\alpha_i=0$ if $i\notin I$. We denote the set of such solutions by $B_{n,I}\subset B_n$. Note that such solutions are in bijective correspondence with solutions $(\alpha',k',p')$ to the Landau equations of $C_n/(E(G)\backslash I)$, the graph with all edges not in $I$ contracted to a point. Therefore, the Landau surface $L_n$ consists of solutions with $\alpha_i\neq0$ for all $i\in\{1,\ldots,n\}$ and parts which are the Landau surfaces of smaller graphs with some external momenta replaced by the appropriate sums of momenta obtained from shrinking the edges corresponding to vanishing $\alpha_i$. Motivated by this observation, we define
\begin{equation}
    L_{n,I}:=\pi(B_{n,I})
\end{equation}
for all $I\subset E(C_n)$, again with
\begin{equation}
    \pi:(\mathbb{C}^n\backslash\{0\})\times\mathbb{C}^D\times T_n\twoheadrightarrow T_n, \qquad (\alpha,k,p)\mapsto p
\end{equation}
the canonical projection. Clearly, we have $L_n=\bigcup_{I\subset E(C_n)}L_{n,I}$.\footnote{Remark: We can also form the sets $\tilde{L}_{n,I}:=\bigcup_{J\supset I}L_{n,J}$. Then, the collection of sets $\tilde{L}_{n,I}$ is partially ordered by inclusion and in fact constitutes a finite simplicial poset. The associated simplicial complex consists of a single simplex. Gluing together all simplices obtained from a graph $C_n$ by permuting the masses $m_1,\ldots,m_n$ along their common faces yields the complex of holocolored one-loop graphs as investigated in \cite{coloredgraphs}.} Furthermore, $L_{n,I}=\emptyset$ whenever $\vert I\vert\leq1$. This decomposition of the Landau surface is useful to organize calculations but also nice from a theoretical point of view due to the following two propositions. Their proof is postponed to Subsection \ref{subsec:cutkosky}, where we have a more convenient description of $L_n$ at our disposal.
\begin{proposition}\label{prop:landau_decomp_1}
    Let $n\geq2$ and $I\subset E(C_n)$ with $\vert I\vert\geq2$. Then, $L_{n,I}$ is a complex analytic submanifold of codimension 1 in $T_n$.
\end{proposition}
\begin{proposition}\label{prop:landau_decomp_2}
    Let $I_1\subsetneq\cdots\subsetneq I_k\subset E(C_n)$. Then, the $L_{n,I_1},\ldots,L_{n,I_k}$ intersect in general position.
\end{proposition}
\paragraph{\textbf{Stratification}}
Let $I=\{i_1,\ldots,i_m\}\subset\{1,\ldots,n\}$ with $i_1<\cdots i_m$ and let us consider the defining equations for $S_{i_1}\cap\cdots\cap S_{i_m}$. The intersection of these $m$ manifolds is given by all $([u:k],p)\in\mathbb{C}\mathbb{P}^D\times(T_n\backslash L_n)$ such that the following $m$ equations hold simultaneously:
\begin{equation}
    (k+uP^{(i)}(p))^2+u^2m_{i}^2=0 \qquad \forall i\in I
\end{equation}
Now, plugging the first equation (the one labeled by $i_1$) into the remaining $m-1$ equations yields the following equivalent set of equations:
\begin{equation}
    (k+uP^{(i_1)}(p))^2+u^2m_{i_1}^2=0
\end{equation}
and 
\begin{equation}
    \begin{split}
        &2u(P^{(i)}(p)-P^{(i_1)}(p))(k+uP^{(i)}(p))\\
        +&u^2((P^{(i)}(p)-P^{(i_1)}(p))^2-m_{i_1}^2+m_i^2)=0
    \end{split}
\end{equation}
for all $i\in I\backslash\{i_1\}$. For notational convenience, we define the holomorphic function
\begin{equation}
    \begin{array}{ccc}
        f_{i,j}:(\mathbb{C}^{D+1}\backslash\{0\})\times T_n &\to & \mathbb{C}\\
        \hspace{1.6cm}((u,k),p) & \mapsto & \substack{\hspace{0.31cm} 2(P^{(j)}(p)-P^{(i)}(p))(k+uP^{(j)}(p)) \\ +u((P^{(j)}(p)-P^{(i)}(p))^2-m_{i}^2+m_{j}^2)}
    \end{array}
\end{equation}
for all $i,j\in\{1,\ldots,n\}$ so that the last $m-1$ equations read
\begin{equation}
    u\cdot f_{i_1,i}((u,k),p)=0 \qquad \forall i\in I\backslash\{i_1\}.
\end{equation}
Note that
\begin{equation}
    f_{i,j}((0,k),p)=2(P^{(j)}(p)-P^{(i)}(p))k
\end{equation}
and that
\begin{equation}\label{eq:strata_p_at_infty}
    P^{(j)}(p)-P^{(i)}(p)=\sum_{l=i}^{j-1}p_l
\end{equation}
for all $i,j\in\{1,\ldots,n\}$ with $j\geq i$. For all $l\in\{1,\ldots,n-1\}$, we also compute the derivatives
\begin{equation}\label{eq:strata_infinity_gradient_1}
    \partial_{p_l}[(k+uP^{(i_1)}(p))^2+u^2m_{i_1}^2]\vert_{u=0}=0
\end{equation}
and for all $i,j\in\{1,\ldots,n\}$ with $j\geq i$ the derivatives
\begin{equation}\label{eq:strata_infinity_gradient_2}
    \partial_{p_l}f_{i,j}((u,k),p)\vert_{u=0}=\begin{cases} 2k & \text{if }i\leq l\leq j-1 \\ 0 & \text{otherwise}.\end{cases}
\end{equation}
Now, we define
\begin{equation}
    B_{I,\text{fin}}:=\left\{([u:k],p)\in\mathbb{C}\mathbb{P}^D\times(T_n\backslash L_n) \;\vert\; \substack{(k+uP^{(i_1)}(p))^2+u^2m_{i_1}^2=0 \\ \forall i\in I\backslash\{i_1\}\,:\,f_{i_1,i}((u,k),p)=0}\right\}
\end{equation}
and
\begin{equation}
    B_{I,\infty}:=\{([u:k],p)\in B_{I,\text{fin}} \;\vert\; u=0\}.
\end{equation}
\begin{lemma}\label{lem:B_is_manifold}
    The sets $B_{I,\text{fin}}$ (resp. $B_{I,\infty}$) define codimension $\vert I\vert$ (resp. codimension $\vert I\vert+1$) complex analytic submanifolds of $\mathbb{C}\mathbb{P}^D\times(T_n\backslash L_n)$.
\end{lemma}
\begin{proof}
    By the Implicit Function Theorem, it suffices to check that the gradients of the defining functions of $B_{I,\text{fin}}$ and $B_{I,\infty}$ are linearly independent. For all points $([u:k],p)\in B_{I,\text{fin}}$ with $u\neq0$, this follows immediately from $p\notin L_n$ and the definition of $L_n$. According to equations \eqref{eq:strata_infinity_gradient_1} and \eqref{eq:strata_infinity_gradient_2}, the derivatives for $B_{I,\text{fin}}$ with respect to $(k,p)$ at $([0:k],p)\in B_{I,\text{fin}}$ (divided by 2) read
    \begin{equation}\label{eq:vectors_B_is_manifold}
        \begin{pmatrix} k \\ 0 \\ \vdots \\ 0 \\ 0 \\ \vdots \\ 0 \\ 0 \\ \vdots \\ 0 \\ 0 \\ \vdots \\ 0 \\ 0 \\ \vdots \end{pmatrix},\quad\begin{pmatrix} P^{(i_2)}(p)-P^{(i_1)}(p) \\ 0 \\ \vdots \\ 0 \\ k \\ \vdots \\ k \\ 0 \\ \vdots \\ 0 \\ 0 \\ \vdots \\ 0 \\ 0 \\ \vdots \end{pmatrix},\quad\begin{pmatrix} P^{(i_3)}(p)-P^{(i_1)}(p) \\ 0 \\ \vdots \\ 0 \\ k \\ \vdots \\ k \\ k \\ \vdots \\ k \\ 0 \\ \vdots \\ 0 \\ 0 \\ \vdots \end{pmatrix},\quad\ldots
    \end{equation}
    Here, it is understood that the $(j+1)$th vector has $i_{j+1}-i_1$ consecutive entries with value $k$ (constituting a total of $(i_{j+1}-i_1)\cdot D$ entries), starting at $i_1\cdot D+1$. Since $k\neq0$, these gradients are clearly linearly independent. For $B_{I,\infty}$, the argument is similar. The only difference is that $B_{I,\infty}$ has the additional defining equation $u=0$ whose gradient is clearly linearly independent of the remaining vectors.
\end{proof}
Now, using our manifolds $B_{I,\text{fin}}$ and $B_{I,\infty}$, we introduce a Whitney stratification on $S$ which allows us to distinguish between the finite and infinite part of $S$. To this end, we first define the sets
\begin{equation}
    \mathcal{T}:=\{(I,x) \;\vert\; I\subset\{1,\ldots,n\},\;I\neq\emptyset,\; x\in\{\text{fin},\infty\}\}
\end{equation}
and
\begin{equation}
    \mathcal{S}:=\{1,\ldots,n\}\times\{\text{fin},\infty\}.
\end{equation}
We equip $\mathcal{S}$ with a partial order $\prec$ defined by
\begin{equation}
    \begin{split}
        &\forall (m_1,x_1),(m_2,x_2)\in\mathcal{S}\,:\, (m_1,x_1)\prec(m_2,x_2)\\
        \quad:\Leftrightarrow\quad &m_2\leq m_1\land (x_1,x_2)\neq(\text{fin},\infty).
    \end{split}
\end{equation}
For any non-empty $I=\{i_1,\ldots,i_m\}\subset\{1,\ldots,n\}$ with $i_1<\cdots<i_m$ we now define
\begin{equation}
    A_{I,\text{fin}}:=B_{I,\text{fin}}-\bigcup_{J\supsetneq I}B_{J,\text{fin}}-H_\infty\times(T_n\backslash L_n),
\end{equation}
as well as
\begin{equation}
    A_{I,\infty}:=B_{I,\infty}-\bigcup_{J\supsetneq I}B_{J,\infty}.
\end{equation}
Note that $A_{I,\text{fin}}\cap A_{J,\infty}=\emptyset$ for all non-empty $I,J\subset\{1,\ldots,n\}$. Also note that
\begin{equation}\label{eq:description_finite_strata}
    A_{I,\text{fin}}=\left\{([u:k],p)\in B_{I,\text{fin}} \;\vert\; \substack{ u\neq0,\;\forall i<i_1\,:\,(k+uP^{(i)}(p))^2+u^2m_i^2\neq0,\\ \forall i\notin I,\, i>i_1\,:\, f_{i_1,i}((u,k),p)\neq0}\right\}
\end{equation}
and
\begin{equation}\label{eq:description_infinite_strata}
    A_{I,\infty}=\left\{([u:k],p)\in B_{I,\infty} \;\vert\; \substack{\forall i<i_1\,:\,(k+uP^{(i)}(p))^2+u^2m_i^2\neq0,\\ \forall i\notin I,\, i>i_1\,:\, f_{i_1,i}((u,k),p)\neq0}\right\}.
\end{equation}
In what follows, we show that the $A_{I,x}$ (where $(I,x)\in\mathcal{T}$) can be used as strata for a Whitney stratification of $S$.
\begin{lemma}\label{lem:closure_lemma}
    For all non-empty $(I,x)\in\mathcal{T}$, we have
    \begin{equation}\label{eq:closure_lemma_1}
        \bar{A}_{I,x}=B_{I,x}.
    \end{equation}
\end{lemma}
\begin{proof}
    Let $(I,x)\in\mathcal{T}$. Clearly, $B_{I,x}$ is a closed set. Thus, it suffices to show that $B_{I,x}\subset\bar{A}_{I,x}$. So let $([u:k],p)\in B_{I,x}$ and let $U\subset B_{I,x}$ be an open neighborhood of $([u:k],p)$. We may assume without loss of generality that $U$ contains a coordinate neighborhood $V$ of $([u:k],p)$. By equations \eqref{eq:description_finite_strata} and \eqref{eq:description_infinite_strata}, the set $A_{I,x}$ is given by all points in $B_{I,x}$ at which a finite set of holomorphic functions does not vanish. By using coordinates, we can view these as holomorphic functions on $V$. Denote these functions by $f_1,\ldots,f_m:V\to\mathbb{C}$. Then, $\prod_{i=1}^mf_i:V\to\mathbb{C}$ is a holomorphic function on $V$ which is not identically zero. Hence it can not vanish on all of $V$. But this means $V$ (and thus also $U$) contains a point $([u':k'],p')\in V$ such that $f_i((u',k'),p)\neq0$ for all $i\in\{1,\ldots,m\}$. This means $([u':k'],p')\in A_{I,x}$, and we conclude that $([u:k],p)$ lies in the closure $\bar{A}_{I,x}$ of $A_{I,x}$.
\end{proof}
Now, we combine the $A_{I,x}$ to form the pieces of a stratification. For all $(m,x)\in\mathcal{S}$ we set
\begin{equation}
    A_{m,x}:=\bigcup_{\substack{I\subset\{1,\ldots,n\} \\ \vert I\vert=m}}A_{I,x}.
\end{equation}
\begin{proposition}\label{prop:S_decomposition}
    The $A_{m,x}$ form an $\mathcal{S}$-decomposition of $S$.
\end{proposition}
\begin{proof}
    Clearly, the collection of the $A_{m,x}$ is locally finite (as there are only finitely many pieces) and every $A_{m,x}$ is a locally closed set. Furthermore, it is not difficult to see that
    \begin{equation}
        \bigcup_{(I,1)\in\mathcal{T}}A_{I,x}=S-H_\infty\times(T_n\backslash L_n) \qquad\text{and}\qquad \bigcup_{(I,\infty)\in\mathcal{T}}A_{I,x}=S\cap(H_\infty\times (T_n\backslash L_n))
    \end{equation}
    so that
    \begin{equation}
        \bigcup_{(m,x)\in\mathcal{S}}A_{m,x}=\bigcup_{(I,x)\in\mathcal{T}}A_{I,x}=S.
    \end{equation}
    The frontier condition is also easy to verify: According to Lemma \ref{lem:closure_lemma}, we have
    \begin{equation}\label{eq:closure_of_union}
        \bar{A}_{m,x}=\bigcup_{\substack{I\subset\{1,\ldots,n\} \\ \vert I\vert=m}}\bar{A}_{I,x}=\bigcup_{\substack{I\subset\{1,\ldots,n\} \\ \vert I\vert=m}}B_{I,x}
    \end{equation}
    for all $(m,x)\in\mathcal{S}$. Since $B_{I,\infty}\subset H_\infty$ and $A_{I,\text{fin}}\cap H_\infty=\emptyset$ for all non-empty $I\subset\{1,\ldots,n\}$, we immediately see that $A_{m_1,x_1}\cap \bar{A}_{m_2,x_2}=\emptyset$ for all $(m_1,x_1),(m_2,x_2)\in\mathcal{S}$ whenever $x_1=\text{fin}$ and $x_2=\infty$. Furthermore, note that
    \begin{equation}
        A_{I,\text{fin}}\cap B_{J,\text{fin}}=A_{I,\infty}\cap B_{J,\text{fin}}=A_{I,\infty}\cap B_{J,\infty}=\emptyset
    \end{equation}
    for all non-empty subsets $I,J\subset\{1,\ldots,n\}$ such that $J$ is not a subset of $I$. In case $J\subset I$, it is also easy to see that
    \begin{equation}
        A_{I,\text{fin}}\subset B_{J,\text{fin}}, \quad A_{I,\infty}\subset B_{J,\text{fin}}, \quad A_{I,\infty}\subset B_{J,\infty}.
    \end{equation}
    Now, applying equation \eqref{eq:closure_of_union} readily yields the desired frontier condition.
\end{proof}
\begin{proposition}\label{prop:strata_manifolds}
    $A_{I,x}$ is a smooth manifold for every $(I,x)\in\mathcal{T}$ and $A_{m,x}$ is a smooth manifold for every $(m,x)\in\mathcal{S}$.
\end{proposition}
\begin{proof}
    By Lemma \ref{lem:B_is_manifold}, we know that the $B_{I,x}$ are smooth (even complex analytic) manifolds. By construction, each $A_{I,x}$ is an open subset of $B_{I,x}$ (with respect to the topology on $B_{I,x}$) and thus, a smooth manifold itself. Since the $A_{I,x}$ are disjoint, we see immediately that the $A_{m,x}$ are smooth manifolds as well (as they are the disjoint union of smooth manifolds).
\end{proof}
\begin{proposition}\label{prop:whitney_stratification}
    The decomposition above is a Whitney stratification of $S$. 
\end{proposition}
\begin{proof}
    By Proposition \ref{prop:S_decomposition}, we know that the $A_{m,x}$ form an $\mathcal{S}$-decomposition of $S$, which consists of smooth manifolds by Proposition \ref{prop:strata_manifolds}. It remains to check Whitney's condition B (see Definition \ref{defn:whitneys_condition_B}) for any pair of strata. All pairs $(A_{I,1},A_{J,1})$ satisfy condition B since the $A_{I,1}$ form the canonical stratification (see equation \eqref{eq:canonical_stratification}) of the manifolds
    \begin{equation}
        S_1-(H_\infty\times(T_n\backslash L_n)),\;\ldots,\;S_n-(H_\infty\times(T_n\backslash L_n)),
    \end{equation}
    which are in general position. For all pairs $(A_{I,\infty},A_{J,\infty})$ and $(A_{I,1},A_{J,\infty})$ with $J\subset I$, it is also easy to see that condition B holds: Locally these pairs look like
    \begin{equation}
        \{x\in\mathbb{C}^a \;\vert\; x_1\neq0\} \quad\text{and}\quad \{x\in\mathbb{C}^a \;\vert\; x_1=\cdots=x_b=0\}
    \end{equation}
    for appropriate $a,b\in\mathbb{N}$ with $a\geq b$. In these coordinates, the condition can be readily verified.
\end{proof}
With our Whitney stratification in place, we can now deduce that $I(C_n)$ defines a holomorphic function outside its Landau surface $L_n$ by showing that the restriction of the projection to any stratum $A_{I,\infty}$ at infinity is always a submersion. Thus, informally speaking, we can ignore the non-general position of the $S_i$ at infinity and can focus the discussion on finite internal momenta.
\begin{theorem}
    The one-loop Feynman integral $I(C_n)$ defines a holomorphic function on $T_n\backslash L_n$.
\end{theorem}
\begin{proof}
    By Proposition \ref{prop:whitney_stratification}, the (connected components of the) $A_{I,x}$ constitute the strata of a Whitney stratification of $S$. Thus, we can apply Thom's Isotopy Theorem \ref{thm:isotopy_theorem}. It is clear that $L_n$ is the set where the projection restricted to the strata of the form $A_{I,1}$ is not a submersion. Hence, it suffices to show that $\pi\vert_{A_{I,\infty}}$ is a submersion for all non-empty $I\subset\{1,\ldots,n\}$. So let $I=\{i_1,\ldots,i_m\}\subset\{1,\ldots,n\}$ be non-empty with $i_1<\cdots<i_m$. The tangent space at $([u:k],p)\in A_{I,\infty}$ can be identified with the orthogonal complement of the vectors
    \begin{equation}
        \begin{pmatrix} k \\ 0 \\ \vdots \\ 0 \end{pmatrix}, \quad \begin{pmatrix} P^{(j)}(p)-P^{(i_1)}(p) \\ (\partial_{p_1}P^{(j)}(p)-\partial_{p_1}P^{(i_1)}(p))\cdot k \\ \vdots \\ (\partial_{p_{n-1}}P^{(j)}(p)-\partial_{p_{n-1}}P^{(i_1)}(p))\cdot k \end{pmatrix},
    \end{equation}
    where $j$ runs over all elements of $I\backslash\{i_1\}$. This space is given by the solution to the homogeneous linear system of equations given by the matrix
    \begin{equation}
        \begin{pmatrix}
            k^T                         & 0      & \cdots & 0      & 0      & \cdots & 0      & 0      & \cdots & 0      & 0      & \cdots & 0      & \cdots & 0      & 0      & \cdots & 0 \\
            (\sum_{l=i_1}^{i_2-1}p_l)^T & 0      & \cdots & 0      & k^T    & \cdots & k^T    & 0      & \cdots & 0      & 0      & \cdots & 0      & \cdots & 0      & 0      & \cdots & 0 \\
            (\sum_{l=i_1}^{i_3-1}p_l)^T & 0      & \cdots & 0      & k^T    & \cdots & k^T    & k^T    & \cdots & k^T    & 0      & \cdots & 0      & \cdots & 0      & 0      & \cdots & 0 \\
            \vdots                      & \vdots & \ddots & \vdots & \vdots & \ddots & \vdots & \vdots & \ddots & \vdots & \vdots & \ddots & \vdots & \ddots & \vdots & \vdots & \ddots & \vdots \\
            (\sum_{l=i_1}^{i_m-1}p_l)^T & 0      & \cdots & 0      & k^T    & \cdots & k^T    & k^T    & \cdots & k^T    & k^T    & \cdots & k^T    & \cdots & k^T    & 0      & \cdots & 0
        \end{pmatrix},
    \end{equation}
    where the columns are the vectors from \eqref{eq:vectors_B_is_manifold}. The differential of the projection restricted to $A_{I,\infty}$ takes a vector $\begin{pmatrix} x \\ p' \end{pmatrix}$ in this tangent space to $p'$. So we need to check if for every $p'\in T_n\backslash L_n$ there is a solution $x\in\mathbb{C}^D$ to the inhomogeneous system of linear equations
    \begin{equation}\label{eq:strata_inhomogeneous_system}
        \begin{pmatrix}
            k^T \\ (\sum_{l=i_1}^{i_2-1}p_l)^T \\ \vdots \\ (\sum_{l=i_1}^{i_m-1}p_l)^T
        \end{pmatrix}\cdot x =
        \begin{pmatrix}
            0 \\ k^T\cdot\sum_{l=i_1}^{i_2-1}p_l' \\ \vdots \\ k^T\cdot\sum_{l=i_1}^{i_m-1}p_l'
        \end{pmatrix}.
    \end{equation}
    We distinguish two cases: If $m\leq D$, we can consider the smaller system
    \begin{equation}
        \begin{pmatrix}\label{eq:strata_inhomogeneous_system_red}
            (\sum_{l=i_1}^{i_2-1}p_l)^T \\ \vdots \\ (\sum_{l=i_1}^{i_m-1}p_l)^T
        \end{pmatrix}\cdot x =
        \begin{pmatrix}
            k^T\cdot\sum_{l=i_1}^{i_2-1}p_l' \\ \vdots \\ k^T\cdot\sum_{l=i_1}^{i_m-1}p_l'
        \end{pmatrix}.
    \end{equation}
    Since $p\in T_n$, the $\sum_{l=i_1}^{i_j-1}p_l$ must be linearly independent and hence, the matrix on the right has the full rank $m-1$. The solution space of \eqref{eq:strata_inhomogeneous_system_red} is thus an affine space of dimension $D-(m-1)\geq1$. In this space, there clearly exists an $x$ such that $k^Tx=0$ so that \eqref{eq:strata_inhomogeneous_system} can indeed be solved. If $m=D+1$ on the other hand (note that we can not have $m>D+1$ since $m\geq n$ and thus $n>D+1$, which implies $p\notin T_n$), we show that the matrix on the left-hand side of \eqref{eq:strata_inhomogeneous_system} has rank $m$, which immediately implies the existence of a solution. First note that in this case we have $I=\{1,\ldots,n\}$ which means $\sum_{l=i_1}^{i_j-1}p_l=\sum_{l=1}^{j-1}p_l$ for all $j\in\{1,\ldots,n\}$. It suffices to show that $k$ is linearly independent of the $\sum_{l=1}^{j-1}p_l$ which is equivalent to $k$ being linearly independent of the $p_j$. To see this, let us suppose the opposite is true. Then, there exist $\lambda_1,\ldots,\lambda_{n-1}\in\mathbb{C}$ (not all equal to zero) such that $k=\sum_{i=1}^{n-1}\lambda_ip_i$. But $([0:k],p)\in A_{I,\infty}$ requires
    \begin{equation}
        (P^{(j)}(p)-P^{(1)}(p))k=\sum_{i=1}^{n-1}\sum_{l=1}^{j-1}\lambda_ip_ip_l\overset{!}=0 \qquad \forall j\in\{2,\ldots,n\}.
    \end{equation}
    But this implies $p\notin T_n$, a contradiction. In conclusion, $\pi\vert_{A_{I,\infty}}$ is indeed a submersion.
\end{proof}
With our compactification in place, we can show the invariance of the holomorphic function defined by $I(C_n)$ with respect to the action \eqref{eq:action_orthogonal_group} of $O(D,\mathbb{C})$. To this end, we consider the following extension of the action \eqref{eq:action_orthogonal_group} to $\mathbb{C}\mathbb{P}^D$:
\begin{equation}
    \tilde{\sigma}:O(D,\mathbb{C})\times\mathbb{C}\mathbb{P}^D\to\mathbb{C}\mathbb{P}^D,\quad (g,[u:z])\mapsto g\cdot[u:z]:=[u:g\cdot z].
\end{equation}
Note that this is well-defined since, if $(u',z')=(\lambda u,\lambda z)$ for some $\lambda\in\mathbb{C}^\times$, we have
\begin{equation}
    g\cdot[u':z']=[\lambda u:g\cdot(\lambda z)]=[\lambda u:\lambda(g\cdot z)]=g\cdot[u:z]
\end{equation}
for all $g\in O(D,\mathbb{C})$. Furthermore, this action is clearly continuous. Note also that the Landau surface $L_n$ is invariant under our action on $T_n$: If $g\in O(D,\mathbb{C})$ and $(\alpha,k,g\cdot p)\in B_n$ is a solution to the Landau equations, then $(\alpha,g^{-1}\cdot k,p)$ is also a solution. Hence, $p\in L_n\;\Leftrightarrow\; g\cdot p\in L_n$ for all $g\in O(D,\mathbb{C})$. In fact this even shows $L_{n,I}=g\cdot L_{n,I}$ for all $I\subset\{1,\ldots,n\}$ and $g\in O(D,\mathbb{C})$. To show the invariance of our integral under $O(D,\mathbb{C})$, we need some information on how the differential forms in question transforms under this group. The following lemma establishes the very simple transformation behavior of $\omega_{n,D}(p)$ under $O(D,\mathbb{C})$:
\begin{lemma}\label{lem:pullback_by_ODC}
    For all $g\in O(D,\mathbb{C})$, we have
    \begin{equation}
        (\tilde{\sigma}(g,\cdot))^\ast\omega_{n,D}(g\cdot p)=\det(g)\cdot\omega_{n,D}(p).
    \end{equation}
\end{lemma}
\begin{proof}
    Obtained from a straightforward calculation.
\end{proof}

\begin{proposition}\label{prop:invariance_under_O(D,C)}
    Let $n\geq2$. The multivalued holomorphic function $I(C_n):T_n\backslash L_n\to\mathbb{C}$ is invariant under the action \eqref{eq:action_orthogonal_group} of $O(D,\mathbb{C})$, i.e., for any $p\in T_n\backslash L_n$ and any $g\in O(D,\mathbb{C})$, we have
    \begin{equation}
        I(C_n)(g\cdot p)=I(C_n)(p)
    \end{equation}
    when evaluating the left and right hand side on the same branch.
\end{proposition}
\begin{proof}
    Now, let $p\in T_n\backslash L_n$ and $g\in O(D,\mathbb{C})$. Let us first suppose that $g$ lies in the connected component of $O(D,\mathbb{C})$ containing the identity $e$. In particular, this means $\det(g)=1$. Then, if $\gamma_g:[0,1]\to O(D,\mathbb{C})$ is a path from $e$ to $g$, we can define a path
    \begin{equation}
        \gamma:[0,1]\to T_n\backslash L_n,\quad s\mapsto \gamma_g(s)\cdot p
    \end{equation}
    from $p$ to $g\cdot p$. Note that this is also well-defined as the Landau equations \eqref{eq:landau_equations_1} and \eqref{eq:landau_equations_2} are invariant under our action of $O(D,\mathbb{C})$ as mentioned above. We know that we can write
    \begin{equation}
        I(C_n)(p)=\int_\Gamma\omega_{n,D}(p)
    \end{equation}
    for an appropriate $D$-cycle $\Gamma\subset\mathbb{C}\mathbb{P}^D\backslash S(p)$. Defining
    \begin{equation}
        \sigma:\mathbb{C}\mathbb{P}^D\times[0,1]\to\mathbb{C}\mathbb{P}^D, \quad ([z],s)\mapsto \gamma_g(s)\cdot[z],
    \end{equation}
    we have $\sigma(\Gamma,s)\cap S(\gamma(s))=\emptyset$ for all $s\in[0,1]$. Indeed, if there was $[u:k]\in\sigma(\Gamma,s)\cap S(\gamma(s))$, there would exist $i\in\{1,\ldots,n\}$ such that
    \begin{equation}
        (k+uP^{(i)}(\gamma(s)))^2+u^2m_i^2=((\gamma_g(s))^{-1}k+uP^{(i)}(p))^2+u^2m_i^2=0.
    \end{equation}
    Since $\sigma([u:(\gamma_g(s))^{-1}k],s)=[u:k]$, this means
    \begin{equation}
        [u:(\gamma_g(s))^{-1}k]\in\sigma(\Gamma,0)\cap S(\gamma(0))=\Gamma\cap S(p),
    \end{equation}
    in contradiction to $\Gamma\cap S(p)=\emptyset$. Thus, by Corollary \ref{cor:extension_along_path}, we can analytically continue $I(C_n)$ from $p$ to $g\cdot p$ along $\gamma$ via
    \begin{equation}
        \begin{split}
            I(C_n)(\gamma(s))&=\int_{\sigma(\Gamma,s)}\omega_{n,D}(\gamma(s))=\int_{\gamma_g(s)\cdot\Gamma}\omega_{n,D}(\gamma_g(s)\cdot p)\\
            &=\int_\Gamma(\sigma(\cdot,s))^\ast\omega_{n,D}(\gamma_g(s)\cdot p)=\int_\Gamma\omega_{n,D}(p)=I(C_n)(p)
        \end{split}
    \end{equation}
    for all $s\in[0,1]$, where we used Lemma \ref{lem:pullback_by_ODC} in the second to last step.\\
    Now, if $g$ is not in the connected component containing $e$, then $g$ can be expressed as a product $g=r\cdot g'$ of an element $g'$ in the connected component containing $e$ and a reflection
    \begin{equation}
        r=\begin{pmatrix} -1 & 0 & \cdots & 0 \\ 0 & 1 & \cdots & 0 \\ \vdots & \vdots & \ddots & \vdots \\
           0 & 0 & \cdots & 1
          \end{pmatrix}.
    \end{equation}
    Thus, it suffices to check the claim for $g=r$. Using $\det(r)=-1$, we obtain
    \begin{equation}
        \begin{split}
            I(C_n)(r\cdot p_\text{in})&=\int_{\mathbb{R}\mathbb{P}^D}\frac{u^{2\lambda-D-1}\cdot\Omega_D}{\prod_{i=1}^n((k+uP^{(i)}(r\cdot p_\text{in}))^2+u^2m_i^2)^{\lambda_i}}\\
            &=-\int_{r\cdot\mathbb{R}\mathbb{P}^D}\frac{u^{2\lambda-D-1}\cdot\Omega_D}{\prod_{i=1}^n((r\cdot k+uP^{(i)}(r\cdot p_\text{in}))^2+u^2m_i^2)^{\lambda_i}}\\
            &=\int_{\mathbb{R}\mathbb{P}^D}\frac{u^{2\lambda-D-1}\cdot\Omega_D}{\prod_{i=1}^n((k+uP^{(i)}(p_\text{in}))^2+u^2m_i^2)^{\lambda_i}}=I(C_n)(p_\text{in})
        \end{split}
    \end{equation}
    for all $p_\text{in}\in(\mathbb{R}^D)^{n-1}$ (where $I(C_n)$ is evaluated on the principal branch). Thus, for any path $\gamma:[0,1]\to T_n\backslash L_n$ from $p_\text{in}$ to $p$, the analytic continuation along the path $\gamma':[0,1]\to T_n\backslash L_n$ from $r\cdot p_\text{in}$ to $r\cdot p$ defined by $\gamma'(s)=r\cdot\gamma(s)$ for all $s\in[0,1]$ yields the same result.
\end{proof}

\subsubsection{\textbf{Remark on the Compactification in \cite{box-graph}}}
The authors of \cite{box-graph}, a paper which was never published but was reprinted in \cite{homfeynman}, employ a different compactification: They consider specifically the box graph $C_4$ which they compactify as follows: The integration domain $\mathbb{R}^D$ can be thought of as part of a $D$-sphere in $\mathbb{R}^{D+1}$, using the stereographic projection. More precisely, we consider the diffeomorphism
\begin{equation}
    f:S^D(-1,0,\ldots,0)\backslash\{0\}\overset{\sim}\to\mathbb{R}^D, \quad (x_0,\ldots,x_D)\mapsto\left(\frac{x_1}{x_0},\ldots,\frac{x_D}{x_0}\right).
\end{equation}
Here, $S^D(-1,0\ldots,0)$ is the real $D$-sphere with radius 1 and center $(-1,0,\ldots,0)$. Applying the pullback $f^\ast$ to the Feynman integral \eqref{eq:one_loop_1}, one obtains
\begin{equation}
    I(C_n)=\int_{S^D(-1,0,\ldots,0)}\frac{x_0^{2\lambda-D-1}\cdot\Omega_D}{\prod_{i=1}^n((k+x_0\cdot P^{(i)}(p))^2+x_0^2m_i^2)^{\lambda_i}}
\end{equation}
after adding the set $\{0\}$ of Lebesgue measure 0 to the integration domain. While this compactifies the integration domain, the natural choice for the complex ambient space, the complex sphere
\begin{equation}
    S_\mathbb{C}^D:=\{(x_0,k)\in\mathbb{C}^{D+1} \;\vert\; (x_0+1)^2+k^2=1\}
\end{equation}

around $(-1,0,\ldots,0)$ with radius 1, is not compact. Thus, the authors view $S_\mathbb{C}^D$ as embedded in $\mathbb{C}\mathbb{P}^{D+1}$ in a second step, i.e., they apply the pullback of the inverse of the inclusion
\begin{equation}
    i:\mathbb{C}^D\hookrightarrow\mathbb{C}\mathbb{P}^{D+1},\qquad z\mapsto[1:z]
\end{equation}
restricted to its image like we do in the first step. The integral they obtain after applying all transformations reads
\begin{equation}\label{eq:alternative_compactification}
    I(C_n)=\int_{S_\mathbb{P}^D}\frac{x_0^{2\lambda-D-1}\cdot\Omega_D}{\prod_{i=1}^n((k+x_0\cdot P^{(i)}(p))^2+x_0^2m_i^2)^{\lambda_i}},
\end{equation}
where
\begin{equation}
    \begin{split}
        S_\mathbb{P}^D:&=\{[u:x_0:k]\in\mathbb{C}\mathbb{P}^{D+1} \;\vert\; (x_0+u)^2+k^2=u^2\}\\
        &=\{[u:x_0:k]\in\mathbb{C}\mathbb{P}^{D+1} \;\vert\; x_0^2+k^2=-2ux_0\}.
    \end{split}
\end{equation}
This compactification comes with significant restrictions with respect to the dimension: The form of equation \eqref{eq:alternative_compactification} suggests that the differential to be integrated might be singular at points $[u:x_0:k]\in S_\mathbb{P}^D$ with $x_0=0$ and $k=0$, which would be an issue. The term $x_0^{2\lambda-D-1}$ could potentially cancel the singular nominator at such points, but if it actually does depends on $D$ and $\lambda$. A careful computation of the relevant differential form in charts shows that this cancellation happens if and only if $D\leq\text{Re}\,\lambda$. Particularly in the case $\lambda_1=\cdots=\lambda_n=1$ one is usually interested in, this is a problem: In this case, we have $\lambda=n$ and for $p\in T_n$ we need $D\geq n-1$. Hence, this compactification can only work if $D=n$ or $D=n-1$. For example, the box graph $C_4$ in 4 dimensions as investigated in \cite{box-graph} can be compactified in this manner, while the triangle graph $C_3$ in 4 dimensions cannot (even though the corresponding integral is convergent). The upside of this compactification is that, after some straightforward manipulations of the integral \eqref{eq:alternative_compactification}, the set on which the differential form we integrate becomes singular can be written as the union of projective hyperplanes. This makes it considerably easier to compute the relevant vanishing classes and intersection indices, a problem which is tackled in \cite{boyling} (where the restriction in application to physics is also mentioned). On the other hand, it seems likely that the ideas used in this computation cannot be generalized to more than one loop.

\subsection{On the Continuation of the Local Section Defined by \texorpdfstring{$\mathbb{R}\mathbb{P}^D$}{RP D}}
Before discussing the discontinuity of $I(C_n)$, we first establish some results on the continuation of the local section defined by the initial integration cycle $\mathbb{R}\mathbb{P}^D$.\\
The next lemma allows us to show that if there is an ambient isotopy realizing a continuation, we can always choose it in a way that fixes the integration cycle at infinity.
\begin{lemma}\label{lem:invariance_at_infinity}
    Let $K\subset T_n$ be a compact set. Then, there exists an open neighborhood $U_\infty\subset\mathbb{C}\mathbb{P}^D$ of $H_\infty\cap\mathbb{R}\mathbb{P}^D$ such that
    \begin{equation}
        U_\infty\cap S(p)=\emptyset \qquad \forall p\in K.
    \end{equation}
\end{lemma}
\begin{proof}
    First, note that for any $p\in T_n$ we have
    \begin{equation}
        Q_i(p)(0,k)=k^2\neq0 \qquad \forall k\in\mathbb{R}^D,\;\forall i\in\{1,\ldots,n\}.
    \end{equation}
    Thus, $H_\infty\cap\mathbb{R}\mathbb{P}^D\cap S(p)=\emptyset$. Now, let $d:\mathbb{C}\mathbb{P}^D\times\mathbb{C}\mathbb{P}^D\to\mathbb{R}_{\geq0}$ be a metric on $\mathbb{C}\mathbb{P}^D$ inducing its standard topology. Then,
    \begin{equation}
        d(H_\infty\cap\mathbb{R}\mathbb{P}^D,S(p))>0
    \end{equation}    
    for all $p\in T_n$. Now, $d(H_\infty\cap\mathbb{R}\mathbb{P}^D,S(\cdot))$ can be viewed as a continuous function on $T_n$ which necessarily takes on a minimum $d_\text{min}>0$ on the compact set $K$. Thus, if we denote by $D([z],d_\text{min})\subset\mathbb{C}\mathbb{P}^D$ the open disc of radius $d_\text{min}$ around $[z]$, we have $D([z],d_\text{min})\cap S(p)=\emptyset$ for all $[z]\in H_\infty\cap\mathbb{R}\mathbb{P}^D$ and $p\in K$. Consequently, the open set
    \begin{equation}
        U_\infty:=\bigcup_{[z]\in H_\infty\cap\mathbb{R}\mathbb{P}^D}D([z],d_\text{min})
    \end{equation}
    fulfills the requirement of the lemma.
\end{proof}
\begin{corollary}
    Let $\gamma:[0,1]\to T_n$ be a path with $\gamma(0)\in(\mathbb{R}^D)^{n-1}$. Suppose there exists an ambient isotopy $\sigma$ of $\mathbb{C}\mathbb{P}^D$ such that $\sigma(\cdot,s)$ takes $S(\gamma(0))$ to $S(\gamma(s))$ for all $s\in[0,1]$. Then, there exists an ambient isotopy $\sigma'$ of $\mathbb{C}\mathbb{P}^D$ adapted to $\gamma$ such that
    \begin{equation}
        \sigma'(H_\infty\cap\mathbb{R}\mathbb{P}^D,s)=H_\infty\cap\mathbb{R}\mathbb{P}^D
    \end{equation}
    for all $s\in[0,1]$.
\end{corollary}
\begin{proof}
    Let $U_\infty$ be an open neighborhood of $H_\infty$ as in Lemma \ref{lem:invariance_at_infinity}. Furthermore, let $U$ be an open neighborhood of $S(\gamma(0))$ disjoint from $U_\infty$. Then, $\sigma$ restricts to an isotopy of $U\cup U_\infty$ such that the track $\hat{\sigma}((U\cup U_\infty)\times[0,1])$ is open. We can modify this ambient isotopy by letting it be the identity on $U_\infty$. Now, by Theorem \ref{thm:isotopy_extension}, there exists an ambient isotopy $\sigma'$ of $\mathbb{C}\mathbb{P}^D$ which agrees with $\sigma$ on the compact set $S(\gamma(0))\cup(H_\infty\cap\mathbb{R}\mathbb{P}^D)$. In particular, $\sigma'(H_\infty\cap\mathbb{R}\mathbb{P}^D,s)=H_\infty\cap\mathbb{R}\mathbb{P}^D$ for all $s\in[0,1]$.
\end{proof}
Now, we discuss a class of path along which we can always analytically continue. For any $I\subset\{0,\ldots,D-1\}$, consider the continuous matrix-valued function $s\mapsto M_I(s)\in M(D\times D,\mathbb{C})$ defined by
\begin{equation}
    (M_I(s))_{ij}:=\begin{cases} e^{i\frac{\pi}{2}s} & \text{if } i=j\in I \\ 1 & \text{if }i=j\notin I \\ 0 & \text{otherwise}. \end{cases}
\end{equation}
Note that we again start our indexing at $0$ instead of 1 and that $M_I(0)=\text{id}_{\mathbb{C}^D}$. Then, for any $p\in(\mathbb{C}^D)^{n-1}$ we can define a path
\begin{equation}
    \gamma_{I,p}:[0,1]\mapsto(\mathbb{C}^D)^{n-1}, \quad s\mapsto (M_I(s)p_1,\ldots,M_I(s)p_{n-1}).
\end{equation}
Furthermore, we define the continuous map
\begin{equation}
    \sigma_{I,p}:\mathbb{C}\mathbb{P}^D\times[0,1]\to\mathbb{C}\mathbb{P}^D, \quad ([u:k],s)\mapsto [u:M_I(s)k].
\end{equation}
\begin{proposition}
    Let $p\in(\mathbb{R}^D)^{n-1}$ and $I\subset\{0,\ldots,D-1\}$. Then, for every $s_1\in(0,1)$ the ambient isotopy $\sigma_{I,p}\vert_{\mathbb{C}\mathbb{P}^D\times[0,s_1]}$ is adapted to $\gamma_{I,p}\vert_{[0,s_1]}$.\footnote{Being technically pedantic, it should be noted that our definition of ambient isotopy only covers the case $s_1=1$. So $\sigma_{I,p}\vert_{\mathbb{C}\mathbb{P}^D\times[0,s_1]}$ and $\gamma_{I,p}\vert_{[0,s_1]}$ are to be understood as appropriate reparametrizations such that they define maps $\mathbb{C}\mathbb{P}^D\times[0,1]\to\mathbb{C}\mathbb{P}^D$ and $[0,1]\to(\mathbb{C}^D)^{n-1}$, respectively.} In particular, the integral $I(C_n)$ can be analytically continued from $p$ to $\gamma_{I,p}(s_1)$ along $\gamma_{I,p}\vert_{[0,s_1]}$.
\end{proposition}
\begin{proof}
    According to Proposition \ref{prop:adapted}, it suffices to show
    \begin{equation}
        \sigma_{I,p}(\mathbb{R}\mathbb{P}^D,s)\cap S(\gamma_{I,p})=\emptyset \qquad \forall s\in[0,s_1].
    \end{equation}
    So let $s\in[0,s_1]$. We write $I=\{i_1,\ldots,i_{\vert I\vert}\}$ and $J:=\{0,\ldots,D-1\}\backslash I=\{j_1,\ldots,j_{\vert J\vert}\}$ with $i_1<\cdots<i_{\vert I\vert}$ and $j_1<\ldots<j_{\vert J\vert}$. Furthermore, we denote
    \begin{equation}
        k_I:=(k_{i_1},\ldots,k_{i_{\vert I\vert}}) \quad\text{and}\quad k_J:=(k_{j_1},\ldots,k_{j_{\vert J\vert}}),
    \end{equation}
    as well as
    \begin{equation}
        P_I^{(l)}(p'):=(P_{i_1}^{(l)}(p'),\ldots,P_{i_{\vert I\vert}}^{(l)}(p')) \quad\text{and}\quad P_J^{(l)}(p'):=(P_{j_1}^{(l)}(p'),\ldots,P_{j_{\vert J\vert}}^{(l)}(p'))
    \end{equation}
    for all $l\in\{1,\ldots,n\}$ and all $p'\in(\mathbb{C}^D)^{n-1}$.
    Now, let $[u:k]\in\mathbb{R}\mathbb{P}^D$. Then, $(u,M_I(s)k)$ is a representative of $\sigma_{I,p}([u:k],s)$ and for any $l\in\{1,\ldots,n\}$ we have
    \begin{equation}
        \begin{split}
            Q_l(\gamma_{I,p}(s))(u,M_I(s)k)&=(M_I(s)k+uM_I(s)P^{(l)}(p))^2+u^2m_l^2\\
            &=e^{i\pi s}(k_I+uP_I^{(l)}(p))^2+(k_J+uP_J^{(l)}(p))^2+u^2m_l^2.
        \end{split}
    \end{equation}
    Now, either $k_I+uP_I^{(l)}(p)=0$ and we have
    \begin{equation}
        Q_l(\gamma_{I,p}(s))(u,M_I(s)k)=(k_J+uP_J^{(l)}(p))^2+u^2m_l^2>0
    \end{equation}
    or $k_I+uP_I^{(l)}(p)\neq0$ in which case $Q_l(\gamma_{I,p}(s))(u,M_I(s)k)$ has nonzero imaginary part. In either case, we conclude $Q_l(\gamma_{I,p}(s))(u,M_I(s)k)\neq0$.
\end{proof}

\subsection{Cutkosky's Theorem in the One-Loop Case}\label{subsec:cutkosky}
Now, we are in a position to compute the discontinuity of the general one-loop integral \eqref{eq:one_loop_2} around simple loops up to the intersection index. This requires us to compute the vanishing sphere.\\
Let us first consider the geometry of the zero locus of $Q_j(p)(1,\cdot)$ at an arbitrary point $p\in(\mathbb{C}^D)^{n-1}$. It is given by all $k\in\mathbb{C}^D$ such that
\begin{equation}
    (k+P^{(j)}(p))^2+m_j^2=0 \qquad\Leftrightarrow\qquad (ik+iP^{(j)}(p))^2=m_j^2.
\end{equation}
This describes a complex $(D-1)$-sphere with radius $m_j$. It contains the real $(D-1)$-sphere
\begin{equation}
    \{k\in\mathbb{C}^D \;\vert\; k+P^{(j)}(p)\in i\mathbb{R}^D,\; (ik+iP^{(j)}(p))^2=m_j^2\}
\end{equation}
to which it deformation retracts. To compute the vanishing sphere, we want to understand the homology of the finite part of $S_1(p)\cap\cdots\cap S_n(p)$. The intersection of $n$ complex $(D-1)$-spheres in general position is a complex $(D-n)$-sphere, which allows us to immediately determine the sought homology groups. But we can be significantly more concrete by employing the Lorentz-invariance, or rather $O(D,\mathbb{C})$-invariance in our setup, of Feynman integrals. To fully take advantage of this invariance, we first prove the following 
\begin{lemma}\label{lem:orthogonal_transformation}
    Let $v_1,\ldots,v_m\in\mathbb{C}^n$ be $m\in\mathbb{N}^\ast$ vectors such that
    \begin{equation}
        \det
        \begin{pmatrix}
            v_1v_1 & v_1v_2 & \cdots & v_1v_m \\
            v_2v_1 & v_2v_2 & \cdots & v_2v_m \\
            \vdots & \vdots & \ddots & \vdots \\
            v_mv_1 & v_mv_2 & \cdots & v_mv_m
        \end{pmatrix}\neq0.
    \end{equation}
    Then, there exists a complex orthogonal matrix $M\in O(n,\mathbb{C})$ such that $(M\cdot v_i)_j\neq0$ for all $j\leq i$ and $(M\cdot v_i)_j=0$ for all $j>i$.
\end{lemma}
\begin{proof}
    We conduct the proof by induction on the number of vectors $m$. The base case $m=1$ can be solved by induction on the dimension $n$. For $n=1$, there is nothing to do. Now, suppose we already know the lemma is true for $m=1$ up to some dimension $n-1$. Let us denote $w:=v_1$. If $\tilde{w}:=(w_2,\ldots,w_n)=0$, there is again nothing to do. By assumption, we have $w^Tw\neq0$. Note that this means we cannot have
    \begin{equation}
        (w_1,\ldots,\hat{w}_i,\ldots,w_n)^T(w_1,\ldots,\hat{w}_i,\ldots,w_n)=0
    \end{equation}
    for all $i\in\{1,\ldots,n\}$ since
    \begin{equation}
        \sum_{i=1}^n(w_1,\ldots,\hat{w}_i,\ldots,w_n)^T(w_1,\ldots,\hat{w}_i,\ldots,w_n)=(n-1)w^Tw\neq0.
    \end{equation}
    The permutation matrices are orthogonal so that we can assume $\tilde{w}^T\tilde{w}\neq0$ without loss of generality. Thus, by the induction hypothesis we can find a complex orthogonal matrix $M_1\in O(n-1,\mathbb{C})$ such that
    \begin{equation}
        M_1\cdot\tilde{w}=\begin{pmatrix} w_2' \\ 0 \\ \vdots \\ 0 \end{pmatrix} \qquad\Rightarrow\qquad \begin{pmatrix} 1 & 0 \\ 0 & M_1\end{pmatrix}\cdot w=\begin{pmatrix} w_1' \\ w_2' \\ 0 \\ \vdots \\ 0\end{pmatrix}
    \end{equation}
    with $w_2'\neq0$. Note that $\begin{pmatrix} 1 & 0 \\ 0 & M_1 \end{pmatrix}$ is also an orthogonal matrix. Now, we define
    \begin{equation}
        M_2:=\frac{1}{\sqrt{{w'_1}^2+{w'_2}^2}}\cdot\begin{pmatrix} w'_1 & w'_2 \\ -w'_2 & w'_1 \end{pmatrix}.
    \end{equation}
    Note that this is well-defined as ${w'_1}^2+{w'_2}^2=w^Tw\neq0$. Then, we have
    \begin{equation}
        M_2^TM_2=\frac{1}{{w'_1}^2+{w'_2}^2}\cdot\begin{pmatrix} {w'_1}^2+{w'_2}^2 & w_1'w_2'-w_2'w_1' \\ w_2'w_1'-w_1'w_2' & {w'_2}^2+{w'_1}^2 \end{pmatrix}=\begin{pmatrix} 1 & 0 \\ 0 & 1 \end{pmatrix}
    \end{equation}
    and
    \begin{equation}
        M_2\cdot\begin{pmatrix} w_1' \\ w_2' \end{pmatrix}=\frac{1}{\sqrt{{w'_1}^2+{w'_2}^2}}\cdot\begin{pmatrix} {w'_1}^2+{w'_2}^2 \\ w_2'w_1'-w_1'w_2' \end{pmatrix}=\begin{pmatrix}\frac{{w'_1}^2+{w'_2}^2}{\sqrt{{w'_1}^2+{w'_2}^2}} \\ 0 \end{pmatrix}\neq0.
    \end{equation}
    We conclude that $M_2$ is orthogonal and that therefore the matrix $\begin{pmatrix} M_2 & 0 \\ 0 & 1_{n-2}\end{pmatrix}$ is also orthogonal, where $1_{n-2}$ is the identity matrix of size $n-2$. Hence the matrix
    \begin{equation}
        M:=\begin{pmatrix} M_2 & 0 \\ 0 & 1_{n-2} \end{pmatrix}\cdot\begin{pmatrix} 1 & 0 \\ 0 & M_1 \end{pmatrix}\quad\in\quad O(n,\mathbb{C})
    \end{equation}
    is orthogonal and fulfills the requirement of the lemma.\\
    Now, suppose that $m>1$. Then, by the induction hypothesis, we can find an orthogonal matrix $M\in O(n,\mathbb{C})$ such that for all $i<m$ we have $(M\cdot v_i)_i\neq0$ and $(M\cdot v_i)_j=0$ for all $j>i$. Now, consider the vector $v':=(x_m,\ldots,x_n)\in\mathbb{C}^{n-m+1}$ obtained by removing the first $m-1$ components of $M\cdot v_m$. Then, there is an orthogonal matrix $M'\in O(n-m+1,\mathbb{C})$ such that $(M'\cdot v')_1\neq0$ and $(M'\cdot v')_j=0$ for all $j>1$. We define
    \begin{equation}
        \tilde{M}:=\begin{pmatrix} 1_{m-1} & 0 \\ 0 & M' \end{pmatrix}\cdot M\quad\in\quad O(n,\mathbb{C}).
    \end{equation}
    Then, $\tilde{M}$ is orthogonal as
    \begin{equation}
        \tilde{M}^T=M^T\cdot\begin{pmatrix} 1_{m-1} & 0 \\ 0 & {M'}^T \end{pmatrix}=M^{-1}\cdot\begin{pmatrix} 1_{m-1} & 0 \\ 0 & {M'}^{-1}\end{pmatrix} = (\begin{pmatrix} 1_{m-1} & 0 \\ 0 & M' \end{pmatrix}\cdot M)^{-1}=\tilde{M}^{-1}.
    \end{equation}
    Furthermore, we clearly have $(\tilde{M}\cdot v_i)_j\neq0$ for all $j\leq i$ and $(\tilde{M}\cdot v_i)_j=0$ for all $j>i$.
\end{proof}
Using this lemma, we can easily compute the center and radius of the complex sphere given by the finite part of the intersection of the $S_1(p),\ldots,S_n(p)$. The system of equations defining $\bigcap_{i=1}^nS_i(p)-H_\infty$ reads
\begin{alignat*}{3}
    &k^2 &&+\;m_1^2 &&=0 \\
    &(k+p_1)^2 &&+\;m_2^2&&=0 \\
    &\hspace{0.8cm}\vdots && \hspace{0.2cm}\vdots && \hspace{0.18cm}\vdots \\
    &\left(k+\sum_{j=1}^{n-1}p_j\right)^2 &&+\;m_n^2 &&=0
\end{alignat*}
in inhomogeneous coordinates. By plugging in $k^2=-m_1^2$ from the first equation into the remaining ones, we see that this system is equivalent to
\begin{alignat*}{3}
    &k^2 && +m_1^2 && =0\\
    &2kp_1+p_1^2 && -m_1^2+m_2^2 &&=0\\
    &\hspace{1cm}\vdots && \hspace{0.18cm}\vdots && \hspace{0.2cm}\vdots\\
    &2k\sum_{j=1}^{n-1}p_j+\left(\sum_{j=1}^{n-1}p_j\right)^2 && -m_1^2+m_n^2 && =0.
\end{alignat*}
By Proposition \ref{prop:invariance_under_O(D,C)} and Lemma \ref{lem:orthogonal_transformation}, we can assume
\begin{equation}
    p_1=\begin{pmatrix} (p_1)_0 \\ 0 \\ \vdots \\ 0\end{pmatrix},\quad p_2=\begin{pmatrix} (p_2)_0 \\ (p_2)_1 \\ 0 \\ \vdots \\ 0\end{pmatrix},\quad \cdots,\quad \begin{pmatrix} (p_n)_0 \\ \vdots \\ (p_n)_{n-1} \\ 0 \\ \vdots \\ 0\end{pmatrix}
\end{equation}
with $(p_i)_{i-1}\neq0$ for all $i\in\{1,\ldots,n\}$ without loss of generality. This allows us to
compute the first $n-1$ components of $k$ recursively: By the second equation, we have
\begin{equation}\label{eq:A0}
    k_0=-\frac{1}{2(p_1)_0}(p_1^2-m_1^2+m_2^2)=:A_0(p)
\end{equation}
and by the $(i+2)$th equation we have
\begin{equation}\label{eq:Ai}
    k_i=-\frac{1}{2(p_{i+1})_i}(2\bar{k}^{(i)}P^{(i+2)}+(P^{(i+2)})^2-m_1^2+m_{i+2}^2)=:A_i(p),
\end{equation}
where
\begin{equation}
    \bar{k}^{(i)}:=(k_0,\ldots,k_{i-1},0,\ldots,0).
\end{equation}
We denote $A^{(n)}(p):=(A_0(p),\ldots,A_{n-2}(p),0,\ldots,0)\in\mathbb{C}^D$. This allows us to conclude that $\bigcap_{i=1}^nS_i(p)-H_\infty$ is given by all solutions to
\begin{equation}\label{eq:An}
    \bar{k}^{(n-1)}=A^{(n)}(p)
\end{equation}
and
\begin{equation}\label{eq:rn}
    (k_{n-1},\ldots,k_{D-1})^2=-m_1^2-(A^{(n)}(p))^2=:r_n^2(p),
\end{equation}
where equation \eqref{eq:rn} is obtained from plugging in \eqref{eq:An} into $k^2+m_1^2=0$. Thus, we see that $\bigcap_{i=1}^nS_i(p)-H_\infty$ is a complex $(D-n)$-sphere around $A^{(n)}(p)$ with radius squared equal to $r_n^2(p)$. The $A_j(p)$ are designed such that
\begin{equation}\label{eq:vanishing_of_A}
    2A^{(n)}(p)P^{(i)}(p)+(P^{(i)}(p))^2+m_i^2-m_1^2=0
\end{equation}
for all $i\in\{1,\ldots,n\}$. Indeed, we have
\begin{equation}
    \begin{split}
        2A^{(n)}(p)P^{(i)}(p)&=2A_{i-1}(p)(p_i)_i+2\sum_{j=0}^{i-2}A_j(p)(p_i)_j+2A^{(n)}(p)P^{(i-1)}\\
        &=-(P^{(i)}(p))^2-m_i^2+m_1^2
    \end{split}
\end{equation}
for all $i\in\{1,\ldots,n\}$.\\
This discussion covers the case in which all $S_1(p),\ldots,S_n(p)$ intersect, but this can easily be generalized to intersections of any subset: Let $I\subsetneq\{1,\ldots,n\}$ and assume $\vert I\vert\geq2$ (otherwise, there is nothing to do). The defining equations for $\bigcap_{i\in I}S_i(p)-H_\infty$ have the same form as in the case $I=\{1,\ldots,n\}$. In fact, they are exactly the equations we would obtain for the cyclic graph with $\vert I\vert$ edges obtained from shrinking every edge not in $I$ to zero length (and thus not $C_{\vert I\vert}$ as we have defined it since the external structure is different). Thus, we can go through the same computation as above, but we replace each momentum $P^{(i)}(p)$ for $i\in I$ by the corresponding sum of momenta. Let $I=\{i_1,\ldots,i_{\vert I\vert}\}$ with $i_1<\cdots<i_{\vert I\vert}$. We shall denote the external momenta $p_I\in(\mathbb{C}^{D})^{\vert I\vert-1}$ given by the external structure of the graph with shrunken edges by
\begin{equation}
    (p_I)_l:=\sum_{j={i_l}}^{i_{l+1}-1}p_j,\qquad \forall l\in\{1,\ldots,n-1-\vert I\vert\}.
\end{equation}
Then, $\bigcap_{i\in I}S_i(p)-H_\infty$ is a complex $(D-\vert I\vert)$-sphere around $A^{(\vert I\vert)}(p_I)$ with radius squared equal to $r_{\vert I\vert}^2(p_I)$.\\
We can also use this description of $\bigcap_{i=1}^nS_i(p)-H_\infty$ to prove that the system
\begin{equation}
    S_1(p)-H_\infty,\ldots,S_n(p)-H_\infty
\end{equation}
has a single pinch point for $p\in L_n$ which is a simple pinch point.
\begin{proposition}\label{prop:simple_pinch}
    Let $I\subset\{1,\ldots,n\}$ be non-empty and let $p\in L_{n,I}$ be a point of codimension 1. Then, there is a single point in $\mathbb{C}\mathbb{P}^D\backslash H_\infty$ at which the manifolds $S_i(p)$ with $i\in I$ are not in general position. Furthermore, this point is a simple pinch point.
\end{proposition}
\begin{proof}
    By the discussion above, we know that there can only be one finite point at which the manifolds $S_i(p)$ with $i\in I$ are not in general position. According to \cite{app-iso}, this is a simple pinch if the corresponding solution to the Landau equations is unique in $\alpha$ up to a homogeneous factor. We may assume $\sum_{i\in I}\alpha_i=1$ without loss of generality. Write $I=\{i_1,\ldots,i_m\}$ with $i_1<\cdots<i_m$. Then, we need to show that $\alpha_{i_2},\ldots,\alpha_{i_m}$ are uniquely determined by the Landau equations. This can easily be verified since $\alpha_{i_2},\ldots,\alpha_{i_m}$ can recursively determined by the equation
    \begin{equation}
        k=-\sum_{i\in I}\alpha_iP^{(i)}(p).
    \end{equation}
\end{proof}
Notice that we have just proven the following
\begin{proposition}[Vanishing Sphere for One-Loop Graphs]\label{prop:vanishing_sphere}
    Let $p\in L_{n,I}$ be a point of codimension 1. Then, the system $\{S_i\}_{i\in I}$ has a simple pinch at $(A^{(\vert I\vert)}(p_I),p)$ and the vanishing sphere can be represented by a deformation retract of
    \begin{equation}
        \tilde{e}=\{k\in\mathbb{C}^D \;\vert\; k=A^{(\vert I\vert)}(p_I')\;,\;\sum_{i=\vert I\vert-1}^{D-1}k_i^2=r_{\vert I\vert}^2(p_I')\}
    \end{equation}
    to the real sphere within $\mathbb{C}\mathbb{P}^D$ for $p'$ close to $p$. In particular, if $p'\in(M^D)^{n-1}$ is a configuration of Minkowski momenta such that $r_{\vert I\vert}^2(p_I')\in\mathbb{R}_{>0}$, we have $\tilde{e}\subset M^D$.
\end{proposition}
Note that the above proof also yields an algorithm to determine the Landau surface for any loop graph which does not require us to solve the Landau equations:
\begin{algorithm}
\caption{The Landau surface $L_n$ can be computed in terms of any kinematic invariant by the following algorithm:}\label{algo:landau_surface}
    \begin{enumerate}
        \item Choose a complete set of kinematic invariants $s_1,\ldots,s_{\frac{n(n-1)}{2}}$.
        \item Compute $A_i(p)$ for $i=0,1,\ldots,n-2$ by using equations \eqref{eq:A0} and \eqref{eq:Ai}.
        \item Compute $r_n^2(p)$ by using equation \eqref{eq:rn}.
        \item Express all components of $p$ in $r_n^2(p)$ in terms of the invariants $s_1,\ldots,s_{\frac{n(n-1)}{2}}$.
        \item The result is a rational function in the $s_1,\ldots,s_{\frac{n(n-1)}{2}}$ which can be solved for one of them.
    \end{enumerate}
\end{algorithm}
\begin{example}\label{ex:vanishing_sphere}
    Returning to our running example $C_2$, we compute $A^{(2)}(p)$ as well as $r_2^2(p)$ in this case. By definition, we have $A_i(p)=0$ for all $i\neq0$. The entry $A_0(p)$ can just be read of from its definition \eqref{eq:A0} as
    \begin{equation}
        A_0(p)=-\frac{1}{2(p_1)_0}((p_1)_0^2-m_1^2+m_2^2).
    \end{equation}
    Thus, we can easily compute
    \begin{equation}
        \begin{split}
            r_2^2(p)&=-m_1^2-A_0^2(p)=-\frac{4(p_1)_0^2m_1^2}{4(p_1)_0^2}-\frac{((p_1)_0^2-m_1^2+m_2^2)^2}{4(p_1)_0^2}\\
            &=-\frac{1}{4(p_1)_0^2}((p_1)_0^4+m_1^4+m_2^4+2(p_1)_0^2m_1^2+2(p_1)_0^2m_2^2-2m_1^2m_2^2)\\
            &=-\frac{\lambda(-(p_1)_0^2,m_1^2,m_2^2)}{4(p_1)_0^2}.
        \end{split}
    \end{equation}
    Here, $\lambda$ is again the Källén function (see equation \eqref{eq:kallen_fc}). Recall that we required $p_1$ to satisfy $(p_1)_i=0$ for all $i>0$. Thus, $(p_1)_0^2=p_1^2$, and we obtain
    \begin{equation}
        r_2^2(p)=-\frac{\lambda(-p_1^2,m_1^2,m_2^2)}{4p_1^2}.
    \end{equation}
    In particular, $r_2^2(p)=0$ if and only if $\lambda(-p_1^2,m_1^2,m_2^2)=0$. Thus, the external momenta $p$ at which $r_2^2$ vanishes are precisely the momenta in the Landau surface of $C_2$ as computed in Example \ref{ex:landau_surface}. This is of course no accident as we have seen that $r_n^2(p)$ is in fact the radius squared of the vanishing sphere.
\end{example}
A priori it might not be easy to perform step (5) in the Algorithm \ref{algo:landau_surface}. But it seems that it is possible to choose appropriate channel variables such that $r_n^2$ can be brought into a form in which the nominator is quadratic in each of the channel variables, making step (5) almost trivial. While we do not prove this statement in this work, we at least see that it works out in non-trivial examples (see Section \ref{sec:examples}).\\
We can also make use of our formula for $r_n^2$ to proof Propositions \ref{prop:landau_decomp_1} and \ref{prop:landau_decomp_2} now, which stated that the $L_{n,I}$ (with $I\subset\{1,\ldots,n\}$ such that $\vert I\vert\geq2$) are codimension 1 complex analytic submanifolds of $T_n$ intersecting in general position:
\begin{proof}
    of Proposition \ref{prop:landau_decomp_1}. It suffices to show that for all $p\in T_n$, the equation $\partial_pr_n^2(p)=0$ implies $r_n^2(p)\neq0$. So let $p\in T_n$ and suppose that $\partial_pr_n^2(p)=0$. Then, in particular $\partial_{p_{n-1}}r_n^2(p)=0$. Since the $A_0,\ldots,A_{n-3}$ do not depend on $p_{n-1}$, this equation reads
    \begin{equation}
        -2A_{n-2}(p)\partial_{p_{n-1}}A_{n-2}(p)=0.
    \end{equation}
    Suppose first that $A_{n-2}(p)=0$. Then, $r_n^2(p)=r_{n-1}^2(p_1,\ldots,p_{n-2})$ and $\partial_pr_n^2(p)=\partial_pr_{n-1}^2(p_1,\ldots,p_{n-2})$, so the claim follows by induction. Now, suppose $\partial_{p_{n-1}}A_{n-2}(p)=0$. Let us denote
    \begin{equation}
        \bar{v}:=(v_0,\ldots,v_{n-3})\;\in\;\mathbb{C}^{n-2}.
    \end{equation}
    for every $v=(v_0,\ldots,v_{D-1})\in\mathbb{C}^D$. Then,
    \begin{equation}
        \partial_{\bar{p}_{n-1}}A_{n-2}(p)=-\frac{1}{2(p_{n-1})_{n-2}}(2\bar{A}^{(n)}(p)+2\bar{P}^{(n)}(p))\overset{!}=0,
    \end{equation}
    which implies $\bar{A}^{(n)}(p)=-\bar{P}^{(n)}(p)$. Furthermore, using equation \eqref{eq:vanishing_of_A}, we compute
    \begin{equation}
        \begin{split}
            &\partial_{(p_{n-1})_{n-2}}A_{n-2}(p)\\
            &\quad=\frac{1}{2(p_{n-1})_{n-2}^2}\left(2A^{(n-1)}(p)P^{(n)}(p)+(P^{(n)}(p))^2-m_1^2+m_n^2\right)-1\\
            &\quad=-\frac{1}{(p_{n-1})_{n-2}}A_{n-2}(p)-1,
        \end{split}
    \end{equation}
    which implies $A_{n-2}(p)=-(p_{n-1})_{n-2}$. Together we obtain $A^{(n)}(p)=-P^{(n)}(p)$. This also means $(P^{(n)}(p))^2=-m_1^2+m_n^2$. This can be used to compute
    \begin{equation}
        r_n^2(p)=-m_1^2-(A^{(n)}(p))^2=-m_1^2-(P^{(n)}(p))^2=-m_n^2\neq0
    \end{equation}
    as claimed.
\end{proof}
\begin{proof}
    of Proposition \ref{prop:landau_decomp_2}. To show that $L_{n,I_1},\ldots,L_{n,I_k}$ intersect in general position, we need to show that $\partial_p r_{\vert I_1\vert}^2(p_{I_1}),\ldots,\partial_p r_{\vert I_k\vert}^2(p_{I_k})$ are linearly independent at every point $p\in\bigcap_{i=1}^kL_{n,I_i}$. We do this by induction on $k$. By applying a cyclic permutation to the vertices and edges of the graph $C_n$, we may assume without loss of generality that $I_k$ contains an element $i\in I_k$ such that $i>j$ for all $j\in I_{k-1}$. Since the $r_{\vert I_1\vert}^2,\ldots,r_{\vert I_{k-1}\vert}^2$ do not depend on $p_{i-1}$, it suffices to show $\partial_{p_{i-1}}r_{\vert I_k\vert}^2(p_{I_k})\neq0$. In the previous proof, we have seen that $\partial_{p_{i-1}}r_{\vert I_k\vert}^2(p_{I_k})=0$ implies $r_{\vert I_k\vert}^2(p_{I_k})\neq0$ so we are done.
\end{proof}
Finally, we are in a position to state and proof Cutkosky's Theorem in the one-loop case. We remind the reader that the integral of interest is
\begin{equation}
    I(C_n)(p)=\int_{\mathbb{R}^D}\frac{d^Dk}{\prod_{i=1}^n((k+P^{(i)}(p))^2+m_i^2)^{\lambda_i}},
\end{equation}
as defined in equation \eqref{eq:one_loop_1}. While all computations in the finite chart work for odd and even dimensions $D$, we recall that the compactification (see equation \eqref{eq:one_loop_2}) as it stands can only be employed for odd $D$ and hence, we need to restrict the theorem to this case for now.
\begin{theorem}[Cutkosky's Theorem for One-Loop Graphs]\label{thm:cutkoskys_thm}
    Let $D$ be odd. Let $I\subset E(C_n)$ and denote $m:=\vert I\vert$. Let $p\in L_{n,I}$ be a point of codimension 1 and $\gamma$ a simple loop around $p$. Let $p'\in M^{D}$ such that $r_n^2(p')$ is small and real. Then, the discontinuity of $I(C_n)$ around $p$ in a neighborhood of $p'$ is given by
    \begin{equation}
        \text{Disc}_{[\gamma]}I(C_n)(p')=N(2\pi i)^m\int_{\tilde{e}}\mathrm{Res}^m[\omega_{n,D}(p')].
    \end{equation}
    In particular for $\lambda_i=1$ for all $i\in I$, we obtain
    \begin{equation}
        \text{Disc}_{[\gamma]}I(C_n)(p')=N(2\pi i)^m\int_{i\mathbb{R}\times\mathbb{R}^{D-1}}\frac{\prod_{j\in I}\delta((k+P^{(j)}(p'))^2+m_j^2)}{\prod_{j\notin I}((k+P^{(j)}(p'))^2+m_j^2)^{\lambda_j}}d^Dk.
    \end{equation}
\end{theorem}
\begin{proof}
    We have seen in Proposition \ref{prop:simple_pinch} that at $p$ there is a single pinch point at $A^{(\vert I\vert)}(p)$. So according to our considerations in Subsubsection \ref{subsubsec:picard_lefschetz} and Theorem \ref{thm:multi_residue_theorem}, we thus have
    \begin{equation}
        \text{Disc}_{[\gamma]}(I(C_n))(p')=N\int_e\omega_{n,D}(p')=N(2\pi i)^m\int_{\tilde{e}}\text{Res}^m[\omega_{n,D}(p')]
    \end{equation}
    where $\tilde{e}$ is the vanishing sphere as computed in Proposition \ref{prop:vanishing_sphere}. Denote by
    \begin{equation}
        i:\mathbb{C}^D\overset{\sim}\to\mathbb{C}\mathbb{P}^D\backslash H_\infty
    \end{equation}
    the natural inclusion with restricted target. The Leray residue commutes with pull-backs (see Proposition \ref{prop:residue_commutes_with_pullbacks}) so that we can write
    \begin{equation}
        \begin{split}
            \text{Disc}_{[\gamma]}(I(C_n))(p')&=N(2\pi i)^m\int_{\tilde{e}}\text{Res}^m[\omega_{n,D}(p')]\\
            &=N(2\pi i)^m\int_{i^{-1}(\tilde{e})}\text{Res}^m[i^\ast\,\omega_{n,D}(p')]
        \end{split}
    \end{equation}
    by removing the plane $H_\infty$ at infinity, which has Lebesgue measure 0, from the integration domain. If now $\lambda_i=1$ for all $i\in I$, we obtain
    \begin{equation}
         \text{Disc}_{[\gamma]}(I(C_n))(p')=N(2\pi i)^m\int_{i\mathbb{R}\times\mathbb{R}^{D-1}}\frac{\prod_{j\in I}\delta((k+P^{(j)}(p'))^2+m_j^2)}{\prod_{j\notin I}((k+P^{(j)}(p'))^2+m_j^2)^{\lambda_j}}d^Dk,
    \end{equation}
    where we used Definition \ref{defn:delta} for the $\delta$-function and the fact that $i^{-1}(\tilde{e})\subset M^D$.
\end{proof}
Comparing our formula for the discontinuity with some of the common literature, it is striking that we replace the propagators which go on-shell at a point in the Landau surface by simple $\delta$s instead of $\delta_+$s, which often appear in texts citing Cutkosky's Theorem. The term $\delta_+((k+P^{(j)}(p))^2+m_j^2)$ contains an additional factor of $\Theta(k_0+P^{(j)}(p))$, forcing the energy flowing through the corresponding edge to be positive. This prescription clearly depends on the arbitrarily chosen orientation of the edges. While it seems to be known by many physicists that the sign of the energy should be chosen in accordance with the corresponding solution of the Landau equations, at the same time there are some potentially confusing formulations in the literature. Thus, we briefly remark on this point: The prescription including the $\delta_+$s without further information does not seem to make sense in the general case. But if we consider singularities corresponding to a set of edges such that removing these edges from the graph yields a new graph with two connected components $G_1$ and $G_2$ (as in the special case of equation (17) in \cite{cutkosky}), then we can make sense of this idea: If we designate one of the connected components as the incoming part and the other one as the outgoing part, there is a preferred orientation of the removed edges, namely the one that points from the incoming part to the outgoing part. The following figure illustrates this situation:
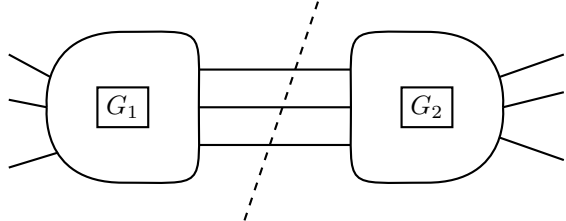
\begin{center}
    \begin{tikzpicture}[thick, baseline={([yshift=-.5ex]current bounding box.center)},vertex/.style={anchor=base,
    circle,fill=black!25,minimum size=18pt,inner sep=2pt}]
        \draw (-0.5,0.7) -- (0.05,0.4);
        \draw (-0.5,0.1) -- (0,0);
        \draw (-0.5,-0.8) -- (0.15,-0.6);
        \node[draw] at (1,0) {$G_1$};
        \draw plot [smooth, tension=2] coordinates { (0,0) (1,1) (2,0) (1,-1) (0,0)};
        \draw (2,0.5) -- (4,0.5);
        \draw (2,0) -- (4,0);
        \draw (2,-0.5) -- (4,-0.5);
        \node[draw] at (5,0) {$G_2$};
        \draw plot [xshift=6cm, rotate=180, smooth, tension=2] coordinates { (0,0) (1,1) (2,0) (1,-1) (0,0)};
        \draw (6.8,0.7) -- (5.95,0.4);
        \draw (6.8,0.2) -- (6,0);
        \draw (6.8,-0.7) -- (5.95,-0.4);
        
        \draw[dashed] (2.6,-1.5) -- (3.6,1.5);
    \end{tikzpicture}
\end{center}
But for more general cuts, there is no preferred orientation and a careful choice has to be made. As mentioned above, the consensus is that the sign fixed by the $\delta_+$s should agree with the signs of the corresponding solution of the Landau equations. Curiously enough, the $\delta_+$ in this sense does not appear explicitly in the general formulation (equation (6)) in the original work \cite{cutkosky} by Cutkosky. He formulates the theorem by writing $\delta_p$ instead of $\delta$ and states \enquote{The subscript $p$ on the delta functions means that only the contributions of the \enquote{proper} root of $q_i^2=M_i^2$ is to be taken}. Here, $q_i$ denotes what we write as $k+P^{(i)}(p)$ and $M_i$ denotes what we write as $m_i$.\\
From our discussion, it is evident that in the one-loop case the vanishing sphere over which we need to integrate lies in a subspace of $\mathbb{C}^D$ with fixed $k_0$ (namely $k_0=A_0(p)$). Thus, the sign of all $k_0+P^{(j)}_0(p)$ is already determined by the $\delta$-functions alone and the $\Theta$ would be superfluous.
\begin{example}
    To conclude our discussion of Cutkosky's Theorem for one-loop graphs, we return to our running example $C_2$ one last time and compute its discontinuity in the case where $\lambda_1=\lambda_2=1$. Note that in this case the corresponding Feynman integral does not converge in $D=4$ space-time dimensions and needs to be renormalized to make sense of it. Since we do not discuss renormalization in this paper, the space-time dimension is assumed to be smaller than 4 so that the integral converges.\\
    In Example \ref{ex:landau_surface}, we saw that the Landau surface $L_2$ of $C_2$ consists of all points $p_1\in\mathbb{C}^D\backslash\{0\}$ such that
    \begin{equation}
        -p_1^2=(m_1+m_2)^2 \qquad\text{or}\qquad -p_1^2=(m_1-m_2)^2
    \end{equation}
    and in Example \ref{ex:vanishing_sphere}, we saw that the vanishing sphere corresponding to such a point is the real sphere inside the complex sphere around $(A_0(p),0,\ldots,0)\in\mathbb{C}^D$ with radius squared equal to $-\frac{\lambda(-p_1^2,m_1^2,m_2^2)}{4p_1^2}$ which needs to be real and positive. Again we may assume $p_1=((p_1)_0,0)$. Thus, it is not difficult to see that $r_2^2(p)$ is real and positive if and only if $(p_1)_0\in i\mathbb{R}$ with either $(\text{Im}((p_1)_0))^2>(m_1+m_2)^2$ or $(\text{Im}((p_1)_0))^2<(m_1-m_2)^2$.\\
    Now, we use Theorem \ref{thm:cutkoskys_thm} to compute the discontinuity of $I(C_2)$: Let $p'\in T_2$ such that $-{p'}^2=(m_1\pm m_2)^2$, let $\gamma:[0,1]\to T_2\backslash L(C_2)$ be a simple loop, consisting of a path from some Euclidean momentum to $p\in T_2$ close to $p'$ such that
    \begin{equation}
        -\frac{1}{4p^2}\lambda\left(-p^2,m_1^2,m_2^2\right)>0
    \end{equation}
    and a small loop around $p'$. Then,
    \begin{equation}
        \begin{split}
            &\text{Disc}_{[\gamma]}I(C_2)(p)\\
            &\quad=N(2\pi i)^2\int_{i\mathbb{R}\times\mathbb{R}^{D-1}}\delta(k^2+m_1^2)\delta((k+p)^2+m_2^2)\mathrm{d}^Dk\\
            &\quad=-N\int_{\mathbb{R}^{D-1}}\frac{2\pi^2}{\sqrt{\vec{k}^2+m_1^2}}\left(\delta\left((i\sqrt{\vec{k}^2+m_1^2}+p_0)^2+\vec{k}^2+m_2^2\right)\right.\\
            &\qquad\left.+\delta\left((-i\sqrt{\vec{k}^2+m_1^2}+p_0)^2+\vec{k}^2+m_2^2\right)\right)\mathrm{d}^{D-1}\vec{k}\\
            &\quad=-N\int_{\mathbb{R}^{D-1}}\frac{2\pi^2}{\sqrt{\vec{k}^2+m_1^2}}\left(\delta\left(2ip_0\sqrt{\vec{k}^2+m_1^2}+p^2-m_1^2+m_2^2\right)\right.\\
            &\qquad\left.+\delta\left(-2ip_0\sqrt{\vec{k}^2+m_1^2}+p^2-m_1^2+m_2^2\right)\right)\mathrm{d}^{D-1}\vec{k}.
        \end{split}
    \end{equation}
    Now, we compute
    \begin{equation}\label{eq:ex_c2_2}
        \begin{split}
            &\pm2ip_0\sqrt{\vec{k}^2+m_1^2}+p^2-m_1^2+m_2^2=0\\
            \Leftrightarrow\quad&\sqrt{\vec{k}^2+m_1^2}=\pm\frac{i}{2p_0}(p^2-m_1^2+m_2^2)\\
            \Rightarrow\quad&\vec{k}^2=-\frac{1}{4p^2}(p^2-m_1^2+m_2^2)^2-m_1^2=-\frac{1}{4p^2}\lambda(-p^2,m_1^2,m_2^2)=r_2^2(p).
        \end{split}
    \end{equation}
    Note that the second line of \eqref{eq:ex_c2_2} can be fulfilled for some $\vec{k}\in\mathbb{R}^{D-1}$ if and only if
    \begin{equation}
        \pm\frac{i}{2p_0}(p^2-m_1^2+m_2^2)\geq m_1.
    \end{equation}
    Since we integrate only over non-negative $\vert\vec{k}\vert$, we can hence write
    \begin{equation}
        \begin{split}
            &\delta\left(\pm2ip_0\sqrt{\vec{k}^2+m_1^2}+p^2-m_1^2+m_2^2\right)\\
            &=\frac{\sqrt{\vec{k}^2+m_1^2}}{2\vert p_0\vert\vert\vec{k}\vert}\Theta\left(\pm\frac{i}{2p_0}(p^2-m_1^2+m_2^2)-m_1\right)\delta\left(\vert\vec{k}\vert-\sqrt{r_2^2(p)}\right)
        \end{split}
    \end{equation}
    under the integral, where $\Theta$ denotes the Heaviside-function.
    \begin{lemma}\label{lem:ex_c2}
        We have
        \begin{equation}
            \begin{split}
                &\Theta\left(\frac{i}{2p_0}(p^2-m_1^2+m_2^2)-m_1\right)+\Theta\left(-\frac{i}{2p_0}(p^2-m_1^2+m_2^2)-m_1\right)\\
                &\quad=\Theta\left(-\frac{\lambda(-p^2,m_1^2,m_2^2)}{4p^2}\right)=\Theta(r_2^2(p)).
            \end{split}
        \end{equation}
    \end{lemma}
    \begin{proof}
        A straightforward computation reveals
        \begin{equation}
            \begin{split}
                &\frac{i}{2p_0}(p^2-m_1^2+m_2^2)\geq m_1\quad\text{or}\quad-\frac{i}{2p_0}(p^2-m_1^2+m_2^2)\geq m_1\\
                &\quad\Leftrightarrow-\frac{(p^2-m_1^2+m_2^2)^2}{4p^2}\geq m_1^2\\
                &\quad\Leftrightarrow-\frac{\lambda(-p^2,m_1^2,m_2^2)}{4p^2}\geq0.
            \end{split}
        \end{equation}
    \end{proof}
    Using Lemma \ref{lem:ex_c2}, the discontinuity can hence be written as
    \begin{equation}
        \begin{split}
            \text{Disc}_{[\gamma]}I(C_2)(p)&=-N\pi^2\int_{\mathbb{R}^{D-1}}\frac{\delta(\vert\vec{k}\vert-\sqrt{r_2^2(p)})}{\vert p_0\vert\vert\vec{k}\vert}\Theta\left(-\frac{\lambda(-p^2,m_1^2,m_2^2)}{4p^2}\right)\mathrm{d}^{D-1}\vec{k}\\
            &=-N\pi^2\frac{2\pi^\frac{D-1}{2}}{\Gamma(\frac{D-1}{2})}\cdot\frac{(r_2^2(p))^\frac{D-3}{2}}{\vert p_0\vert}\\
            &=-\frac{N\pi^\frac{D+3}{2}}{2^{D-4}\Gamma(\frac{D-1}{2})}\cdot\frac{(\lambda(-p^2,m_1^2,m_2^2))^{\frac{D-3}{2}}}{(-p^2)^{\frac{D}{2}-1}}.
        \end{split}
    \end{equation}
    It should be noted that it is known by physicists that while $-(p')^2=(m_1+m_2)^2$ yields a genuine discontinuity on the principal branch, the point $-(p')^2=(m_1-m_2)^2$ (called the \textit{pseudo-threshold}) does not. Thus, we expect the intersection index $N$ to vanish in this case.
\end{example}

\subsection{Discontinuity around Points of Higher Codimension}
Aside from the somewhat pathological colinear kinematic configurations outside of $T_n$ where the determinant of the matrix containing all products of external momenta vanish, we have discussed how to compute the discontinuity along any simple loop around a point of codimension 1 in the Landau surface of a one-loop Feynman graph. This does, however, not exhaust all possible loops in the space of non-pathological external momenta minus the Landau surface. We have seen in Proposition \ref{prop:landau_decomp_1} and \ref{prop:landau_decomp_2} that the $L_{n,I}$ are codimension 1 submanifold intersecting in general position. Thus, a point $p\in L_n$ in the intersection of more than one $L_{n,I}$ is a point of higher codimension. But we also know that $T_n\backslash L_n$ is spanned by simple loops (see Proposition \ref{prop:simple_loops}), so that it is possible to decompose a loop \enquote{around} $p$ into simple loops.
\begin{proposition}\label{prop:decomposition_of_loops}
    Let $p\in L_n$ be a point of codimension $k\in\mathbb{N}^\ast$ and $\gamma:[0,1]\to T$ be a simple loop around $p$. Then, $\gamma$ can be decomposed as
    \begin{equation}
        \gamma=\gamma_1\cdots\gamma_k,
    \end{equation}
    where the $\gamma_i:[0,1]\to T$ are powers of simple loops around the codimension 1 parts of $L_n$ intersecting at $p$.
\end{proposition}
\begin{proof}
    Denote $m:=(n-1)D$. According to Proposition \ref{prop:landau_decomp_2}, the point $p$ lies in the intersection of $k$ codimension 1 parts of $L$ which intersect in general position. Thus, locally the intersection looks like
    \begin{equation}
        X_{m,k}:=\mathbb{C}^m\backslash\{(z_1,\ldots,z_D)\in\mathbb{C}^m \;\vert\; \exists i\in\{1,\ldots,k\}:z_i=0\}.
    \end{equation}
    Note that
    \begin{equation}
        \begin{array}{lcc}
            H:X_{m,k}\times[0,1] & \to & X_{m,k}\\
            ((z_1,\ldots,z_m),t) & \mapsto & (z_1,\ldots,z_k,(1-t)z_{k+1},\ldots,(1-t)z_m)
        \end{array}
    \end{equation}
    is a deformation retract from $X_{m,k}$ to (a subspace isomorphic to) $X_{k,k}$, so that
    \begin{equation}
        \pi_1(X_{m,k},(z_1,\ldots,z_m))\simeq\pi_1(X_{k,k},(z_1,\ldots,z_k))
    \end{equation}
    for any $(z_1,\ldots,z_m)\in\mathbb{C}^m$. Now, note that
    \begin{equation}
        \pi:X_{k,k}\to X_{k-1,k-1},\quad (z_1,\ldots,z_k)\mapsto(z_1,\ldots,z_{k-1})
    \end{equation}
    defines a Serre fibration with fiber
    \begin{equation}
        \pi^{-1}(\{z_1,\ldots,z_{k-1}\})=\{(z_1,\ldots,z_k)\in X_{k,k} \;\vert\; z_k\neq0\}
    \end{equation}
    homotopy equivalent to the circle $S^1$ for any $(z_1,\ldots,z_{k-1})\in X_{k-1,k-1}$. Hence, we obtain a short exact sequence
    \begin{equation}
        1\to\pi_1(S^1)\to\pi_1(X_{k,k})\to\pi_1(X_{k-1,k-1})\to1
    \end{equation}
    contained in the long exact sequence of the fibration. Thus,
    \begin{equation}
        \pi_1(X_{k,k})\simeq\pi_{X_{k-1,k-1}}\oplus\pi_1(S^1)
    \end{equation}
    and by induction we obtain
    \begin{equation}
        \pi_1(X_{k,k})\simeq\mathbb{Z}^k
    \end{equation}
    for all $k\in\mathbb{N}^\ast$, where the factors of $\mathbb{Z}$ are generated by simple loops around the $k$ codimension 1 parts of $L_n$ intersecting at $p$. This proves the claim.
\end{proof}
This allows us in principle to compute the discontinuity around any loop by Proposition \ref{prop:loop_product}: Given a loop $\gamma$ of interest, we have to find a decomposition $\gamma=\gamma_1\cdots\gamma_k$ into simple loops whose existence is guaranteed by Proposition \ref{prop:decomposition_of_loops}. Then, by applying Proposition \ref{prop:loop_product} sufficiently often, we can reduce the computation of $\text{Disc}_{[\gamma]}$ to the computation of $\text{Disc}_{[\gamma_i]}$ for all $i\in\{1,\ldots,k\}$. The latter case is covered by the general theory.

\section{More Advanced Examples}\label{sec:examples}
To illustrate the results from the last section and to check the results therein by comparing them with the common literature, we look at the two examples $C_3$ and $C_4$ in more detail. In both examples, we focus the discussion on the case where all analytic regulators $\lambda_i$ are set to 1.

\subsection{The Triangle Graph \texorpdfstring{$C_3$}{C3}}
First, we consider the triangle graph
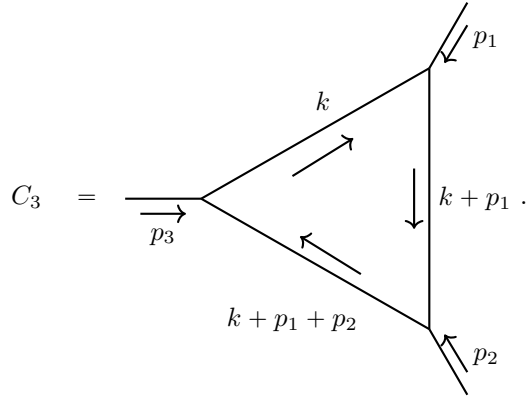
\begin{center}
    $C_3\quad=\quad$\begin{tikzpicture}[thick, baseline={([yshift=-.5ex]current bounding box.center)},vertex/.style={anchor=base,
    circle,fill=black!25,minimum size=18pt,inner sep=2pt}]
        \draw (-2,0) to (1,1.73);
        \draw (-3,0) to (-2,0);
        \draw [<-] (-2.2,-0.2) -- (-2.8,-0.2);
        \draw node at (-2.5,-0.5) {$p_3$};

        \draw [->] (-0.8,0.3) to (0,0.8);
        \draw node at (-0.4,1.3) {$k$};

        \draw (1,1.73) to (1,-1.73);
        \draw (1.5,2.6) to (1,1.73);
        \draw [<-] (1.2,1.8) -- (1.5,2.3);
        \draw node at (1.75,2.1) {$p_1$};

        \draw [<-] (0.8,-0.4) to (0.8,0.4);
        \draw node at (1.6,0) {$k+p_1$};

        \draw (1,-1.73) to (-2,0);
        \draw (1.5,-2.6) to (1,-1.73);
        \draw [<-] (1.2,-1.8) -- (1.5,-2.3);
        \draw node at (1.75,-2.1) {$p_2$};

        \draw [->] (0.1,-1.0) to (-0.7,-0.5);
        \draw node at (-0.8,-1.6) {$k+p_1+p_2$};
    \end{tikzpicture}.
\end{center}
In projective form, the corresponding Feynman integral reads
\begin{equation}
    \begin{split}
        &I(C_3)(p)\\
        =&\int_{\mathbb{R}\mathbb{P}^D}\frac{u^{2\lambda-D-1}\cdot\Omega_D}{(k^2+u^2m_1^2)^{\lambda_1}((k+up_1)^2+u^2m_2^2)^{\lambda_2}((k+u(p_1+p_2))^2+u^2m_3^2)^{\lambda_3}}.
    \end{split}
\end{equation}
Having discussed $C_2$ already, we know that there are discontinuities associated with any pair of edges put on mass-shell. The three corresponding reduced graphs are 
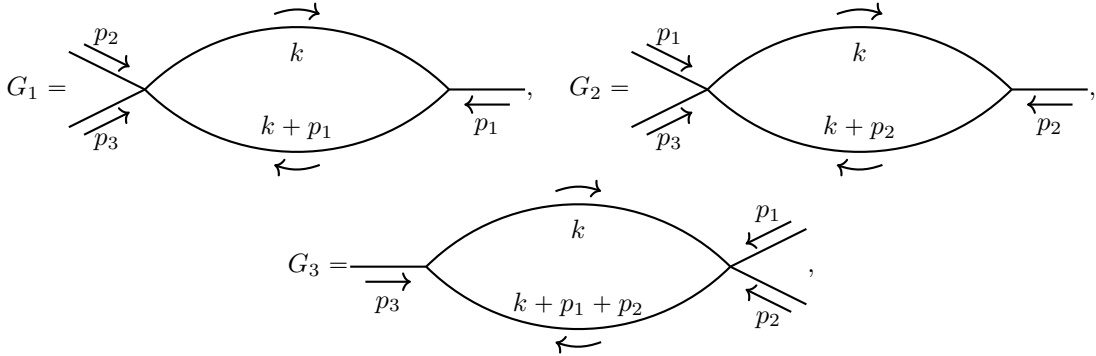
\begin{center}
    $G_1=$\begin{tikzpicture}[thick, baseline={([yshift=-.5ex]current bounding box.center)},vertex/.style={anchor=base,
    circle,fill=black!25,minimum size=18pt,inner sep=2pt}]
        \draw (-2,0) to [out=45, in=135] (2,0);
        \draw [->] (-2.8,0.6) -- (-2.2,0.3);
        \draw [->] (-2.8,-0.6) -- (-2.2,-0.3);
        \draw node at (-2.5,0.7) {$p_2$};
        \draw node at (-2.5,-0.7) {$p_3$};
        \draw (-2,0) to [out=-45, in=-135] (2,0);
        \draw [->] (-0.3,1.0) to [out=20, in=160] (0.3,1.0);
        \draw node at (0,0.5) {$k$};
        \draw [<-] (-0.3,-1.0) to [out=-20, in=-160] (0.3,-1.0);
        \draw node at (0,-0.5) {$k+p_1$};
        \draw (-3,0.5) -- (-2,0);
        \draw (-3,-0.5) -- (-2,0);
        \draw (2,0) -- (3,0);
        \draw [<-] (2.2,-0.2) -- (2.8,-0.2);
        \draw node at (2.5,-0.5) {$p_1$};
    \end{tikzpicture},
    \quad$G_2=$\begin{tikzpicture}[thick, baseline={([yshift=-.5ex]current bounding box.center)},vertex/.style={anchor=base,
    circle,fill=black!25,minimum size=18pt,inner sep=2pt}]
        \draw (-2,0) to [out=45, in=135] (2,0);
        \draw [->] (-2.8,0.6) -- (-2.2,0.3);
        \draw [->] (-2.8,-0.6) -- (-2.2,-0.3);
        \draw node at (-2.5,0.7) {$p_1$};
        \draw node at (-2.5,-0.7) {$p_3$};
        \draw (-2,0) to [out=-45, in=-135] (2,0);
        \draw [->] (-0.3,1.0) to [out=20, in=160] (0.3,1.0);
        \draw node at (0,0.5) {$k$};
        \draw [<-] (-0.3,-1.0) to [out=-20, in=-160] (0.3,-1.0);
        \draw node at (0,-0.5) {$k+p_2$};
        \draw (-3,0.5) -- (-2,0);
        \draw (-3,-0.5) -- (-2,0);
        \draw (2,0) -- (3,0);
        \draw [<-] (2.2,-0.2) -- (2.8,-0.2);
        \draw node at (2.5,-0.5) {$p_2$};
    \end{tikzpicture},
    $G_3=$\begin{tikzpicture}[thick, baseline={([yshift=-.5ex]current bounding box.center)},vertex/.style={anchor=base,
    circle,fill=black!25,minimum size=18pt,inner sep=2pt}]
        \draw (-2,0) to [out=45, in=135] (2,0);
        \draw [->] (-2.8,-0.2) -- (-2.2,-0.2);
        \draw node at (-2.5,-0.5) {$p_3$};
        \draw (-2,0) to [out=-45, in=-135] (2,0);
        \draw [->] (-0.3,1.0) to [out=20, in=160] (0.3,1.0);
        \draw node at (0,0.5) {$k$};
        \draw [<-] (-0.3,-1.0) to [out=-20, in=-160] (0.3,-1.0);
        \draw node at (0,-0.5) {$k+p_1+p_2$};
        \draw (-3,0) -- (-2,0);
        \draw (2,0) -- (3,0.5);
        \draw (2,0) -- (3,-0.5);
        \draw [->] (2.8,0.6) -- (2.2,0.3);
        \draw [->] (2.8,-0.6) -- (2.2,-0.3);
        \draw node at (2.5,0.7) {$p_1$};
        \draw node at (2.5,-0.7) {$p_2$};
    \end{tikzpicture},
\end{center}
and they have the following Landau surfaces:
\begin{equation}
    \begin{split}
        L_{2,\{3\}}&=\{(p_1,p_2)\in T_3 \;\vert\; -p_1^2=(m_1\pm m_2)^2\}\\
        L_{2,\{1\}}&=\{(p_1,p_2)\in T_3 \;\vert\; -p_2^2=(m_2\pm m_3)^2\}\\
        L_{2,\{2\}}&=\{(p_1,p_2)\in T_3 \;\vert\; -(p_1+p_2)^2=(m_1\pm m_3)^2\}
    \end{split}
\end{equation}
Due to the $S_3$-symmetry of the graph $C_3$ (with respect to the action permuting the vertices), it suffices to compute the discontinuity along one of the surfaces $L_{2,\{1\}},L_{2,\{2\}},L_{2,\{3\}}$. So let $p'\in L_{2,\{3\}}$, let $\gamma:[0,1]\to T_3\backslash L(C_3)$ be a simple loop around $p'$ and let $p=(p_1,p_2)\in T_3\backslash L(C_3)$ such that
\begin{equation}
    r_2^2(p_1)=-\frac{1}{4p_1^2}\lambda(-p_1^2,m_1^2,m_2^2)\in\mathbb{R}_+.
\end{equation}
Let us now restrict our attention to the case $\lambda_1=\lambda_2=\lambda_3=1$. Then, according to Cutkosky's Theorem \ref{thm:cutkoskys_thm}, we get
\begin{equation}
    \begin{split}
        &\text{Disc}_{[\gamma]}I(C_3)(p)\\
        &\quad=N(2\pi i)^2\int_{i\mathbb{R}\times\mathbb{R}^{D-1}}\frac{\delta(k^2+m_1^2)\delta((k+p_1)^2+m_2^2)}{(k+p_1+p_2)^2+m_3^2}\mathrm{d}^Dk\\
        &\quad=-N\int_{\mathbb{R}^{D-1}}\frac{2\pi^2}{\sqrt{\vec{k}^2+m_1^2}}\\
        &\qquad\times\sum_{\tau\in\{-1,1\}}\frac{\delta(2i\tau(p_1)_0\sqrt{\vec{k}^2+m_1^2}+p_1^2-m_1^2+m_2^2)}{2i\tau((p_1)_0+(p_2)_0)\sqrt{\vec{k}^2+m_1^2}+2\vec{k}\vec{p}_2+(p_1+p_2)^2-m_1^2+m_3^2}\mathrm{d}^{D-1}\vec{k}.
    \end{split}
\end{equation}
As in our calculations regarding $C_2$, we have
\begin{equation}
    \begin{split}
        &\delta\left(\pm2i(p_1)_0\sqrt{\vec{k}^2+m_1^2}+p_1^2-m_1^2+m_2^2\right)\\
        &\quad=\frac{\sqrt{\vec{k}^2+m_1^2}}{2\vert(p_1)_0\vert\vert\vec{k}\vert}
        \Theta\left(\pm\frac{i}{2(p_1)_0}(p_1^2-m_1^2+m_2^2)-m_1\right)\delta\left(\vert\vec{k}\vert-\sqrt{r_2^2(p_1)}\right).
    \end{split}
\end{equation}
We set
\begin{equation}
    B:=(p_1+p_2)^2-m_1^2+m_3^2-\frac{(p_1)_0+(p_2)_0}{(p_1)_0}(p_1^2-m_1^2+m_2^2)
\end{equation}
and obtain
\begin{equation}
    \begin{split}
        &\text{Disc}_{[\gamma]}I(C_3)(p)=\frac{-\pi^2N}{\vert(p_1)_0\vert\sqrt{r_2^2(p_1)}}\int_{S^{D-1}}\mathrm{d}\Omega_{D-1}\int_{-1}^1\mathrm{d}\cos(\varphi)\\
        &\quad\times\frac{\sum_{\tau\in\{-1,1\}}\Theta(\tau\frac{i}{2(p_1)_0}(p_1^2-m_1^2+m_2^2)-m_1)}{-\frac{(p_1)_0+(p_2)_0}{(p_1)_0}(p_1^2-m_1^2+m_2^2)+2\sqrt{r_2^2(p_1)}\vert\vec{p}_2\vert\cos(\varphi)+(p_1+p_2)^2-m_1^2+m_3^2}\\
        &\quad-\frac{\pi^\frac{D+3}{2}}{\Gamma(\frac{D-1}{2})\vert(p_1)_0\vert\vert\vec{p}_2\vert r_2^2(p_1)}\left(\ln(B+2\sqrt{r_2^2(p_1)}\vert\vec{p}_2\vert)-\ln(B-2\sqrt{r_2^2(p_1)}\vert\vec{p}_2\vert)\right),
    \end{split}
\end{equation}
where we again used Lemma \ref{lem:ex_c2} to evaluate the sum of the two Heaviside-functions to 1.\\
Now we discuss the case where all three edges are put on shell. We can easily compute an equation for the Landau surface by the procedure from Section \ref{sec:one_loop}: According to our recursive formula \eqref{eq:Ai}, we have
\begin{equation}
    A_1(p)=-\frac{1}{2(p_2)_1}\left((p_1+p_2)^2-m_1^2+m_3^2-\frac{(p_1)_0+(p_2)_0}{(p_1)_0}(p_1^2-m_1^2+m_2^2)\right).
\end{equation}
Note that $A_1(p)=-\frac{1}{2(p_2)_1}B$. We set
\begin{equation}
    B_1:=(p_1+p_2)^2-m_1^2+m_3^2 \quad\text{and}\quad B_2:=\frac{(p_1)_0+(p_2)_0}{(p_1)_0}(p_1^2-m_1^2+m_2^2)
\end{equation}
and compute
\begin{equation}
    r_3^2(p)=r_2^2(p)-A_1^2(p)=-\frac{\lambda(-p_1^2,m_1^2,m_2^2)}{4p_1^2}-\frac{B_1^2-2B_1B_2+B_2^2}{4(p_2)_1^2}.
\end{equation}
At this point, we could end the calculation: We have successfully reduced the description of the Landau surface $L_{3,\{1,2,3\}}$ to a single equation. However, it is very useful for physicists to be able to solve this equation for the $O(D,\mathbb{C})$-invariant products. In our case $C_3$, we do this for
\begin{equation}
    s:=-p_1^2, \quad t:=-p_2^2,\quad u:=-(p_1+p_2)^2.
\end{equation}
Due to the symmetry of the graph, it suffices to solve for one of the variables and we choose $s$ here. The solution solved for $t$ or $u$ can then be obtained by simply permuting the vertices of the graph $C_3$. To this end, we first write
\begin{equation}
    B_2^2=\frac{((p_1)_0+(p_2)_0)^2}{p_1^2}(\lambda(-p_1^2,m_1^2,m_2^2)-4m_1^2p_1^2)
\end{equation}
to obtain
\begin{equation}
    \begin{split}
        r_3^2(p)&=-\frac{1}{4p_1^2(p_2)_1^2}((p_1+p_2)^2\lambda(-p_1^2,m_1^2,m_2^2)+p_1^2B_1^2-2p_1^2B_1B_2\\
        &\quad-4p_1^2((p_1)_0+(p_2)_0)^2m_1^2)\\
        &=-\frac{1}{4p_1^2(p_2)_1^2}(-u(s+(m_1+m_2)^2)(s+(m_1-m_2)^2)\\
        &\quad-s(u+m_1^2-m_3)^2-m_1^2(s-t+u)^2\\
        &\quad¸+(u+m_1^2-m_3^2)(s-t+u)(s+m_1^2-m_2^2))\\
        &=-\frac{1}{4(p_1)_0^2(p_2)_1^2}(-m_3^2s^2+c\cdot s+d)
    \end{split}
\end{equation}
with
\begin{equation}
    c:=2m_1^2t+2m_2^2u-(t+m_2^2-m_3^2)(u+m_1^2-m_3^2)
\end{equation}
and
\begin{equation}
    d:=-m_1^2(t-u)^2-u(m_1^2-m_2^2)^2+(m_1^2-m_2^2)(-t+u)(u+m_1^2-m_3^2).
\end{equation}
We see that the denominator of $r_3^2(p)$ is a polynomial of degree 3 in $s,t,u$ which is quadratic in each of the channel variables separately. In particular, we can view it as a quadratic polynomial in $s$. This allows us to conclude that $r_3^2(p)=0$ if and only if
\begin{equation}
    \begin{split}
        s=&\frac{1}{2m_3^2}\left(c\pm\sqrt{c^2+4m_3^2d}\right)\\
        =&{}\frac{1}{2m_3^2}(-(t-m_2^2)(u-m_1^2)+m_3^2(t+u+m_1^2+m_2^2)-m_3^4\\
        {}&{}+\sqrt{m_2^4+(t-m_3^2)^2-2m_2^2(t+m_3^2)((u-m_3^2)^2-2m_1^2(u+m_3^2)+m_1^4)}).
    \end{split}
\end{equation}
This concludes the computation of the Landau surface $L_3$.\\
Turning to the computation of the discontinuity of $I(C_3)$ along simple loops around codimension 1 points in $L_{3,\{1,2,3\}}$, assume again that $\gamma_1=\gamma_2=\gamma_3=1$. Note that in this case the integral $I(C_3)(p)$ converges absolutely in $D=4$ space-time dimensions for $p$ close to Euclidean momenta according to the power counting criterion. Let $p\in L_{3,\{1,2,3\}}$ and let $\gamma:[0,1]\to T_3\backslash L_3$ be simple loop around $p$. We can compute the corresponding discontinuity using Theorem \ref{thm:cutkoskys_thm}. For the actual calculation, it is convenient to again choose $p_1$ and $p_2$ such that $(p_i)_j=0$ for $j\geq i$ which we can do by Lemma \ref{lem:orthogonal_transformation}. By our previous calculation, we have
\begin{equation}
    \begin{split}
        &\text{Disc}_{[\gamma]}(I(C_3))(p)\\
        &\quad=\frac{-i\pi^3N}{\vert(p_1)_0\vert\sqrt{r_2^2(p_1)}}\int_{S^{D-1}}\mathrm{d}\Omega_{D-1}\int_{-1}^1\mathrm{d}\cos(\varphi)\delta(2\sqrt{r_2^2(p_1)}\vert\vec{p}_2\vert\cos(\varphi)+B).
    \end{split}
\end{equation}
The integral over the $\delta$-function evaluates to $\frac{1}{2\sqrt{r_2^2(p_1)}\vert\vec{p}_2\vert}$ if and only if
\begin{equation}
    \frac{A_1(p)}{\sqrt{r_2^2(p_1)}}\in[-1,1].
\end{equation}
This is always true for any $p$ such that $r_3^2(p)=r_2^2(p_1)-A_1^2(p)>0$. Thus
\begin{equation}
    \text{Disc}_{[\gamma]}(I(C_3))(p)=\frac{-iN\pi^\frac{D+5}{2}}{\Gamma(\frac{D-1}{2})\vert(p_1)_0\vert\vert(p_2)_1\vert r_2^2(p_1)}.
\end{equation}

\subsection{The Box Graph \texorpdfstring{$C_4$}{C4}}
Finally, we consider the graph
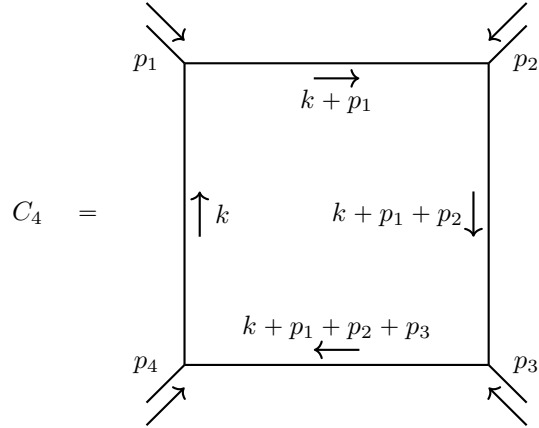
\begin{center}
    $C_4\quad=\quad$\begin{tikzpicture}[thick, baseline={([yshift=-.5ex]current bounding box.center)},vertex/.style={anchor=base,
    circle,fill=black!25,minimum size=18pt,inner sep=2pt}]
        \draw (-2,-2) to (-2,2);
        \draw (-2.5,-2.5) to (-2,-2);
        \draw [->] (-2.5,-2.8) -- (-2,-2.3);
        \draw node at (-2.5,-2) {$p_4$};

        \draw [->] (-1.8,-0.3) to (-1.8,0.3);
        \draw node at (-1.5,0) {$k$};

        \draw (-2,2) to (2,2);
        \draw (-2.5,2.5) to (-2,2);
        \draw [->] (-2.5,2.8) -- (-2,2.3);
        \draw node at (-2.5,2) {$p_1$};

        \draw [->] (-0.3,1.8) to (0.3,1.8);
        \draw node at (0,1.5) {$k+p_1$};

        \draw (2,2) to (2,-2);
        \draw (2.5,2.5) to (2,2);
        \draw [->] (2.5,2.8) -- (2,2.3);
        \draw node at (2.5,2) {$p_2$};

        \draw [->] (1.8,0.3) to (1.8,-0.3);
        \draw node at (0.8,0) {$k+p_1+p_2$};

        \draw (2,-2) to (-2,-2);
        \draw (2.5,-2.5) to (2,-2);
        \draw [->] (2.5,-2.8) -- (2,-2.3);
        \draw node at (2.5,-2) {$p_3$};
        
        \draw [->] (0.3,-1.8) to (-0.3,-1.8);
        \draw node at (0,-1.5) {$k+p_1+p_2+p_3$};
    \end{tikzpicture}
\end{center}
The corresponding Feynman integral in projective space reads
\begin{equation}
    \begin{split}
        I(C_4)(p)&=\int_{\mathbb{R}\mathbb{P}^D}\frac{u^{2\lambda-D-1}\cdot\Omega_D}{(k^2+u^2m_1^2)^{\lambda_1}((k+up_1)^2+u^2m_2^2)^{\lambda_2}}\\
        &\quad\times\frac{1}{(k+u(p_1+p_2))^2+u^2m_3^2)^{\lambda_3}((k+u(p_1+p_2+p_3))^2+u^2m_4^2)^{\lambda_4}}.
    \end{split}
\end{equation}
As before, we can obtain parts of the Landau surface by considering the reduced graphs of $C_4$. There are four graphs with one edge contracted to zero length
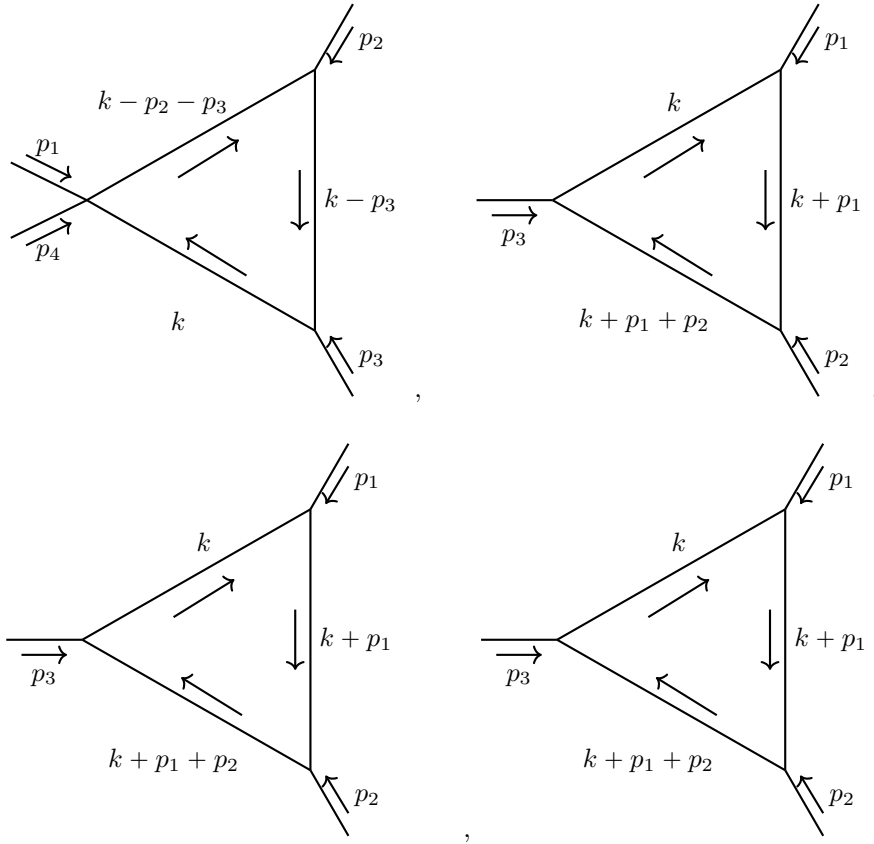
\begin{center}
    \begin{tikzpicture}[thick]
        \draw (-2,0) to (1,1.73);
        \draw (-3,0.5) -- (-2,0);
        \draw (-3,-0.5) -- (-2,0);
        \draw [->] (-2.8,0.6) -- (-2.2,0.3);
        \draw [->] (-2.8,-0.6) -- (-2.2,-0.3);
        \draw node at (-2.5,0.7) {$p_1$};
        \draw node at (-2.5,-0.7) {$p_4$};

        \draw [->] (-0.8,0.3) to (0,0.8);
        \draw node at (-1,1.3) {$k-p_2-p_3$};

        \draw (1,1.73) to (1,-1.73);
        \draw (1.5,2.6) to (1,1.73);
        \draw [<-] (1.2,1.8) -- (1.5,2.3);
        \draw node at (1.75,2.1) {$p_2$};

        \draw [<-] (0.8,-0.4) to (0.8,0.4);
        \draw node at (1.6,0) {$k-p_3$};

        \draw (1,-1.73) to (-2,0);
        \draw (1.5,-2.6) to (1,-1.73);
        \draw [<-] (1.2,-1.8) -- (1.5,-2.3);
        \draw node at (1.75,-2.1) {$p_3$};

        \draw [->] (0.1,-1.0) to (-0.7,-0.5);
        \draw node at (-0.8,-1.6) {$k$};
    \end{tikzpicture}
    ,\qquad
    \begin{tikzpicture}[thick]
        \draw (-2,0) to (1,1.73);
        \draw (-3,0) to (-2,0);
        \draw [<-] (-2.2,-0.2) -- (-2.8,-0.2);
        \draw node at (-2.5,-0.5) {$p_3$};

        \draw [->] (-0.8,0.3) to (0,0.8);
        \draw node at (-0.4,1.3) {$k$};

        \draw (1,1.73) to (1,-1.73);
        \draw (1.5,2.6) to (1,1.73);
        \draw [<-] (1.2,1.8) -- (1.5,2.3);
        \draw node at (1.75,2.1) {$p_1$};

        \draw [<-] (0.8,-0.4) to (0.8,0.4);
        \draw node at (1.6,0) {$k+p_1$};

        \draw (1,-1.73) to (-2,0);
        \draw (1.5,-2.6) to (1,-1.73);
        \draw [<-] (1.2,-1.8) -- (1.5,-2.3);
        \draw node at (1.75,-2.1) {$p_2$};

        \draw [->] (0.1,-1.0) to (-0.7,-0.5);
        \draw node at (-0.8,-1.6) {$k+p_1+p_2$};
    \end{tikzpicture},\vspace{0.5cm}
    \begin{tikzpicture}[thick]
        \draw (-2,0) to (1,1.73);
        \draw (-3,0) to (-2,0);
        \draw [<-] (-2.2,-0.2) -- (-2.8,-0.2);
        \draw node at (-2.5,-0.5) {$p_3$};

        \draw [->] (-0.8,0.3) to (0,0.8);
        \draw node at (-0.4,1.3) {$k$};

        \draw (1,1.73) to (1,-1.73);
        \draw (1.5,2.6) to (1,1.73);
        \draw [<-] (1.2,1.8) -- (1.5,2.3);
        \draw node at (1.75,2.1) {$p_1$};

        \draw [<-] (0.8,-0.4) to (0.8,0.4);
        \draw node at (1.6,0) {$k+p_1$};

        \draw (1,-1.73) to (-2,0);
        \draw (1.5,-2.6) to (1,-1.73);
        \draw [<-] (1.2,-1.8) -- (1.5,-2.3);
        \draw node at (1.75,-2.1) {$p_2$};

        \draw [->] (0.1,-1.0) to (-0.7,-0.5);
        \draw node at (-0.8,-1.6) {$k+p_1+p_2$};
    \end{tikzpicture}
    \qquad,
    \begin{tikzpicture}[thick]
        \draw (-2,0) to (1,1.73);
        \draw (-3,0) to (-2,0);
        \draw [<-] (-2.2,-0.2) -- (-2.8,-0.2);
        \draw node at (-2.5,-0.5) {$p_3$};

        \draw [->] (-0.8,0.3) to (0,0.8);
        \draw node at (-0.4,1.3) {$k$};

        \draw (1,1.73) to (1,-1.73);
        \draw (1.5,2.6) to (1,1.73);
        \draw [<-] (1.2,1.8) -- (1.5,2.3);
        \draw node at (1.75,2.1) {$p_1$};

        \draw [<-] (0.8,-0.4) to (0.8,0.4);
        \draw node at (1.6,0) {$k+p_1$};

        \draw (1,-1.73) to (-2,0);
        \draw (1.5,-2.6) to (1,-1.73);
        \draw [<-] (1.2,-1.8) -- (1.5,-2.3);
        \draw node at (1.75,-2.1) {$p_2$};

        \draw [->] (0.1,-1.0) to (-0.7,-0.5);
        \draw node at (-0.8,-1.6) {$k+p_1+p_2$};
    \end{tikzpicture},
\end{center}
and 6 graphs with two edges contracted to zero length
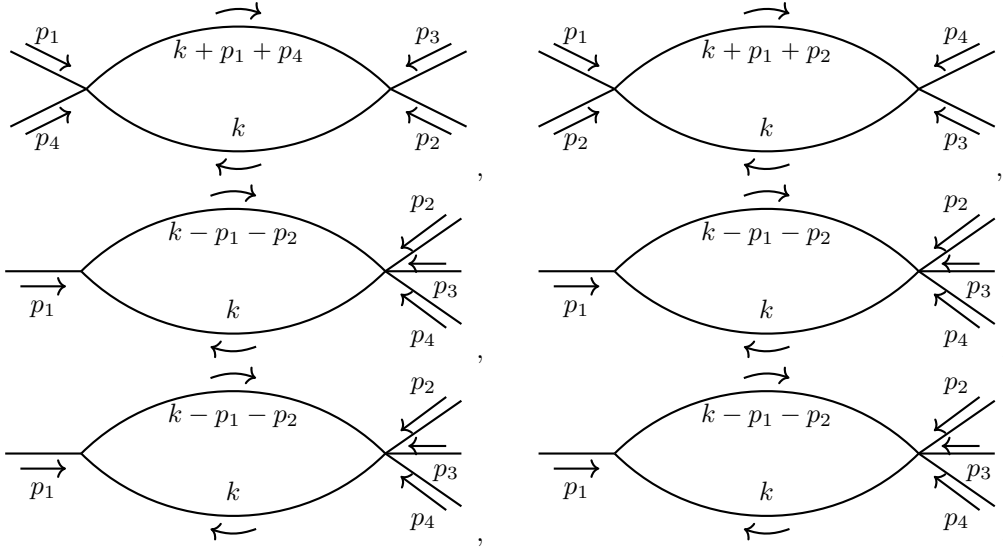
\begin{center}
    \begin{tikzpicture}[thick]
        \draw (-2,0) to [out=45, in=135] (2,0);
        \draw (-2,0) to [out=-45, in=-135] (2,0);
        \draw [->] (-0.3,1.0) to [out=20, in=160] (0.3,1.0);
        \draw node at (0,0.5) {$k+p_1+p_4$};
        \draw [<-] (-0.3,-1.0) to [out=-20, in=-160] (0.3,-1.0);
        \draw node at (0,-0.5) {$k$};
        
        \draw (-3,0.5) -- (-2,0);
        \draw (-3,-0.5) -- (-2,0);
        \draw [->] (-2.8,0.6) -- (-2.2,0.3);
        \draw [->] (-2.8,-0.6) -- (-2.2,-0.3);
        \draw node at (-2.5,0.7) {$p_1$};
        \draw node at (-2.5,-0.7) {$p_4$};
        
        \draw (2,0) -- (3,0.5);
        \draw (2,0) -- (3,-0.5);
        
        \draw [<-] (2.2,-0.3) -- (2.8,-0.6);
        \draw [<-] (2.2,0.3) -- (2.8,0.6);
        \draw node at (2.5,-0.7) {$p_2$};
        \draw node at (2.5,0.7) {$p_3$};
    \end{tikzpicture}
    ,\qquad
    \begin{tikzpicture}[thick]
        \draw (-2,0) to [out=45, in=135] (2,0);
        \draw (-2,0) to [out=-45, in=-135] (2,0);
        \draw [->] (-0.3,1.0) to [out=20, in=160] (0.3,1.0);
        \draw node at (0,0.5) {$k+p_1+p_2$};
        \draw [<-] (-0.3,-1.0) to [out=-20, in=-160] (0.3,-1.0);
        \draw node at (0,-0.5) {$k$};
        
        \draw (-3,0.5) -- (-2,0);
        \draw (-3,-0.5) -- (-2,0);
        \draw [->] (-2.8,0.6) -- (-2.2,0.3);
        \draw [->] (-2.8,-0.6) -- (-2.2,-0.3);
        \draw node at (-2.5,0.7) {$p_1$};
        \draw node at (-2.5,-0.7) {$p_2$};
        
        \draw (2,0) -- (3,0.5);
        \draw (2,0) -- (3,-0.5);
        
        \draw [<-] (2.2,-0.3) -- (2.8,-0.6);
        \draw [<-] (2.2,0.3) -- (2.8,0.6);
        \draw node at (2.5,-0.7) {$p_3$};
        \draw node at (2.5,0.7) {$p_4$};
    \end{tikzpicture},
    \begin{tikzpicture}[thick]
        \draw (-2,0) to [out=45, in=135] (2,0);
        \draw [->] (-2.8,-0.2) -- (-2.2,-0.2);
        \draw node at (-2.5,-0.5) {$p_1$};
        \draw (-2,0) to [out=-45, in=-135] (2,0);
        \draw [->] (-0.3,1.0) to [out=20, in=160] (0.3,1.0);
        \draw node at (0,0.5) {$k-p_1-p_2$};
        \draw [<-] (-0.3,-1.0) to [out=-20, in=-160] (0.3,-1.0);
        \draw node at (0,-0.5) {$k$};
        \draw (-3,0) -- (-2,0);
        \draw (2,0) -- (3,-0.7);
        \draw (2,0) -- (3,0);
        \draw (2,0) -- (3,0.7);
        \draw [->] (2.8,0.75) -- (2.2,0.3);
        \draw [->] (2.8,-0.75) -- (2.2,-0.3);
        \draw [->] (2.8,0.1) -- (2.3,0.1);
        \draw node at (2.5,0.9) {$p_2$};
        \draw node at (2.8,-0.25) {$p_3$};
        \draw node at (2.5,-0.9) {$p_4$};
    \end{tikzpicture}
    ,\qquad
    \begin{tikzpicture}[thick]
        \draw (-2,0) to [out=45, in=135] (2,0);
        \draw [->] (-2.8,-0.2) -- (-2.2,-0.2);
        \draw node at (-2.5,-0.5) {$p_1$};
        \draw (-2,0) to [out=-45, in=-135] (2,0);
        \draw [->] (-0.3,1.0) to [out=20, in=160] (0.3,1.0);
        \draw node at (0,0.5) {$k-p_1-p_2$};
        \draw [<-] (-0.3,-1.0) to [out=-20, in=-160] (0.3,-1.0);
        \draw node at (0,-0.5) {$k$};
        \draw (-3,0) -- (-2,0);
        \draw (2,0) -- (3,-0.7);
        \draw (2,0) -- (3,0);
        \draw (2,0) -- (3,0.7);
        \draw [->] (2.8,0.75) -- (2.2,0.3);
        \draw [->] (2.8,-0.75) -- (2.2,-0.3);
        \draw [->] (2.8,0.1) -- (2.3,0.1);
        \draw node at (2.5,0.9) {$p_2$};
        \draw node at (2.8,-0.25) {$p_3$};
        \draw node at (2.5,-0.9) {$p_4$};
    \end{tikzpicture},
    \begin{tikzpicture}[thick]
        \draw (-2,0) to [out=45, in=135] (2,0);
        \draw [->] (-2.8,-0.2) -- (-2.2,-0.2);
        \draw node at (-2.5,-0.5) {$p_1$};
        \draw (-2,0) to [out=-45, in=-135] (2,0);
        \draw [->] (-0.3,1.0) to [out=20, in=160] (0.3,1.0);
        \draw node at (0,0.5) {$k-p_1-p_2$};
        \draw [<-] (-0.3,-1.0) to [out=-20, in=-160] (0.3,-1.0);
        \draw node at (0,-0.5) {$k$};
        \draw (-3,0) -- (-2,0);
        \draw (2,0) -- (3,-0.7);
        \draw (2,0) -- (3,0);
        \draw (2,0) -- (3,0.7);
        \draw [->] (2.8,0.75) -- (2.2,0.3);
        \draw [->] (2.8,-0.75) -- (2.2,-0.3);
        \draw [->] (2.8,0.1) -- (2.3,0.1);
        \draw node at (2.5,0.9) {$p_2$};
        \draw node at (2.8,-0.25) {$p_3$};
        \draw node at (2.5,-0.9) {$p_4$};
    \end{tikzpicture}
    ,\qquad
    \begin{tikzpicture}[thick]
        \draw (-2,0) to [out=45, in=135] (2,0);
        \draw [->] (-2.8,-0.2) -- (-2.2,-0.2);
        \draw node at (-2.5,-0.5) {$p_1$};
        \draw (-2,0) to [out=-45, in=-135] (2,0);
        \draw [->] (-0.3,1.0) to [out=20, in=160] (0.3,1.0);
        \draw node at (0,0.5) {$k-p_1-p_2$};
        \draw [<-] (-0.3,-1.0) to [out=-20, in=-160] (0.3,-1.0);
        \draw node at (0,-0.5) {$k$};
        \draw (-3,0) -- (-2,0);
        \draw (2,0) -- (3,-0.7);
        \draw (2,0) -- (3,0);
        \draw (2,0) -- (3,0.7);
        \draw [->] (2.8,0.75) -- (2.2,0.3);
        \draw [->] (2.8,-0.75) -- (2.2,-0.3);
        \draw [->] (2.8,0.1) -- (2.3,0.1);
        \draw node at (2.5,0.9) {$p_2$};
        \draw node at (2.8,-0.25) {$p_3$};
        \draw node at (2.5,-0.9) {$p_4$};
    \end{tikzpicture}.
\end{center}
Using our calculations for $C_2$, we immediately obtain
\begin{equation}
    \begin{split}
        L_{4,\{1,2\}}\quad&=\quad\{p\in T_4 \;\vert\; -p_3^2=(m_3\pm m_4)^2\},\\
        L_{4,\{1,3\}}\quad&=\quad\{p\in T_4 \;\vert\; -(p_1+p_2)^2=(m_2\pm m_4)^2\},\\
        L_{4,\{1,4\}}\quad&=\quad\{p\in T_4 \;\vert\; -p_2^2=(m_2\pm m_3)^2\},\\
        L_{4,\{2,3\}}\quad&=\quad\{p\in T_4 \;\vert\; -(p_1+p_2+p_3)^2=(m_1\pm m_4)^2\},\\
        L_{4,\{2,4\}}\quad&=\quad\{p\in T_4 \;\vert\; -(p_2+p_3)^2=(m_1\pm m_3)^2\},\\
        L_{4,\{3,4\}}\quad&=\quad\{p\in T_4 \;\vert\; -p_1^2=(m_1\pm m_2)^2\}.
    \end{split}
\end{equation}
Similar to our calculations for $C_3$, we obtain
\begin{equation}
    \begin{split}
        L_{4,\{1,2,3\}}&=\{p\in T_4 \;\vert\; r_3^2(p_1,p_2,p_3+p_4)=0\}-\bigcup_{\substack{I\subset\{1,2,3,4\} \\ \vert I\vert=2}}L_{4,I},\\
        L_{4,\{1,2,4\}}&=\{p\in T_4 \;\vert\; r_3^2(p_1,p_2+p_3,p_4)=0\}-\bigcup_{\substack{I\subset\{1,2,3,4\} \\ \vert I\vert=2}}L_{4,I},\\
        L_{4,\{1,3,4\}}&=\{p\in T_4 \;\vert\; r_3^2(p_1+p_2,p_3,p_4)=0\}-\bigcup_{\substack{I\subset\{1,2,3,4\} \\ \vert I\vert=2}}L_{4,I},\\
        L_{4,\{2,3,4\}}&=\{p\in T_4 \;\vert\; r_3^2(p_1+p_4,p_2,p_3)=0\}-\bigcup_{\substack{I\subset\{1,2,3,4\} \\ \vert I\vert=2}}L_{4,I}.
    \end{split}
\end{equation}
Now, we compute $L_{4,\{1,2,3,4\}}$: We have
\begin{equation}
    r_4^2(p)=r_3^2(p)-A_2^2(p)
\end{equation}
with
\begin{equation}
    \begin{split}
        A_2(p)=-\frac{1}{2(p_3)_2}(&A_0(p)(p_1+p_2+p_3)_0+A_1(p)(p_1+p_2+p_3)_1\\
        +&(p_1+p_2+p_3)^2-m_1^2+m_4^2).
    \end{split}
\end{equation}
Similarly to our example $C_3$ above, we have reduced the task of computing $L_{4,\{1,2,3,4\}}$ to solving a single equation: $r_4^2(p)=0$. Again it is useful to solve this equation for the channel variables. We set
\begin{equation}
    s:=p_1^2,\; t:=p_2^2,\; u:=p_3^2,\; v:=(p_1+p_2)^2,\; w:=(p_1+p_3)^2,\; x:=(p_2+p_3)^2.
\end{equation}
However, in this case the task is a bit more tricky and we have to distinguish some cases. While calculating $C_3$, it was enough to solve for one of the channel variables due to the $S_3$-symmetry of the underlying graph. Since $C_4$ does not enjoy a corresponding $S_4$-symmetry, we have to solve for say $s$, $v$ and $w$ to be able to cover all results. The nominator of $r_4^2(p)$ is again quadratic in each channel variable separately. Hence, we have
\begin{equation}
    y=\frac{1}{2a_y}\Bigg(-b_y\pm\sqrt{b_y^2-4a_yc_y}\Bigg)
\end{equation}
for all $y\in\{s,t,u,v,w,x\}$ where the coefficients $a_y,b_y,c_y$ can be read off from $r_4^2(p)$ directly. But $r_4^2(p)$ is a pretty messy expression so that we do not perform the calculation by hand. Instead we used a Mathematica program. The results obtained by solving $r_4^2(p)=0$ for one of the channel variables are equally messy and might not seem particularly useful. For completeness, we include them anyway. For $s$, we obtain
\begin{align}
    \begin{autobreak}
        a_s=
        \frac{1}{16}(m_2^4
        -2m_2^2 \left(2m_3^2-m_4^2+t+u\right)
        +4m_3^4
        -4m_3^2 \left(m_4^2+t+u-x\right)
        +\left(-m_4^2+t+u\right)^2)
    \end{autobreak}
\end{align}
\begin{align}
    \begin{autobreak}
        b_s=
        \frac{1}{8}(m_2^2(m_3^2 (-t-3 u+v+3 w+2 x)
        +m_4^2 (2 u-v-w-x)
        -2 t^2
        +t (-4 u+v+2 (w+x))
        -2 u^2
        +u (4 v+w+x)-v x)
        +m_3^2(2 m_3^2 (t+u-v-w-x)
        +2 x (2 t+2 u-v-w)
        -(3 t+u) (t+u-v-w)-2 x^2)
        +m_4^2 \left(m_3^2 (-t-3 u+3 v+w+2 x)+t (-2 v+w+x)+v (u+x)\right)
        +m_2^4 (t+2 u-w-x)
        +m_1^2(m_2^2(-t-3 u+x)
        +m_4^2 (3 t+u-x)
        +(t-u) \left(-2 m_3^2+t+u-x\right))
        -m_4^4 (t+v)
        -x \left(t^2+t (u-v)+u v\right)
        +t (t+u) (t+u-v-w))
    \end{autobreak}
\end{align}
\begin{align}
    \begin{autobreak}
        c_s=
        \frac{1}{16}(m_2^4 \left(t^2+4 t u-2 t (w+x)+4 u^2-4 u (v+w+x)+(w+x)^2\right)
        +m_1^4 \left(t^2-2 t (u+x)+(u-x)^2\right)
        +2 m_1^2(-m_2^2 \left(t^2+t (u-w-2 x)+2 u^2-u (2 v+w+3 x)+x (w+x)\right)
        +m_3^2 \left(-t^2+t (v+w+x)+u^2-u (v+w+x)+x (w-v)\right)
        +m_4^2 \left(t^2+t (3 u-v-2 w-x)+v (x-u)\right)+t^3-t^2 (v+w+2 x)
        +t \left(-u^2+u (v+w-2 x)+x (2 v+w+x)\right)+v x (u-x))
        -2 m_2^2(m_4^2 \left(t^2+t (2 u-v-w-x)+v (x-w)\right)
        +m_3^2 (x (t-u-v+w)-(t-2 u+w) (t+u-v-w))
        +t^3+t^2 (3 u-v-2 (w+x))
        +t \left(2 u^2-4 u v-3 u (w+x)+v (w+2 x)+(w+x)^2\right)
        +v \left(-2 u^2+2 u (v+w+x)-x (w+x)\right))
        +\left(m_3^2 (-t-u+v+w)+m_4^2 (t-v)+t (t+u-v-w)+x (v-t)\right){}^2)
    \end{autobreak}
\end{align}
For $v$, we obtain
\begin{equation}
    a_v=\frac{1}{16} \left(\left(-m_3^2+m_4^2+t-x\right){}^2-4 m_2^2 u\right)
\end{equation}
\begin{align}
    \begin{autobreak}
        b_v=
        \frac{1}{8}(m_2^2(m_3^2 (s-t+2 u-w+x)
        +m_4^2 (-s+t+w-x)
        -x (s+2 (t+u)-w)
        +s t
        +4 s u
        +t^2
        +4 t u
        -t w
        +2 u^2
        -2 u w
        +x^2)
        +m_4^2 (m_3^2 (3 s+2 t+u-w)
        +s (-2 t+u+x)
        +t (-2 t-u+w+2 x))
        +m_3^2 (m_3^2 (-(2 s+t+u-w))
        +s (3 t+u-2 x)
        +x (-2 t-u+w)
        +2 t (t+u-w))
        -m_4^4 (s+t)
        +m_1^2 (m_3^2 (t-u-x)
        -m_4^2 (t+u-x)
        +2 m_2^2 u
        -t^2
        +t (u+2 x)
        +x (u-x))
        -2 m_2^4 u
        +s x (t-u)
        -s t (t+u)
        -t (t-x) (t+u-w-x))
    \end{autobreak}
\end{align}
\begin{align}
    \begin{autobreak}
        c_v=
        \frac{1}{16}(-2 m_2^2(m_3^2(2 s^2
        +s (t+3 u-3 w-2 x)
        -t^2
        +t (u+x)
        +(u-w) (2 u-w-x))
        +(s+t+2 u-w-x) \left(m_4^2 (t-s)+s (t+u)+t (t+u-w-x)\right))
        -2 m_4^2(m_3^2 \left(2 s^2+s (t+3 u-w-2 x)+t (t+u-w)\right)
        +(s-t) (s (t+u)+t (t+u-w-x)))
        -4 m_3^2 s^2 t
        -4 m_3^2 s^2 u
        +4 m_3^2 s^2 x
        +4 m_3^4 s^2
        +2 m_1^2(m_2^2(-(s (t+3 u-x)
        +t^2
        +t (u-w-2 x)
        +(u-x) (2 u-w-x)))
        +m_3^2 (x (t-u+w)-(t-u) (2 s+t+u-w))
        +m_4^2 (s (3 t+u-x)+t (t+3 u-2 w-x))
        +s (t-u) (t+u-x)
        +t \left(x (w-2 (t+u))+(t-u) (t+u-w)+x^2\right))
        -6 m_3^2 s t^2
        +m_2^4 (s+t+2 u-w-x)^2
        -8 m_3^2 s t u
        +6 m_3^2 s t w
        +8 m_3^2 s t x
        +4 m_3^4 s t
        +m_4^4 (s-t)^2
        -2 m_3^2 s u^2
        +2 m_3^2 s u w
        +8 m_3^2 s u x
        +4 m_3^4 s u
        -4 m_3^2 s w x
        -4 m_3^4 s w
        -4 m_3^2 s x^2
        -4 m_3^4 s x
        -2 m_3^2 t^3
        +m_1^4 \left(t^2-2 t (u+x)+(u-x)^2\right)
        -4 m_3^2 t^2 u
        +4 m_3^2 t^2 w
        +2 m_3^2 t^2 x
        +m_3^4 t^2
        -2 m_3^2 t u^2
        +4 m_3^2 t u w
        +2 m_3^2 t u x
        +2 m_3^4 t u
        -2 m_3^2 t w^2
        -2 m_3^2 t w x
        -2 m_3^4 t w
        +m_3^4 u^2
        -2 m_3^4 u w
        +m_3^4 w^2
        +(s (t+u)+t (t+u-w-x))^2)
    \end{autobreak}
\end{align}
Finally, for $w$ we obtain
\begin{equation}
    a_w=\frac{1}{16} \left(-2 m_2^2 \left(m_3^2+t\right)+\left(t-m_3^2\right){}^2+m_2^4\right)
\end{equation}
\begin{align}
    \begin{autobreak}
        b_w=
        \frac{1}{8}(m_2^2 (m_4^2 (-s+t+v)
        +m_3^2 (3 s+3 u-v-x)
        +2 s t
        +s u
        +2 t^2
        +3 t u
        -t v
        -2 t x
        -2 u v
        +v x)
        +m_3^2 (m_3^2 (-(2 s+t+u-v))
        +s (3 t+u-2 x)
        +2 t (t+u-v)
        +x (v-t))
        +m_2^4 (-(s+t+2 u-x))
        +m_4^2 \left(m_3^2 (s+t-v)+t (s-t+v)\right)
        +m_1^2 (m_2^2 (t+u-x)
        +m_3^2 (t-u+x)
        +t \left(-2 m_4^2-t+u+x\right))
        -t \left(s (t+u)+t^2+t (u-v-x)+v x\right))
    \end{autobreak}
\end{align}
\begin{align}
    \begin{autobreak}
        c_w=
        \frac{1}{16}(\left(t^2-2 (u+x) t+(u-x)^2\right) m_1^4
        +2((t-v) x^2
        +(s (u-t)-2 t (t+u)+(2 t+u) v) x
        -\left(t^2+s t+u t+2 u^2+x^2+3 s u-2 u v-(s+2 t+3 u) x\right) m_2^2
        -((t-u) (2 s+t+u-v)+(-t+u+v) x) m_3^2
        +\left(t^2+3 s t+3 u t-v t+s u-u v-(s+t-v) x\right) m_4^2
        +(t-u) ((s+t) (t+u)-t v)) m_1^2
        +\left(s^2+2 t s+4 u s+t^2+4 u^2+x^2+4 t u-4 u v-2 (s+t+2 u) x\right) m_2^4
        +4 s^2 m_3^4
        +t^2 m_3^4
        +u^2 m_3^4
        +v^2 m_3^4
        +4 s t m_3^4
        +4 s u m_3^4
        +2 t u m_3^4
        -4 s v m_3^4
        -2 t v m_3^4
        -2 u v m_3^4
        -4 s x m_3^4
        +\left(s^2-2 (t+v) s+(t-v)^2\right) m_4^4
        +((s+t) (t+u)-t v)^2
        +(t-v)^2 x^2
        -2 t^3 m_3^2
        -6 s t^2 m_3^2
        -2 s u^2 m_3^2
        -2 t u^2 m_3^2
        -2 t v^2 m_3^2
        -4 s x^2 m_3^2
        -4 s^2 t m_3^2
        -4 s^2 u m_3^2
        -4 t^2 u m_3^2
        -8 s t u m_3^2
        +4 t^2 v m_3^2
        +6 s t v m_3^2
        +2 s u v m_3^2
        +4 t u v m_3^2
        +4 s^2 x m_3^2
        +2 t^2 x m_3^2
        +2 v^2 x m_3^2
        +8 s t x m_3^2
        +8 s u x m_3^2
        +2 t u x m_3^2
        -4 s v x m_3^2
        -4 t v x m_3^2
        -2 u v x m_3^2
        -2((t+u) s^2
        +(2 t-u) v s
        -(t+v) x s
        +\left(2 s^2+(t+3 u-3 v-2 x) s+(t-v) (t+u-v)\right) m_3^2
        +(t-v) ((t-v) x-t (t+u-v))) m_4^2
        -2 \left(t (t-v) (t+u-v)+s \left(t^2+(u-v) t+u v\right)\right) x
        -2 m_2^2(t^3
        +(3 u-v-2 x) t^2
        +(2 u-x) (u-2 v-x) t
        +(2 s^2
        +(t+3 u-v-2 x) s
        -t^2
        +2 u (u-v)
        -(u+v) x
        +t (u+v+x)) m_3^2
        +((s-t+v) x-(s-t) (s+t+2 u-v)) m_4^2
        +s^2 (t+u)
        +v \left(-2 u^2+2 (v+x) u-x^2\right)
        +s \left(2 t^2+(4 u-v-2 x) t+2 u^2+v x-u (4 v+x)\right)))
    \end{autobreak}
\end{align}
These results agree with the results of \cite{landaudisc}, which cover the case $m_1=\cdots=m_4$ and $s=t=u=(p_1+p_2+p_3)^2$.\footnote{For easier comparison: In \cite{landaudisc}, our variables $v$ and $x$ are called $s$ and $t$ respectively.}

\section{Conclusion and Outlook}\label{sec:conclusion}
We have seen that the techniques to treat singular integrals developed in works such as \cite{app-iso} and \cite{pham} can be applied to one-loop Feynman integrals to rigorously state and prove Cutkosky's Theorem on the discontinuity of Feynman integrals on the principal branch at Minkowski momenta. This settles a claim appearing regularly in the literature (see for example \cite{cutkosky} or \cite{abreu}) in this particular case, up to the computation of the relevant intersection index. The latter will be done in a follow-up paper.\\
Of course the goal is to eventually prove Cutkosky's Theorem in the general multi-loop case. There seems to be no straightforward way to apply the program to graphs with more than loop. Problems occur already at the first stage of compactification since the naive compactification via projective space does not work. What one could do instead is consider Feynman integrals as iterated integrals (one iteration for each loop-momentum) and then apply the generalization of the techniques in \cite{pham} to the case where the integrand itself is also ramified. This is a non-trivial task, however, and future research has to show if this is a valid route.

\section*{Acknowledgement}
I would like to express my gratitude toward everyone who supported me during the time this paper came to be, professionally or otherwise. In particular, I thank Dirk Kreimer, Marko Berghoff, my family, Nora Tabea Sibert and Christian Hammermeister.

\end{document}